\documentclass[11pt,a4paper]{article} 
\usepackage[utf8]{inputenc}
\usepackage[english]{babel} 
\usepackage{amsmath}
\usepackage{amssymb}
\usepackage{amsthm}
\usepackage{mathrsfs}
\usepackage{graphicx}
\usepackage{mathtools}
\usepackage{tikz}
\usepackage{xcolor}
\usepackage{ulem}
\usepackage{caption}
\usepackage[colorlinks=true,linkcolor=red]{hyperref}
\usepackage{tabularx}
\usepackage{authblk}
\usepackage[a4paper, top=3.2cm, bottom=4.2cm, left=3.2cm, right=3.2cm]{geometry}
\usepackage{graphicx}
\usepackage{hyperref}

\title{Nonradial perturbations of static charged wormholes}

\begin{document}

\author[1]{Jose Luis Blázquez-Salcedo\thanks{jlblaz01@ucm.es}}
\author[1]{Luis Manuel González-Romero\thanks{mgromero@ucm.es}}
\author[1]{Fech Scen Khoo\thanks{fkhoo@ucm.es}}
\author[2]{Jutta Kunz\thanks{jutta.kunz@uni-oldenburg.de}}
\author[1]{Pablo Navarro Moreno\thanks{panava03@ucm.es}}
\affil[1]{Departamento de Física Teórica and IPARCOS, Facultad de Ciencias Físicas, Universidad Complutense de Madrid, 28040 Madrid, Spain}
\affil[2]{Institut für Physik, Universität Oldenburg, Postfach 2503, D-26111 Oldenburg, Germany}

\date{\today}

\maketitle

\begin{abstract}
We investigate the nonradial quasinormal-mode spectrum of static charged Ellis--Bronnikov wormholes in Einstein--Maxwell theory minimally coupled to a phantom scalar field. 
The background solutions are known in closed form and comprise three classes: subcritical, critical and supercritical, which all approach the extremal Reissner--Nordstr\"om geometry at the boundary of their domain of existence. 
We derive the linear perturbation equations for axial and polar sectors, including the coupled gravitational, electromagnetic and phantom-scalar degrees of freedom, and compute the corresponding spectra by means of a Chebyshev spectral method. 
The uncharged limit reproduces the known Ellis--Bronnikov spectrum and exhibits the expected electromagnetic isospectrality. 
For charged configurations we track the axial and polar branches across the three families of solutions and identify the effect of the charge on the damping times and oscillation frequencies. 
In particular, we find that charge can substantially reduce damping rates as the extremal Reissner--Nordstr\"om limit is approached. 
We also uncover a nonradial polar instability, most clearly visible in the fundamental $l=2$ branch for sufficiently large wormhole masses. 
This instability is distinct from the familiar radial Ellis--Bronnikov instability and shows that the nonradial sector imposes additional constraints on the dynamical viability of charged wormholes.
\end{abstract}

\section{Introduction}
\label{sec:Introduction}

Although still hypothetical, wormholes are considered interesting black hole mimickers\cite{Damour:2007ap} whose properties are widely studied.
Of particular relevance are, of course, their observational signatures.
Thus, gravitational lensing by wormholes has been a topic studied for decades
\cite{Cramer:1994qj,Safonova:2001vz,Perlick:2003vg,Nandi:2006ds,Abe:2010ap,Toki:2011zu,Nakajima:2012pu,Tsukamoto:2012xs,Kuhfittig:2013hva,Bambi:2013nla,Takahashi:2013jqa,Tsukamoto:2016zdu}, and in more recent years it has focused on the investigation of the shadows of wormholes \cite{Bambi:2013nla,Nedkova:2013msa,Ohgami:2015nra,Shaikh:2018kfv,Gyulchev:2018fmd,Bouhmadi-Lopez:2021zwt,Guerrero:2022qkh,Huang:2023yqd,Takahashi:2026hfp}, 
which makes the study of accretion disks around wormholes and the emission of electromagnetic radiation associated with quasi-periodic oscillations compelling as well
\cite{Harko:2008vy,Harko:2009xf,Bambi:2013jda,Zhou:2016koy,Lamy:2018zvj,Deligianni:2021ecz,Deligianni:2021hwt,Gjorgjieski:2025rjy}.
Another pertinent topic yielding observational signatures is the study of gravitational radiation from wormholes \cite{Konoplya:2005et,Kim:2008zzj,Konoplya:2010kv,Konoplya:2016hmd,Bueno:2017hyj,Volkel:2018hwb,Aneesh:2018hlp,Konoplya:2018ala,Blazquez-Salcedo:2018ipc,Konoplya:2019hml,Churilova:2019qph,Jusufi:2020mmy,Bronnikov:2021liv,Gonzalez:2022ote,Azad:2022qqn}.
This is also the field of the present study.

In general relativity (GR) the Ellis-Bronnikov (EB) wormholes represent static, spherically symmetric solutions that are known in closed form and owe their existence to the presence of a phantom scalar field \cite{Ellis:1973yv,Bronnikov:1973fh}.
The radial instability of static EB wormholes is well-known \cite{Shinkai:2002gv,Gonzalez:2008wd,Gonzalez:2008xk,Cremona:2018wkj,Xu:2025jad}, and the study of their quasinormal mode (QNM) spectrum revealed interesting features, like a threefold isospectrality of the massless wormholes \cite{Azad:2022qqn}.

Static EB wormholes have been generalized to include rotation, but closed form solutions exist only for slowly rotating EB wormholes \cite{Kashargin:2007mm,Kashargin:2008pk}.
Rapidly rotating EB wormholes are known numerically, though \cite{Kleihaus:2014dla,Chew:2016epf,Volkov:2021blw}.
The attenuation of the radial instability of wormholes in the presence of rotation conjectured earlier \cite{Matos:2010pcd} was addressed in recent years, together with the QNM spectrum of rotating wormholes \cite{Azad:2023iju,Azad:2024axu,Khoo:2024yeh,Azad:2025kay}.
However, the fate of the instability for rather fast rotation has not yet been clarified, since the background solutions were not known in closed form, yielding accumulating inaccuracies in the determination of the modes.

Angular momentum and electromagnetic charge have important traits in common, at least when the properties of black holes are considered.
However, charge allows for static spherically symmetric solutions, while rotation breaks spherical symmetry.
We may therefore consider charged wormholes as a toy model for capturing significant properties of rotating wormholes.
Thus, it is of interest to study the instability and QNM spectrum of charged EB wormholes.
Thanks to their spherical symmetry, these charged EB wormholes are known in closed form \cite{Gonzalez:2009hn}\footnote{
A two-parameter solution of the same Einstein-Maxwell-phantom-scalar theory, which is not globally regular, 
was discussed in \cite{Turimov:2025zxy}.
}.

A recent study investigated perturbations of charged EB-type wormholes in a different Einstein-Maxwell-phantom theory featuring a non-minimal scalar-Maxwell coupling $Z^{-1}(\phi)F_{\mu\nu}F^{\mu\nu}$ \cite{Wu:2025wlz}. 
Its background solutions and perturbation equations differ from those considered here, and its polar analysis uses an approximate closure rather than the fully coupled polar system studied here.

Clearly, when the charge is set to zero, the ordinary EB wormholes are recovered.
For finite values of the charge, however, one finds three types of solutions that have been classified as subcritical, critical and supercritical \cite{Gonzalez:2009hn}.
In all cases, the families of wormholes approach an extremal Reissner-Nordstr\"om (eRN) black hole as their limiting solution.
Gonzalez \textit{et al.} \cite{Gonzalez:2009hn} already performed a partial analysis of the fate of the unstable radial mode.
To have the complete picture, we recently re-examined this problem and found that close to the limiting solution the instability was basically quenched and the mode became long-lived \cite{Blazquez-Salcedo:2025dit}.

Here we investigate the full spectrum of QNMs of the charged EB wormholes.
To obtain the modes, we employ a spectral method that we have applied previously to wormholes and black holes \cite{Khoo:2024yeh,Blazquez-Salcedo:2025dit,Blazquez-Salcedo:2023hwg,Blazquez-Salcedo:2024oek,Blazquez-Salcedo:2024dur,Khoo:2024agm}.
We present the theoretical framework in section 2, and discuss the background solutions.
In section 3 we discuss the general set of perturbations for the wormholes.
Section 4 briefly recalls the spectral method.
The numerical results for the spectrum are then presented in section 5 for each type of the charged EB wormholes.
In addition, we display the limiting case of uncharged EB wormholes, covering a larger mass range than before \cite{Azad:2023iju}.
We conclude in section 6 and provide some tables with the modes in the Appendix.

\section{Theoretical framework}
\label{sec:TheoreticalFramework}

We consider the following Einstein-Maxwell action coupled with a phantom scalar field in geometrical units,
\begin{equation}
    \mathcal{S}=\frac{1}{16\pi}\int d^4x\sqrt{-g}\left[R-F^2+2(\nabla\varphi)^2\right] \, ,
    \label{action}
\end{equation}
where $R$ is the curvature scalar, $F$ is the electromagnetic tensor, and $\varphi$ is the phantom scalar field. 
Varying the action (\ref{action}) with respect to the metric $g$, the gauge field $A$, and the scalar field $\varphi$ gives rise to
the following equations of motion, respectively,
\begin{equation}\label{Einstein_eqs}
    R_{\mu\nu}=8\pi T^{\rm{EM}}_{\mu\nu}-2\nabla_{\mu}\varphi\nabla_{\nu}\varphi \, ,
\end{equation}
\begin{equation}\label{Maxwell_eqs}
    \nabla^{\mu}F_{\mu\nu}=0 \, , \,\,\,\,\,\,\,\,\, \nabla_{[\sigma}F_{\mu\nu]}=0 \, ,
\end{equation}
\begin{equation}\label{scalar_field_eq}
    \Box\varphi=0\, ,
\end{equation}
where $T^{\rm{EM}}_{\mu\nu}=(F_{\mu}^{\,\,\,\sigma}F_{\nu\sigma}-\frac{1}{4}g_{\mu\nu}F^2)/4\pi$.

\subsection{Background solutions}

The static, charged wormholes were first studied in \cite{Gonzalez:2009hn}.
The background is given by
\begin{equation}\label{metric_b}
    ds^2=-G(r)dt^2+G(r)^{-1}\left[dr^2+(r^2+r_0^2)(d\theta^2+\sin^2{\theta}\,d\phi^2)\right]\, ,
\end{equation}
for the scalar field,
\begin{equation}\label{scalar_b}
    \varphi=\varphi_1\arctan{\left(\frac{r}{r_0}\right)}+\varphi_0 \, ,
\end{equation}
and 
\begin{equation}\label{F_exp}
    F=\frac{2Q_e}{r^2+r_0^2}G(r)dt\wedge dr+2Q_m\sin{\theta}d\theta\wedge d\phi\, ,
\end{equation}
where $r_0$ is a scale factor and $Q_e$ and $Q_m$ are the electric and magnetic charges respectively.
In what follows, we set the magnetic charge  to zero, hence focusing on the purely electric setup.

Depending on the parameter $\Lambda$, there are three distinct classes of solutions, known as subcritical, critical, and supercritical wormholes,
\begin{equation}
G(r)
= \left\{
\begin{array}{ccc} 
 {\sigma^2} \left[\cosh{(\Lambda y)}-\gamma_1\frac{\sinh{(\Lambda y)}} {\Lambda}\right]^{{-2}}  \, ,
& \Lambda > 0 \, , &\rm{subcritical} ,\\[5pt]
 {\sigma^2} (1-\gamma_1y)^{{-2}} \, ,
& \Lambda = 0 \, , &\rm{critical}, \\[2pt]
{\sigma^2} \left[ \cos(\mu y) - \gamma_1\frac{\sin(\mu y)}{\mu} \right]^{{-2}} \, ,
& {\Lambda=i\mu\, ,\,\mu >0} \, , &\rm{supercritical},
\end{array} 
\right.
\label{G_exp}
\end{equation}
where $y=\arctan{({r}/{r_0})}$. 
The parameter $\sigma$ ensures that the spacetime is asymptotically flat, i.e. $g_{tt}=-1$, $g_{rr}=1$ when $r$ approaches infinity. 

For global wormhole solutions, the following conditions should be satisfied, 
\begin{equation}
\left.
\begin{array}{cc}
  {\Lambda < |\gamma_1|} \,\,\, \hbox{ and } \,\,\,
  \frac{\tanh\left(\Lambda\frac{\pi}{2}\right)}{\Lambda} |\gamma_1| < 1 \, ,
& \hbox{subcritical},\\[5pt]
  \frac{\pi}{2}|\gamma_1| < 1 \, ,
& \hbox{critical},\\[2pt]
  \mu < 1 \,\,\, \hbox{ and } \,\,\,
  \frac{\tan\left(\mu\frac{\pi}{2}\right)}{\mu} |\gamma_1| < 1 \, ,
& \hbox{supercritical}.
\end{array} 
\right.
\label{cases2}
\end{equation}
Global wormhole solutions are determined only by three free parameters, 
$r_0$, $\gamma_1$ and $\Lambda$ (or $\mu$),
with the remaining
charges 
determined by these three parameters.
For instance, we arrive at the uncharged EB wormhole solution  
when $\Lambda=\gamma_1$.

For the phantom scalar field $\varphi$, the scalar charge $Q_s$ is identified asymptotically using
$\varphi_0=-\pi Q_s/2 r_0$ and $\varphi_1=Q_s/r_0$, meaning that

\begin{equation}
Q_s = \left\{
\begin{array}{cc}
r_0\sqrt{1+\Lambda^2} \, , & \rm{subcritical},\\[1ex]
r_0 \, , & \rm{critical} ,\\[1ex]
r_0\sqrt{1-\mu^2} \, , & \rm{supercritical}.
\end{array}
\right.
\label{Qs_exp}
\end{equation}
On the other hand, the parameter $\gamma_1$ in the background functions is given by
\begin{equation}\label{gamma_exp}
    \gamma_1=\frac{1}{r_0}\sqrt{Q_s^2+\sigma^2 Q_e^2-r_0^2} \, .
\end{equation}
This parameter encodes the three fundamental parameters of the charged wormhole, i.e., $Q_e, Q_s$ and $r_0$.
In fact, $\gamma_1$ is known as the asymmetry parameter \cite{Gonzalez:2009hn} where $\gamma_1=0$ corresponds to the symmetric wormholes.
The total mass of the wormhole can be calculated asymptotically from $g_{tt}\approx-1+2M/r$ when $r\to\infty$.
The throat of the wormhole is found when $\partial_r g_{\theta\theta}|_{r_{min}}=0$, and hence the area of the throat is determined by $A_T=4\pi g_{\theta\theta}(r_{min})$.

\begin{figure}[t!]
\begin{center}
\includegraphics[width=6.5cm,angle=-90]{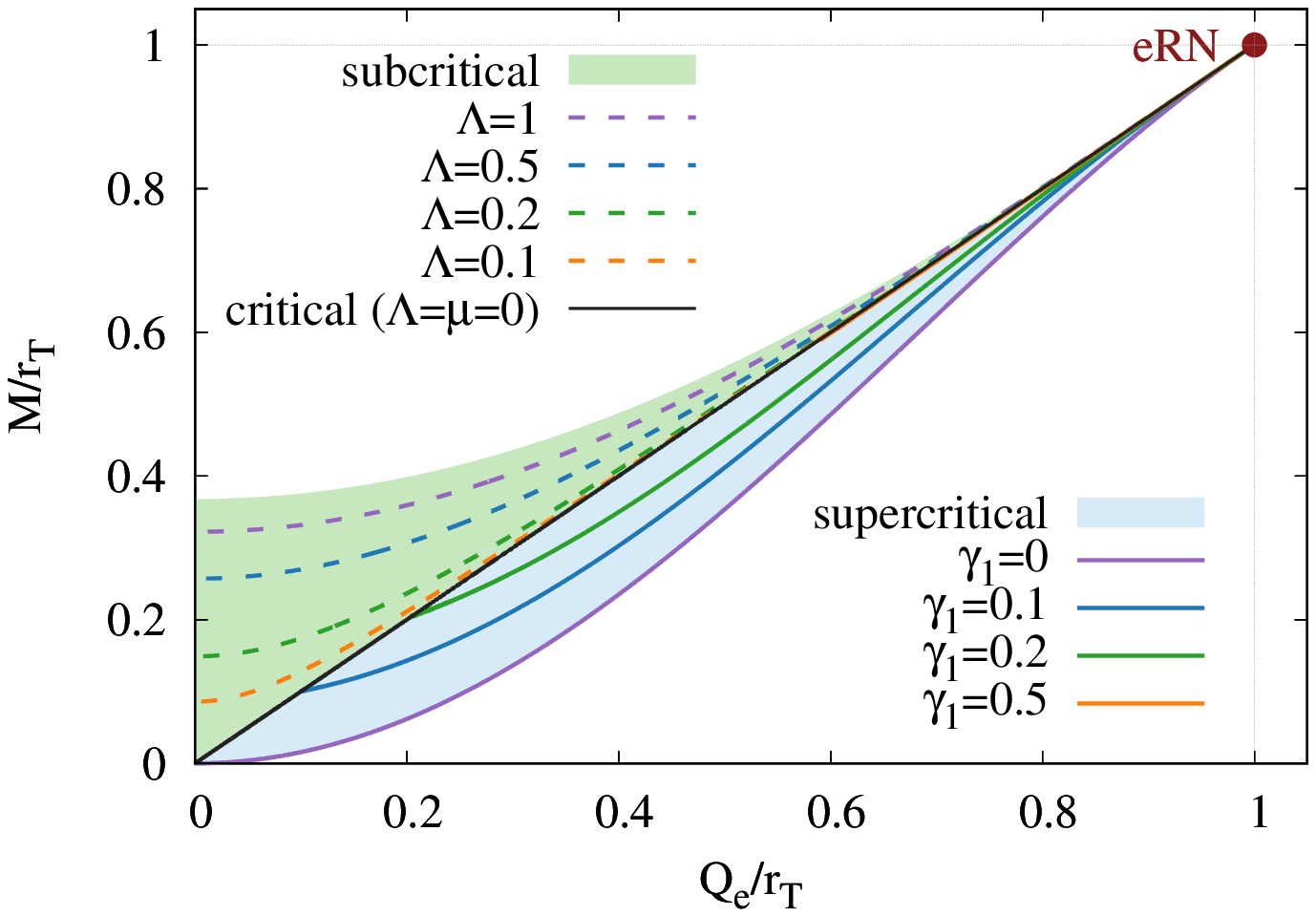}
\vspace*{-0.7cm}
\end{center}
\caption{
Domain of existence of the static charged wormholes where the supercritical and subcritical wormholes are separated by the critical wormholes. 
Several branches of solutions are shown in colors for fixed values of the parameters. All the branches end at the extremal Reissner-Nordström black hole (eRN).
}
\label{fig1}
\end{figure}

As a result, in the following we will characterize the wormholes by three physical parameters,
the electric charge $Q_e$, the mass $M$ and the throat radius $r_T=\sqrt{A_T/4\pi}$.
Figure \ref{fig1} shows the domain of existence of the electrically charged wormholes, described by relations between their mass and their electric charge, both scaled by the throat radius.
The critical wormholes that satisfy $Q_e=M$ (or $\Lambda=\mu=0$) separate the other two classes of wormholes from each other, where the domain is bounded by the supercritical wormholes, i.e., $Q_e>M$ from the bottom, and from the top by the subcritical wormholes given by $Q_e<M$. 
Each solid curve within the region of the supercritical wormholes is given by a value of the parameter $\gamma_1$, while each dashed  curve in the subcritical family is parametrized by $\Lambda$.
Additionally, these physical quantities are bounded by $Q_e=M=r_T$. 
At this point, the wormhole configuration approaches a limiting configuration that is an eRN black hole, as the wormhole throat radius tends to the location of the black hole horizon.

\begin{figure*}[h!]
\begin{center}
\mbox{
\resizebox{\textwidth}{!}
{
    \includegraphics[angle=-90]{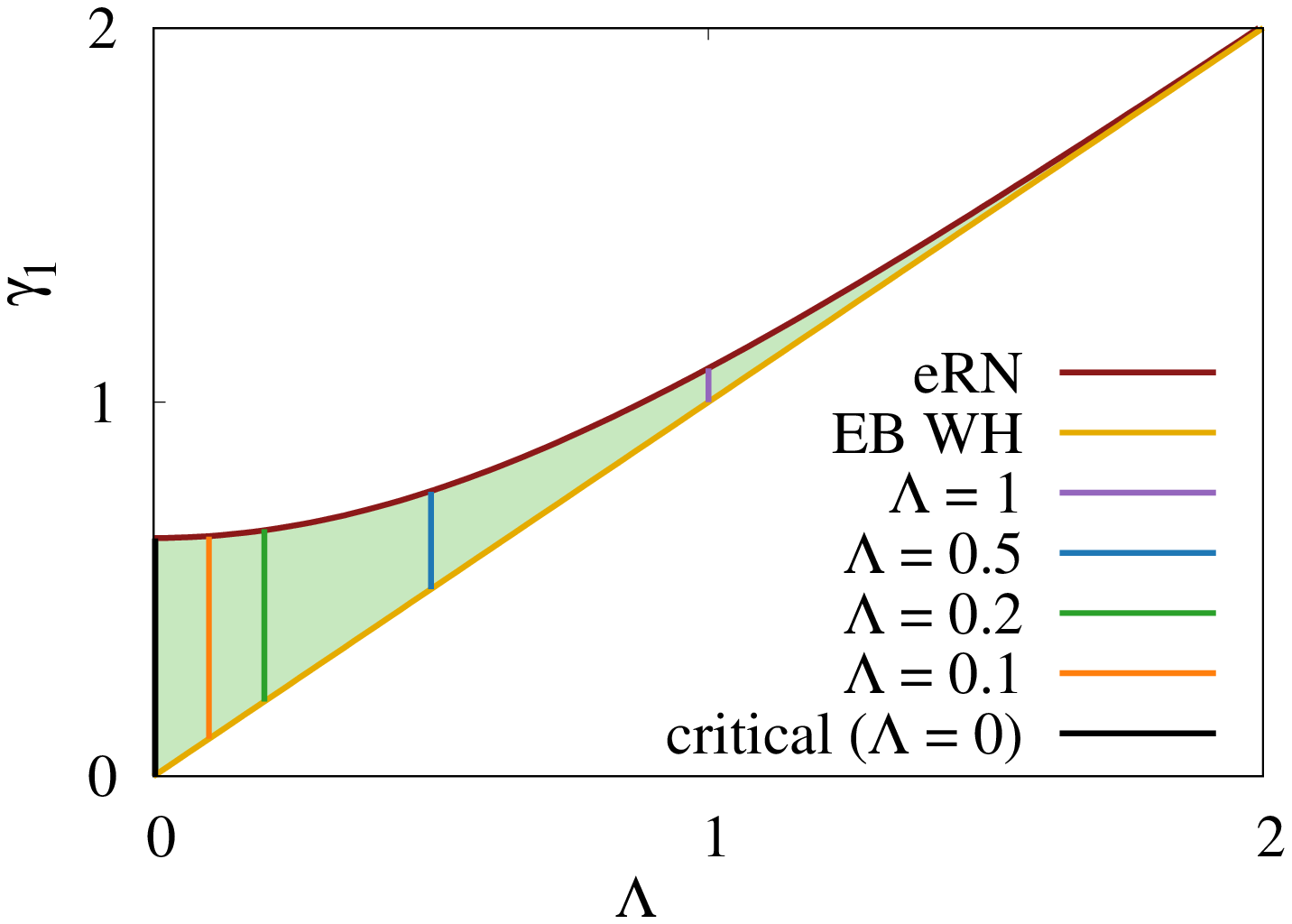}
    \includegraphics[angle=-90]{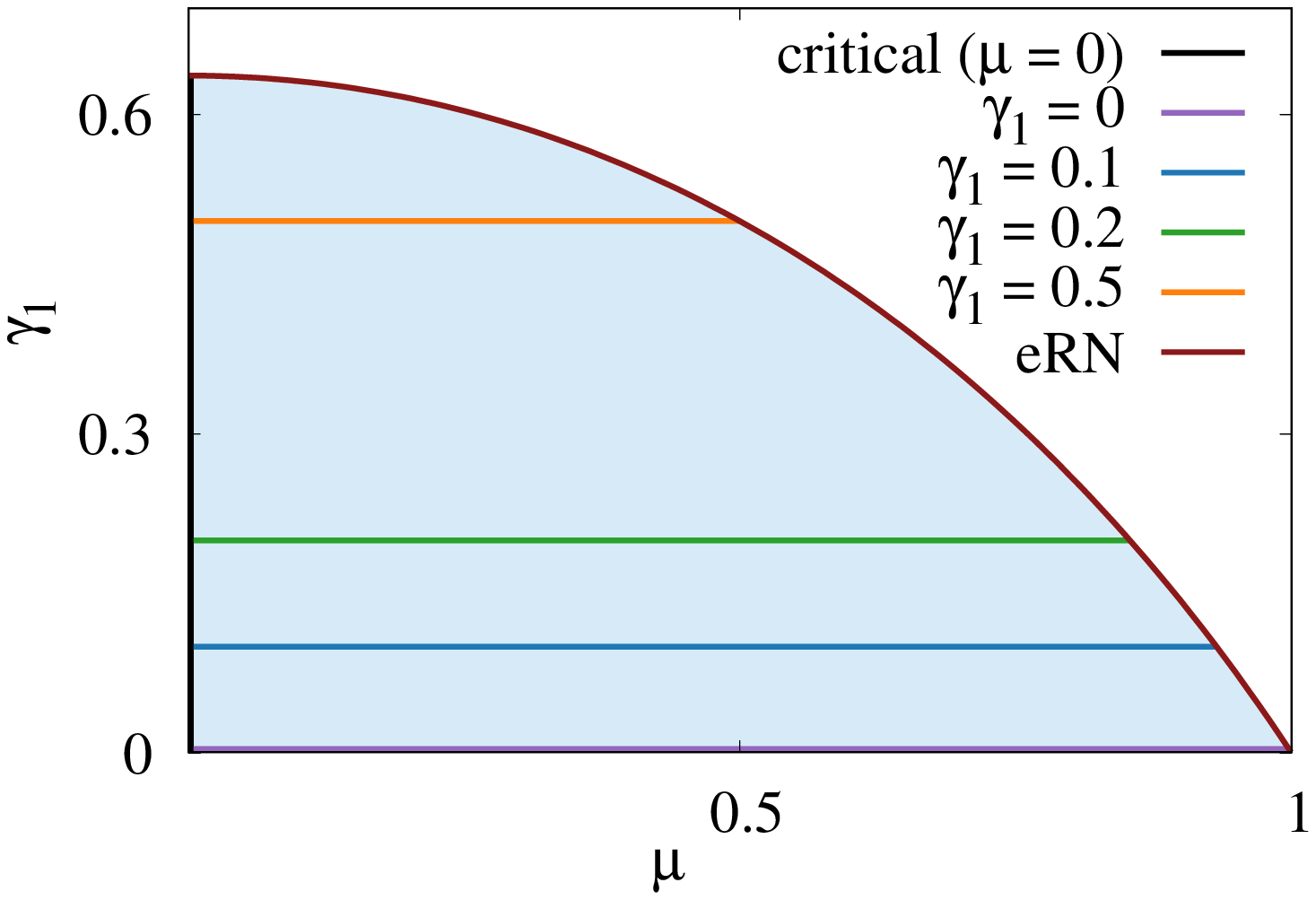}
}
}
\vspace*{-0.7cm}
\end{center}
\caption{
Another representation of the domain of existence of the static charged wormholes, 
separately for the subcritical and supercritical wormholes (shaded region). 
Left: Subcritical wormholes. 
The Ellis-Bronnikov (EB) wormholes are given by the linear relation between the parameters $\gamma_1$ and $\Lambda$. 
Configurations with different values of $\Lambda$ are shown in vertical lines, where they end at a maximal $\gamma_1$ value which denotes the eRN black hole.
Right: Supercritical wormholes. 
Similarly, the domain is bounded by the eRN black holes and the critical wormholes.
Configurations with different values of $\gamma_1$ are shown in horizontal lines.
}
\label{fig:Parameters}
\end{figure*}

On the other hand, we show in Figure \ref{fig:Parameters} another representation of the domain of existence using the parameters $\gamma_1, \Lambda, \mu$ of the background functions.
In the left panel, we exhibit the relation between $\gamma_1$ and $\Lambda$ for the subcritical wormholes, and in the right panel for $\gamma_1$ and $\mu$ for the supercritical wormholes.
In the left panel, the curve $\gamma_1=\Lambda$ represents the uncharged EB wormholes.
Hence, the domain of the subcritical wormholes is bounded by the uncharged EB wormholes, the charged critical wormholes and the eRN black holes.
In the right panel, the domain of the supercritical wormholes is bounded by the charged critical wormholes, the symmetric wormholes ($\gamma_1=0$) and the extremal Reissner-Nordstr\"om black holes.
In both panels, one finds at the origin the massless and uncharged EB wormhole.

\section{Perturbation theory}
\label{sec:PerturbationTheory}

Here we present the perturbations of the background fields.
These can be further categorized into axial, radial and polar perturbations.
We derive the equations for these perturbations and discuss the boundary conditions.

We perturb the metric, the electromagnetic (EM) gauge potential and the phantom field as follows,
\begin{equation}
    g_{\mu\nu}=g_{\mu\nu}^0+\delta g_{\mu\nu} \, ,
\end{equation}
\begin{equation}
    A_{\mu}=A_{\mu}^0+\delta A_{\mu} \, ,
\end{equation}
\begin{equation}
    \varphi=\varphi^0+\delta\varphi \, ,
\end{equation}
where the superscript $0$ denotes the background solution given by \eqref{metric_b}-\eqref{F_exp}. 
The perturbations will be decomposed into their axial and polar components in order to split the problem into two pieces. 
We fix the gauge by employing the well-known Regge-Wheeler gauge. 
Due to this choice, we are left with two axial perturbation functions and four polar perturbation functions as will become clear in the next sections.
The spherical symmetry allows us to decompose the ansatz of the perturbations into spherical harmonics, thus decoupling the angular dependence.
This leads to a system of ordinary differential equations (ODEs) instead of partial differential equations (PDEs). 
The symmetry of the problem simplifies the spectral method substantially, since only a one-dimensional grid is required.
Furthermore, the time-dependent part of the damped oscillations can be written as $\sim e^{-i\omega t}$ with $\omega\in \mathbb{C}$. 
Thus, $\omega_R=\text{Re}(\omega)$ denotes the frequency of a mode, while $\omega_I=\operatorname{Im}(\omega)$ determines the damping or growth rate.
Notice that, with this convention, negative values $\omega_I<0$ represent stable modes, whereas positive values $\omega_I>0$ reveal unstable modes. 
We define the characteristic timescale as $\tau=1/|\omega_I|$. 
It represents a damping time for stable modes and a growth time for unstable modes.

\subsection{Axial perturbations}

The axial perturbations of the metric are given by
\begin{equation}\label{deltag_axial}
    \delta g_{\mu\nu}(t,r,\theta,\phi)=\begin{pmatrix}
0 & 0 & -h_0(r)\frac{\partial_{\phi}Y_{lm}(\theta,\phi)}{\sin{(\theta)}} & h_0(r)\sin{(\theta)}\partial_{\theta}Y_{lm}(\theta,\phi)\\
0 & 0 & -h_1(r)\frac{\partial_{\phi}Y_{lm}(\theta,\phi)}{\sin{(\theta)}} & h_1(r)\sin{(\theta)}\partial_
{\theta}
Y_{lm}(\theta,\phi)\\
* & * & 0 & 0\\
* & * & 0 & 0
\end{pmatrix}e^{-i\omega t} \,,
\end{equation}
where $*$ stands for the symmetric components, and an implicit sum over the multipoles $l$ is assumed. 
This ansatz is in the Regge-Wheeler gauge \cite{Regge:1957td}.
Owing to the spherical symmetry of the background, the spectrum is independent of the azimuthal number $m$. 
We may therefore set $m=0$ without loss of generality, in which case $Y_{l0}(\theta,\varphi)\propto P_l(\cos\theta)$.

On the other hand, the perturbation of the EM gauge field can be written as
\begin{equation}
    \delta A(t,r,\theta,\phi)=-W_1(r)\frac{\partial_{\phi}Y_{lm}(\theta,\phi)}{\sin{(\theta)}}e^{-i\omega t}d\theta+W_1(r)\sin{(\theta)}\partial_{\theta}Y_{lm}(\theta,\phi)e^{-i\omega t}d\phi \,,
\end{equation}
while the phantom field perturbations decouple and thus do not contribute.

After computing the field equations \eqref{Einstein_eqs}-\eqref{scalar_field_eq} up to first order in perturbation theory and performing some algebraic manipulations, we find the following set of ODEs
\begin{equation}\label{eq_h0_axial}
\begin{split}
    h_0'\:= & \:\:\frac{1}{r^2+r_0^2}\left(-\frac{G'}{G}r^2-\frac{G'}{G}r_0^2+2r\right)h_0-4A'W_1+\frac{i}{\omega(r^2+r_0^2)}\biggl(2rGG' \\
    & + (r^2+r_0^2)(GG''-G'^2-2GA'^2-\omega^2)+(l^2+l-2)G^2\biggr)h_1\,,
\end{split}
\end{equation}
\begin{equation}\label{eq_h1_axial}
    h_1'=-\frac{G'}{G}h_1-\frac{i\omega}{G^2}h_0 \,,
\end{equation}
\begin{equation}\label{eq_W1_axial}
    \begin{split}
    W_1''\:= & \:\:\left(-\frac{A''}{G}+\frac{A'}{G^2(r^2+r_0^2)}\left((r^2+r_0^2)G'-2rG\right)\right)h_0\\
    & -\frac{iA'}{\omega}\left(G''-2A'^2-\frac{G'^2}{G}+\frac{2G'r}{r^2+r_0^2}+\frac{(l+2)(l-1)G}{r^2+r_0^2}\right)h_1\\
    & +\left(\frac{4A'^2}{G}+\frac{l(l+1)}{r^2+r_0^2}-\frac{\omega^2}{G^2}\right)W_1-\frac{G'}{G}W_1'\,.
\end{split}
\end{equation}
Note that from now on, the dependence of the functions on the radial coordinate is not explicitly shown, whenever the system is completely decoupled from temporal and angular coordinates. 
Also note that the prime $'$ denotes the derivative w.r.t.~$r$, and $A'=\partial_r A^0_{t}(r)$.

Next, we study the asymptotic behavior of the solutions at $r=\pm\infty$. 
Thus, we expand the solutions in power series at both radial infinities. 
Since we want outgoing waves at both sides, we define a modified tortoise coordinate that takes care of the sign at both infinities simultaneously as
\begin{equation}\label{tortoise}
    \frac{dr^*}{dr}=\frac{r}{G(r)\sqrt{r^2+r_0^2}} \, .
\end{equation}
The expansion of the perturbations reads
\begin{equation}
    h_0(r)=re^{i\omega r^*}\left(a_0+\frac{a_1}{r}+\dots\right) \,,
\end{equation}
\begin{equation}
    h_1(r)=re^{i\omega r^*}\left(b_0+\frac{b_1}{r}+\dots\right) \,,
\end{equation}
\begin{equation}
    W_1(r)=e^{i\omega r^*}\left(c_0+\frac{c_1}{r}+\dots\right) \,.
\end{equation}

We have only given the expansions at $+\infty$, but they hold,
with different coefficients,
at $-\infty$, as well. 
Asymptotically, only two constants are free, for example $a_0$ and $c_0$ (the amplitudes of the gravitational and electromagnetic perturbations), while the other constants are fixed via algebraic relations by the two free amplitudes.
In any case, the previous expansions allow us to identify the dominant term in the asymptotic behavior of the functions. In order to avoid divergences when numerically computing the solutions, it is convenient to reparametrize the perturbation functions by factorizing this behavior. 
Thus, the system of equations \eqref{eq_h0_axial}-\eqref{eq_W1_axial} can be conveniently rewritten
by introducing the following reparametrization of the perturbation functions:
$h_0 = \sqrt{r^2+r_0^2}e^{i\omega r^*}\hat{h}_0$, 
$h_1 = \sqrt{r^2+r_0^2}e^{i\omega r^*}\hat{h}_1$
and 
$W_1 = e^{i\omega r^*}\hat{W_1}$.

\subsection{Radial perturbations}

Before considering the higher multipole polar perturbations, we first study the $l=0$ radial perturbations, since the instability of the EB wormholes resides in the radial sector. 

The radial perturbations of the metric are given by
\begin{equation}
    \delta g_{\mu\nu}(t,r,\theta,\phi)=\begin{pmatrix}
G(r)F_0(r) & 0 & 0 & 0\\
0 & \frac{1}{G(r)}F_1(r) & 0 & 0\\
0 & 0 & \frac{r^2+r_0^2}{G(r)}F_2(r) & 0\\
0 & 0 & 0 & \frac{(r^2+r_0^2)\sin^2(\theta)}{G(r)}F_2(r)
\end{pmatrix}e^{-i\omega t} \,.
\end{equation}
The perturbations of the EM field and the scalar field can be cast as
\begin{eqnarray}\label{deltaA_radial}
\delta A(t,r,\theta,\phi)&=& e^{-i\omega t} \left( V_0(r)dt  
+ V_1(r)dr   
\right) \, ,
\ \ \ \\
\delta\varphi(t,r,\theta,\phi)&=&u(r)e^{-i\omega t} \, .
\end{eqnarray}

We fix the gauge freedom by setting $F_0+F_1-2F_2=0$ and $V_1=F_0(A_t^{0})'/\omega$. 
This choice allows us to get rid of $F_1$ and $V_0$, decreasing the order of the resulting system of differential equations, which is then given by
\begin{equation}\label{eq_F0_radial}
\begin{split}
    F_0'\:= & \:\:\frac{1}{(r^2+r_0^2)G'-2rG}\left((r^2+r_0^2)G''-(r^2+r_0^2)\frac{G'^2}{G}+rG'-\frac{2Gr_0^2}{r^2+r_0^2}\right)F_0\\
    & +\frac{1}{(r^2+r_0^2)G'-2rG}\left(-2(r^2+r_0^2)G''+\frac{2(r^2+r_0^2)G'^2}{G}-4rG'+4(r^2+r_0^2)A'^2\right.\\
    & \left.+\frac{2Gr_0^2}{r^2+r_0^2}-\frac{2\omega^2(r^2+r_0^2)}{G}\right)F_2-\frac{4GQ_s}{(r^2+r_0^2)G'-2rG}\left(u'+\frac{r}{r^2+r_0^2}u\right) \,,
\end{split}
\end{equation}
\begin{equation}\label{eq_F2_radial}
    F_2'=\left(\frac{G'}{2G}-\frac{r}{r^2+r_0^2}\right)F_0+\frac{r}{r^2+r_0^2}F_2+\frac{2Q_s}{r^2+r_0^2}u \,,
\end{equation}
\begin{equation}\label{eq_u_radial}
    u''=-\frac{2r}{r^2+r_0^2}u'-\frac{\omega^2}{G^2}u \,.
\end{equation}

Following a procedure similar to the axial case, we study asymptotically the outgoing wave solutions of this system. Using the tortoise coordinate (\ref{tortoise}),
the expansion at infinity can be cast as
\begin{equation}
    F_0(r)=re^{i\omega r^*}\left(d_0+\frac{d_1}{r}+\dots\right) \,,
\end{equation}
\begin{equation}
    F_2(r)=e^{i\omega r^*}\left(e_0+\frac{e_1}{r}+\dots\right) \,,
\end{equation}
\begin{equation}
    u(r)=\frac{1}{r}e^{i\omega r^*}\left(f_0+\frac{f_1}{r}+\dots\right) \,.
\end{equation}
For strictly unstable solutions, it is well known that the previous system can be written as a single master equation by setting $u=0$.
In that case the resulting potential becomes divergent at the throat, requiring regularization of the master equation (see e.g. \cite{Gonzalez:2009hn, Bronnikov:2012ch, Torii:2013xba}). 
However, by integrating directly the above system of ODEs (\ref{eq_F0_radial})-(\ref{eq_u_radial}), which is regular at the throat, this problem is avoided. 
As with the axial case, we factorize the divergent behavior by reparametrizing $(F_0, F_2, u)=(\sqrt{r^2+r_0^2}e^{i\omega r^*}\hat{F}_0, e^{i\omega r^*}\hat{F}_2, \frac{1}{\sqrt{r^2+r_0^2}}e^{i\omega r^*}\hat{u})$.

\subsection{Polar perturbations}

The polar perturbations of the metric are given by
\begin{equation}
    \delta g_{\mu\nu}(t,r,\theta,\phi)=\begin{pmatrix}
G(r)H_0(r) & H_1(r) & 0 & 0\\
H_1(r) & \frac{1}{G(r)}H_2(r) & 0 & 0\\
0 & 0 & \frac{r^2+r_0^2}{G(r)}K(r) & 0\\
0 & 0 & 0 & \frac{(r^2+r_0^2)\sin^2(\theta)}{G(r)}K(r)
\end{pmatrix}Y_{lm}(\theta,\phi)e^{-i\omega t} \,,
\end{equation}
the corresponding perturbation of the gauge field is
\begin{equation}\label{deltaA_polar}
\begin{split}
    \delta A(t,r,\theta,\phi)= &\;V_0(r)Y_{lm}(\theta,\phi)e^{-i\omega t}dt+V_1(r)Y_{lm}(\theta,\phi)e^{-i\omega t}dr\\
    & +W_0(r)\partial_{\theta}Y_{lm}(\theta,\phi)e^{-i\omega t}d\theta+W_0(r)\partial_{\phi}Y_{lm}(\theta,\phi)e^{-i\omega t}d\phi \,, 
\end{split}
\end{equation}
and the scalar field perturbation is
\begin{equation}
    \delta\varphi(t,r,\theta,\phi)=u(r)Y_{lm}(\theta,\phi)e^{-i\omega t} \,.
\end{equation}

After manipulating the equations of motion, we arrive at the following system of ODEs for the polar perturbations
\begin{equation}\label{eq_H0_polar}
\begin{split}
    H_0'\:= & \:\:\left(-\frac{3G'}{2G}+\frac{r}{r^2+r_0^2}\right)H_0+\frac{(r^2+r_0^2)G'-rG}{G(r^2+r_0^2)}K-\frac{2Q_s}{r^2+r_0^2}u\\
    & +\frac{i}{2\omega}\left(-G''+2A'^2+\frac{l(l+1)G}{r^2+r_0^2}+\frac{G'^2-2\omega^2}{G}-\frac{2G'r}{r^2+r_0^2}\right)H_1+\frac{4A'}{G}F_1 \,,
\end{split}
\end{equation}
\begin{equation}\label{eq_H2_polar}
    H_2=H_0 \,,
\end{equation}
\begin{equation}\label{eq_H1_polar}
    H_1'=-\frac{i\omega}{G}\left(H_0+K\right)-\frac{G'}{G}H_1+4A'F_2 \,,
\end{equation}
\begin{equation}\label{eq_K_polar}
\begin{split}
    K'\:= & \:\:\frac{2Gr-G'(r^2+r_0^2)}{2G(r^2+r_0^2)}H_0+\frac{(r^2+r_0^2)G'-rG}{G(r^2+r_0^2)}K+\frac{2Q_s}{r^2+r_0^2}u\\
    & +\frac{i}{2\omega}\left(-G''+2A'^2+\frac{l(l+1)G}{r^2+r_0^2}+\frac{G'^2}{G}-\frac{2G'r}{r^2+r_0^2}\right)H_1 \,,
\end{split}
\end{equation}
\begin{equation}\label{eq_u_polar}
\begin{split}
    u''\:= & \:\:-\frac{G'Q_s}{G(r^2+r_0^2)}H_0+\frac{4A'Q_s}{G(r^2+r_0^2)}F_1-\frac{2r}{r^2+r_0^2}u'\\
    & +\left(\frac{l(l+1)}{r^2+r_0^2}-\frac{4Q_s^2}{(r^2+r_0^2)^2}-\frac{\omega^2}{G^2}\right)u \,,
\end{split}
\end{equation}
\begin{equation}\label{eq_F1_polar}
    F_1'=A'K+\frac{i\left(l(l+1)G^2-\omega^2(r^2+r_0^2)\right)}{\omega(r^2+r_0^2)}F_2 \,,
\end{equation}
\begin{equation}\label{eq_F2_polar}
    F_2'=-\frac{G'}{G}F_2-\frac{i\omega}{G^2}F_1 \,.
\end{equation}
A combination of the remaining non-trivial field equations yields the following relation between the functions,
\begin{eqnarray}
\label{alg_rel}
    && \biggl(-2G(r^2+r_0^2)G''+2(r^2+r_0^2)G'^2-4G'Gr-2G(r^2+r_0^2)A'^2 \\
    && +(4-l^2-l)G^2+\omega^2(r^2+r_0^2)\biggr)H_0-3i\omega\left(G'(r^2+r_0^2)-\frac{4}{3}rG\right)H_1  \nonumber \\
    && + \biggl(2G(r^2+r_0^2)G''
   -2(r^2+r_0^2)G'^2+4G'Gr-4G(r^2+r_0^2)A'^2 
    \nonumber \\
    && +2(l+2)(l-1)G^2-2\omega^2(r^2+r_0^2)\biggr)K 
    +12GA'(r^2+r_0^2)F_1' 
    \nonumber \\
    && -3\left(G'(r^2+r_0^2)-\frac{2}{3}rG\right)GH_0'+\biggl(G'(r^2+r_0^2)-4Gr\biggr)GK' \nonumber \\
    && -G^2(r^2+r_0^2)H_0''-8G^2Q_su'-2i(r^2+r_0^2)\omega GH_1'+12i\omega GA'(r^2+r_0^2)F_2=0 \,. \nonumber 
\end{eqnarray}

Once again we use the tortoise coordinate (\ref{tortoise}) to study the asymptotically outgoing solution.
The expansion at infinity is
\begin{equation}
    H_0(r)=re^{i\omega r^*}\left(g_0+\frac{g_1}{r}+\dots\right)\,,
\end{equation}
\begin{equation}
    H_1(r)=re^{i\omega r^*}\left(i_0+\frac{i_1}{r}+\dots\right) \,,
\end{equation}
\begin{equation}
    K(r)=e^{i\omega r^*}\left(j_0+\frac{j_1}{r}+\dots\right) \,,
\end{equation}
\begin{equation}
    u(r)=\frac{1}{r}e^{i\omega r^*}\left(k_0+\frac{k_1}{r}+\dots\right) \,,
\end{equation}
\begin{equation}
    F_1(r)=re^{i\omega r^*}\left(l_0+\frac{l_1}{r}+\dots\right) \,,
\end{equation}
\begin{equation}
    F_2(r)=re^{i\omega r^*}\left(m_0+\frac{m_1}{r}+\dots\right) \,.
\end{equation}
In this expansion, only three amplitudes are free (related to the gravitational, electromagnetic and scalar degrees of freedom), for example $g_0$, $k_0$ and $l_0$, while the rest of the constants are given in terms of these three via algebraic relations.

The expansion allows us to define the following reparametrization that factorizes the asymptotically divergent behavior of the functions:
\begin{eqnarray}
   && (H_0,H_1,K,u,F_1,F_2) =  \\ 
   && e^{i\omega r^*} \times
   \left(
   \sqrt{r^2+r_0^2}\hat{H}_0, \sqrt{r^2+r_0^2}\hat{H}_1, \hat{K}, \frac{1}{\sqrt{r^2+r_0^2}}\hat{u}, \sqrt{r^2+r_0^2}\hat{F}_1, \sqrt{r^2+r_0^2}\hat{F}_2
   \right) \,.
     \nonumber
\end{eqnarray}

As a final point, we turn to the issue of isospectrality of the axial and polar perturbations.
This feature can be understood in the following way. 
Since in the uncharged case there is no electric charge, it follows that $A(r)=0$ and, therefore, the EM perturbations decouple from the gravitational and scalar ones. 
One arrives at the conclusion that Eq.~\eqref{eq_W1_axial} is equivalent to Eqs.~\eqref{eq_F1_polar} and \eqref{eq_F2_polar} after combining them into a single ODE and replacing, for example, $F_2(r)\to \hat{W}_1(r)=G^{-1}(r)F_2(r)$. 
It follows that
\begin{equation}\label{eq_W1_isos}
    \hat{W}_1''=\left(\frac{l(l+1)}{r^2+r_0^2}-\frac{\omega^2}{G^2}\right)\hat{W}_1-\frac{G'}{G}\hat{W}_1' \,,
\end{equation}
which agrees with Eq.~\eqref{eq_W1_axial} for $Q_e=0$. 
This result implies that the spectrum of both axial and polar EM perturbations is the same in the uncharged case, as we will show later. 
Finally, we rewrite Eq.~\eqref{eq_W1_isos} as a Schrödinger-like equation. 
Defining the tortoise coordinate as 
$\frac{dR}{dr}=G^{-1}(r)$ 
we arrive by 
a straightforward manipulation at
\begin{equation}\label{eq_Schrodinger}
\frac{d^2\hat{W}_1}{d{R}^{2}}+\left(\omega^2-V_l(r)\right)\hat{W}_1=0 \,,
\end{equation}
with the potential $V_l(r)$ given by
\begin{equation}
    V_l(r)=G(r)^2\frac{l(l+1)}{r^2+r_0^2} \,.
\end{equation}

\section{Spectral method}
\label{sec:Spectral method}

In this section we cover the basics of the numerical algorithm employed to solve the field equations and compute the QNMs of different types of configurations of the charged EB wormholes. 
First of all, it is convenient to compactify the radial coordinate. We thus choose the compactification $\mathcal{A}:\mathbb{R}\to\left[-1,1\right]$ to be
\begin{equation}\label{compactification}
    x=\frac{2}{\pi}\arctan{\left(\frac{r}{r_0}\right)} \,,
\end{equation}
and $r_0$ can be set to $1$ in the numerical calculations without loss of generality.
Then, we proceed to decompose the perturbation functions into Chebyshev polynomials
\begin{equation}\label{spectral_decomp}
    F_i(x)=\sum_{k=0}^{N_p-1}C_{i,k}T_k(x) \,,
\end{equation}
where $N_p$ is the size of the grid, 
$F_i=\{\hat h_0,\hat h_1,\hat u,\dots\}$,
$C_{i,k}$ are the coefficients of the decomposition for the $i$-th function, and $T_k(x)$ denotes the Chebyshev polynomials of the first kind. 
The constants $C_{i,k}$ are obtained by solving the ODEs. 

The next step is to discretize the domain $x\in\left[-1,1\right]$, and a possible discretization is given by
\begin{equation}
    x_j=\cos{\left(\frac{\pi j}{N_p+1}\right)} \,,\:\:\:\:\:\: j=1,\dots,N_p \,.
\end{equation}

After substituting \eqref{spectral_decomp} into the corresponding equations, we are left with $3\times N_p$ algebraic equations for the axial perturbations, $3\times N_p$ for the radial perturbations, and $6\times N_p$ for the polar perturbations. 
Each problem can be written as
\begin{equation}
    \left(\mathcal{M}_0+\mathcal{M}_1\omega+\mathcal{M}_2\omega^2\right)C=0 \,.
\end{equation}
This is a quadratic eigenvalue problem where $\omega$ is the eigenvalue. 
We solve this problem using the Advanpix Multiprecision Computing Toolbox \cite{Advanpix} (see \cite{Khoo:2024yeh,Blazquez-Salcedo:2023hwg,Blazquez-Salcedo:2024oek, Blazquez-Salcedo:2024dur} for more details). 

The spectral method allows us to compute a number of modes with high precision.
In particular, the fundamental $n=0$ mode and a few excited modes with small $n$ are obtained with a numerical accuracy of order $10^{-8}$--$10^{-6}$.
We note that the accuracy decreases for higher excited modes and for larger values of $\Lambda$ and $\gamma_1$, but remains better than $10^{-3}$.

\section{Numerical results}
\label{sec:NumericalResults}

In this section we present the results obtained with the spectral method as explained in Section \ref{sec:Spectral method},
first for the EB wormholes, then for the critical, subcritical, and supercritical charged wormholes.
The Appendix contains some tables with a selection of the numerical results shown in this section.

\subsection{Ellis-Bronnikov wormholes}

The uncharged case ($\gamma_1=\Lambda$) corresponds to the EB wormholes. 
It is easy to see from Eq.~\eqref{Qs_exp} and Eq.~\eqref{gamma_exp} that this limit leads to $Q_e=0$. 
By inserting $\gamma_1=\Lambda$ into the axial equations \eqref{eq_h0_axial}-\eqref{eq_W1_axial}, the radial equations \eqref{eq_F0_radial}-\eqref{eq_u_radial} and the polar equations \eqref{eq_H0_polar}-\eqref{eq_F2_polar}, one finds that the EM field decouples from the metric equations as expected for a vanishing EM field. 
It is straightforward to see that $W_1$ disappears from Eq.~\eqref{eq_h0_axial} and Eq.~\eqref{eq_h1_axial} because $A^0_t\sim Q_e$. 

By solving the systems of differential equations via the spectral method, we have been able to not only recover the results of \cite{Blazquez-Salcedo:2018ipc,Azad:2022qqn} but also to calculate QNMs for larger masses and, more importantly, more excited states, revealing a rather rich spectrum\footnote{{Our analysis suggests that the unstable modes identified in \cite{Batic:2025hgp} may be spurious artifacts introduced by numerical error.}}.

\subsubsection{Axial modes}

\begin{figure*}[p!]
\begin{center}
\mbox{
\includegraphics[width=1.2\textwidth,angle=-90]{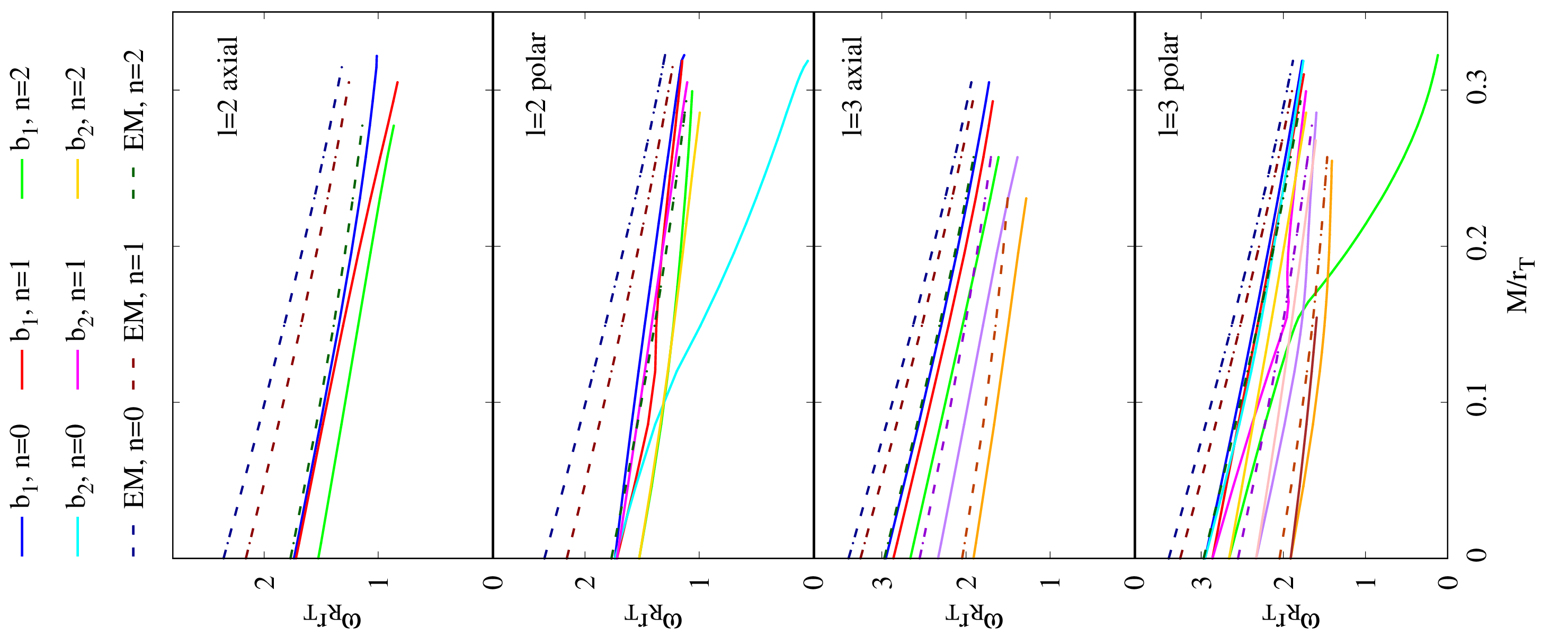}
\includegraphics[width=1.2\textwidth,angle=-90]{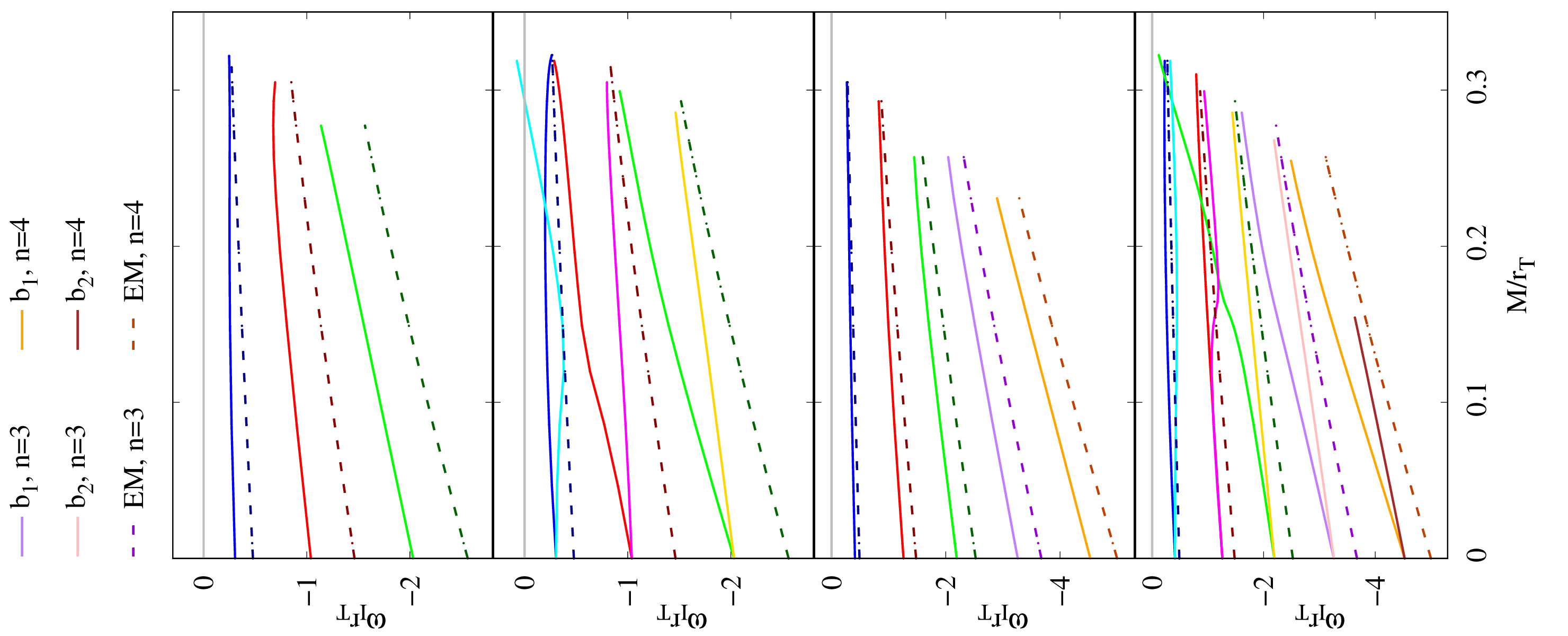}
}
\vspace*{-0.5cm}
\end{center}
\caption{
$l=2,3$ axial and polar modes for EB wormholes. 
From top to bottom the rows correspond to: $l=2$ axial, $l=2$ polar, $l=3$ axial, $l=3$ polar modes. 
Left panels show the real part of the QNMs. 
Right panels exhibit the imaginary part. 
In each panel we show, with different colors, the fundamental modes ($n=0$), and a few excitations ($n>0$).
}
\label{fig:axial_polar_massive_WH}
\end{figure*}

Figure~\ref{fig:axial_polar_massive_WH} displays the axial and polar QNM branches for $l=2$ and $l=3$ as functions of the dimensionless mass $M/r_T$. 
The left column shows the dimensionless oscillation frequency $\omega_R r_T$, while the right column shows the dimensionless imaginary part $\omega_I r_T$.
In this section, we focus on the axial modes, which are shown in row 1 of Figure \ref{fig:axial_polar_massive_WH} for the  $l=2$ multipole and in row 3 for the $l=3$ multipole.
In each panel we show with solid lines the axial gravitational modes, and with dashed lines the electromagnetic modes. We show with different colors the fundamental modes ($n=0$) and a few excitations ($n>0$).

\begin{figure*}[h!]
\begin{center}
\mbox{
\resizebox{\textwidth}{!}
{
    \includegraphics[width=1.2\textwidth,angle=-90]{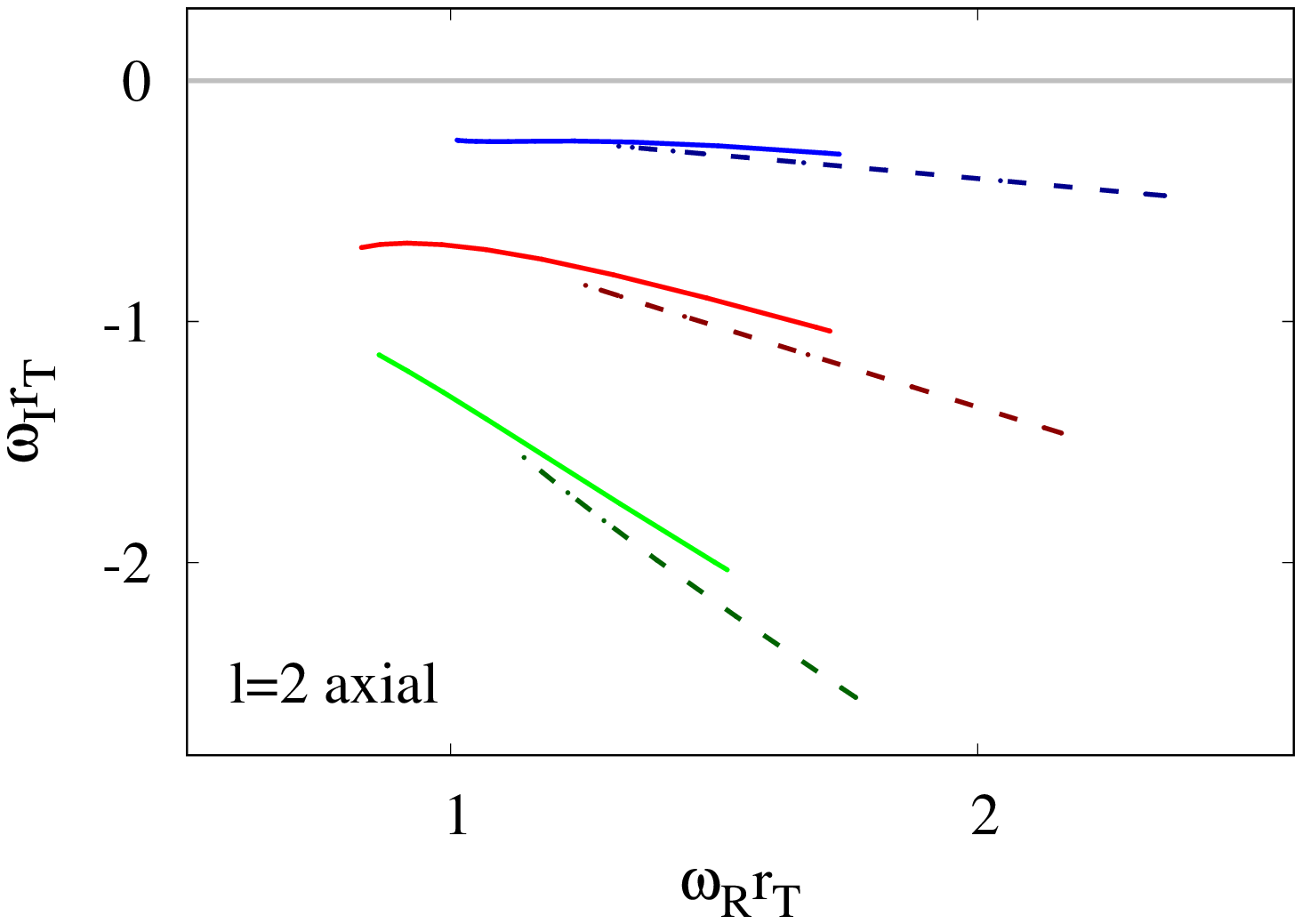}
    \includegraphics[width=1.2\textwidth,angle=-90]{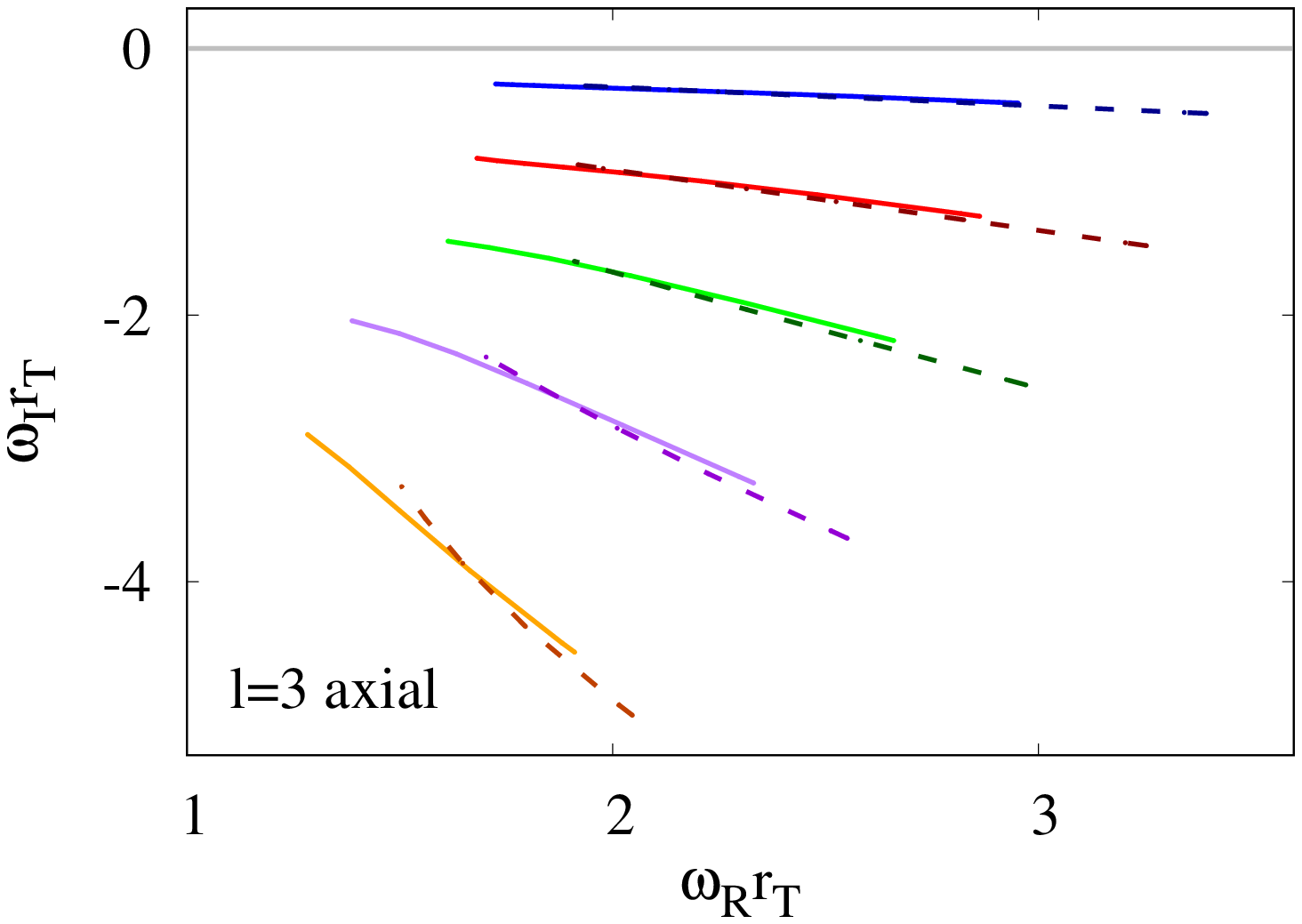}
}
}
\vspace*{-0.7cm}
\end{center}
\caption{
Axial modes for EB wormholes.
$\text{Im}(\omega)$ vs $\text{Re}(\omega)$ for $l=2$ (left) and $l=3$ (right).
}
\label{fig:axial_spectrum_massive_WH}
\end{figure*}

In general we note that the imaginary part of the modes increases with mass, as previously shown \cite{Blazquez-Salcedo:2018ipc}\footnote{Here we scale with $r_T$, instead of $r_0$ \cite{Blazquez-Salcedo:2018ipc}, which is a more physical quantity.}. 
Recalling that, for stable modes, $|\omega_I|$ represents the inverse of the damping time, we define the fundamental mode as the one with the largest damping time, i.e., the lowest $|\text{Im}(\omega)|$. 
On the other hand, we observe that the real part of the modes decreases with mass, which is again in good agreement with previous studies. 
Alternatively, we show in Figure \ref{fig:axial_spectrum_massive_WH} the obtained axial spectrum for these EB wormholes in the complex plane. Clearly all the axial modes calculated are stable (meaning negative imaginary part in our convention), and our calculations do not show any other unstable modes in the axial spectrum.

\begin{figure*}[h!]
\begin{center}
\mbox{
\resizebox{\textwidth}{!}{
    \includegraphics[angle=-90]{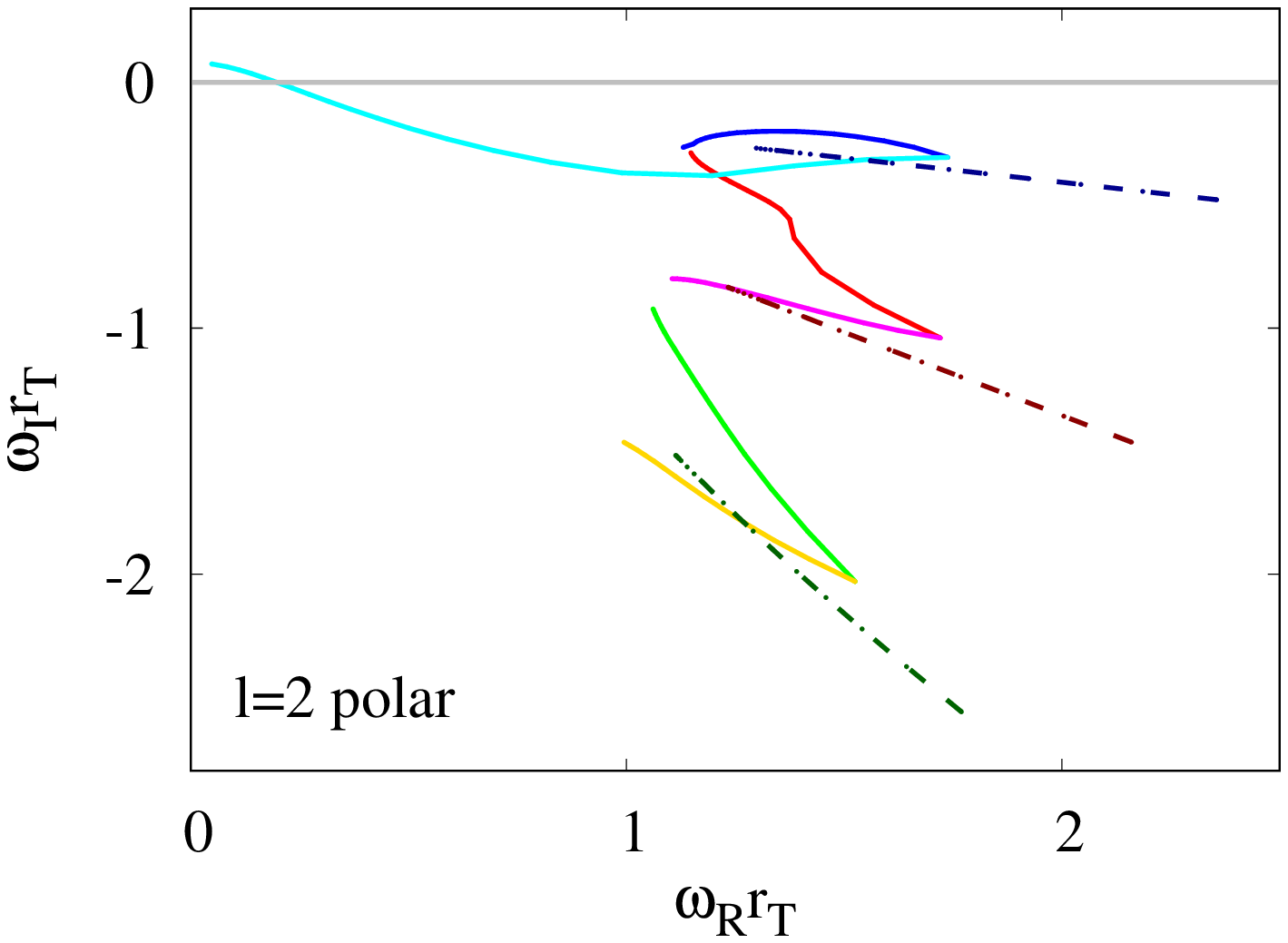}
    \includegraphics[angle=-90]{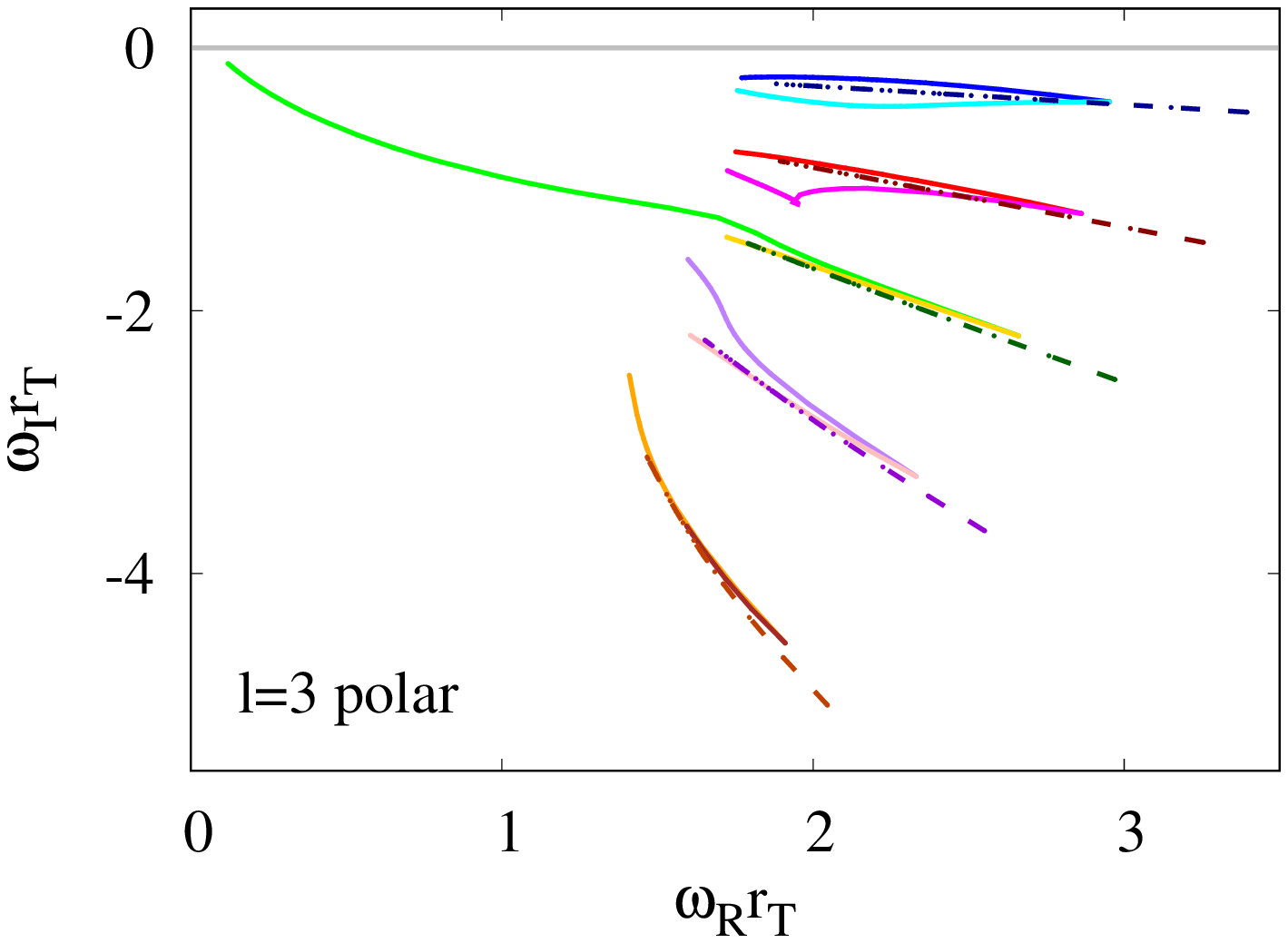}
}
}
\vspace*{-0.7cm}
\end{center}
\caption{
Polar modes for EB wormholes.
$\text{Im}(\omega)$ vs $\text{Re}(\omega)$ for $l=2$ (left) and $l=3$ (right).
}
\label{fig:polar_spectrum_massive_WH}
\end{figure*}

\begin{figure*}[h!]
\begin{center}
\mbox{
\resizebox{\textwidth}{!}{
    \includegraphics[angle=-90]{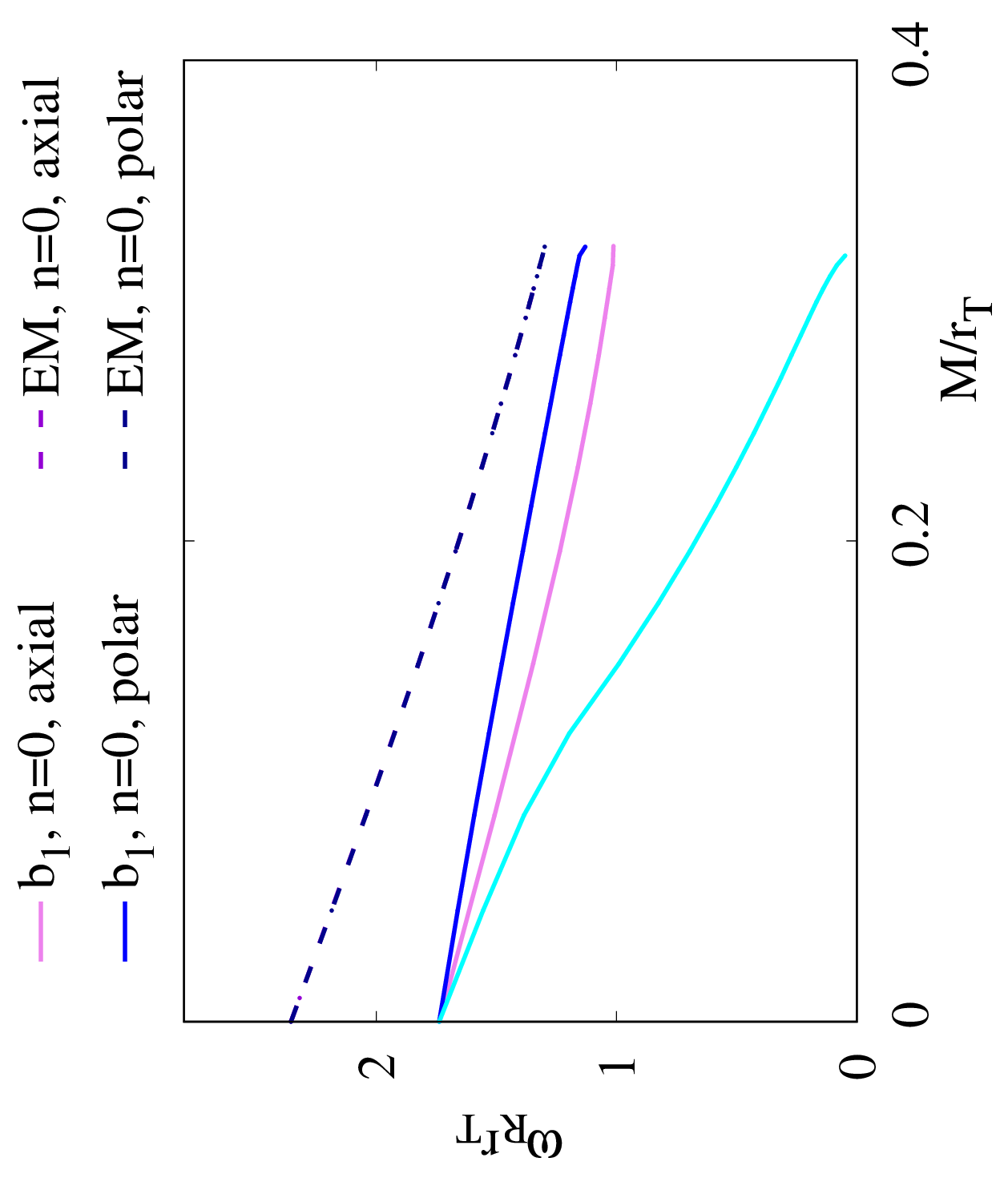}
    \includegraphics[angle=-90]{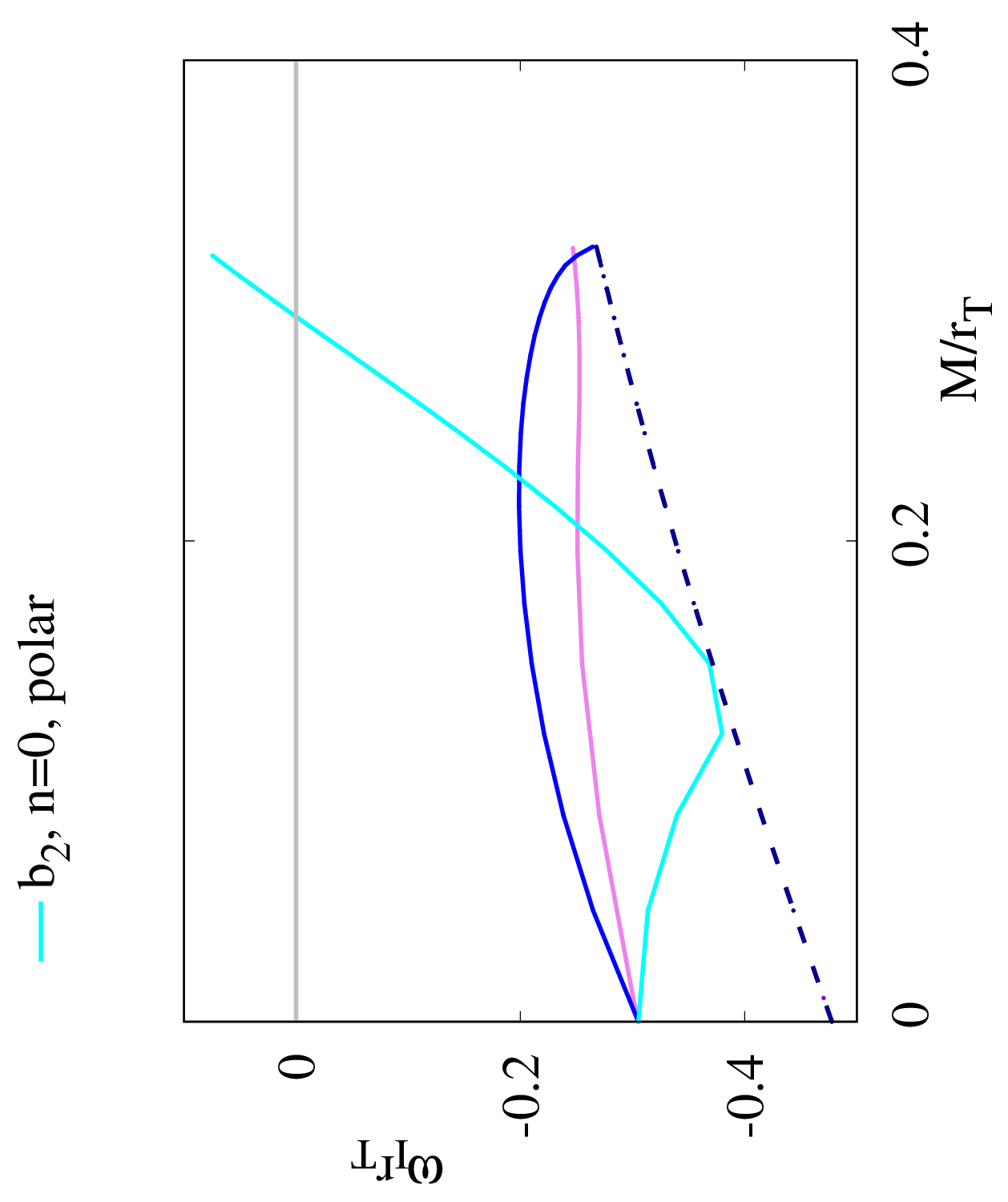}
}
}
\vspace*{-0.7cm}
\end{center}
\caption{
Isospectrality of $l=2$ axial and polar modes for EB wormholes. 
The EM (dashed) curves overlap for these uncharged configurations.}
\label{fig:Isospectrality}
\end{figure*}

\subsubsection{Polar modes}

Let us continue now with the polar modes.
The $l=2$ modes are shown in row 2 of Figure \ref{fig:axial_polar_massive_WH} and the $l=3$ modes are shown in row 4.
An overall similar behavior to that of the axial modes is found for both $\omega_R$ and $\omega_I$. 
The real part decreases, in general, with $M/r_T$ while the imaginary part increases with it. 
However, also important differences arise w.r.t.~the results for the axial spectrum. 
The first feature to discuss is the presence of an additional branch of polar modes compared to the axial spectrum.
Since the scalar field behaves as a polar function, we have an additional polar branch of modes in the spectrum (for each pair of $l,n$).
We denote these polar branches $b_1$ and $b_2$, as we cannot identify them as gravitational-led or scalar-led, without being able to connect them to a unique decoupling limit. In other words, in both branches the gravitational perturbations and the scalar perturbations are strongly coupled together.

Furthermore, we see in Figure \ref{fig:axial_polar_massive_WH} that, in the massless limit, there exists a degeneracy of the polar $b_1$ and $b_2$ modes. 
In fact, all three modes, the axial $b_1$, and the polar $b_1$ and $b_2$, coincide in this limit \cite{Azad:2022qqn}. 
When the mass of the wormhole is increased, the degeneracy is broken, and we obtain three distinct branches.
For example, the fundamental polar branches are shown in blue and cyan in rows 2 and 4 of Figure \ref{fig:axial_polar_massive_WH}, while the corresponding fundamental axial branch is shown in blue in rows 1 and 3.

Another feature we notice is that some of the curves cross or repel each other. 
For example, in Figure \ref{fig:axial_polar_massive_WH}, we observe that the fundamental $l=2$ polar branches $b_1$ and $b_2$ repel each other for small masses.
But their imaginary parts $\omega_I$ cross when $M/r_T\approx0.23$. 
Thus, the $n=0$ $b_2$ branch becomes the longest-lived mode of the spectrum. 
Such a behavior has also been observed in e.g. regular black holes \cite{Khoo:2025qjc}.

Surprisingly, the imaginary part of this polar $l=2$ $b_2$ branch continues to rise further, such that it crosses zero.
Consequently, the fundamental $l=2$ $b_2$ branch becomes unstable for $M/r_T\gtrapprox 0.3$.
This nonradial instability was not found in previous studies of the minimally coupled EB family, where smaller values of the mass were considered. 
Instabilities have also been reported for charged wormholes in a distinct Einstein-Maxwell-phantom theory with a non-minimal scalar-Maxwell coupling \cite{Wu:2025wlz}.
However, the underlying background solutions, perturbation equations, and instability mechanism differ from those considered here.
We note that this $l=2$ instability is not yet seen for $l=3$, where, however, the imaginary part $\omega_I$ of the 2nd excited $b_1$ mode is seen to cross the 1st excited and the fundamental modes on its rise toward zero.
This suggests that it will also become unstable above a critical mass.
The real parts $\omega_R$ of these two modes, the fundamental $l=2$ $b_2$ branch and the 2nd excited $l=3$ $b_1$ branch, also exhibit analogous behavior.
Both decrease rapidly toward zero, while crossing all modes in between.
The corresponding polar spectra in the complex plane are shown in 
Figure \ref{fig:polar_spectrum_massive_WH}.

For the imaginary parts $\omega_I$ in the polar sector we further observe that the 1st excited $l=2$ $b_1$ mode tends toward the fundamental EM mode and the fundamental $l=2$ $b_1$ mode for high values of the mass.  
Likewise, the 2nd excited $l=2$ $b_1$ mode approaches the 1st excited EM mode and the 1st excited $l=2$ $b_2$ mode.
An analogous behavior is found for the imaginary part $\omega_I$ of the $l=3$ polar modes. 

Finally, in Figure \ref{fig:Isospectrality} we show together the axial and polar modes. The figure 
exhibits the isospectrality of the EM modes (note that both dashed curves overlap perfectly, see also Tables \ref{table1} and \ref{table2}).
As noted above, the axial and two polar QNMs coincide when $M/r_T=0$  \cite{Azad:2022qqn}.
But here we have moreover shown that the EM modes are isospectral \textit{for all} $M/r_T$ when $Q_e=0$.

\subsubsection{Radial modes}

\begin{figure*}[h!]
\begin{center}
\mbox{
\resizebox{\textwidth}{!}{
    \includegraphics[angle=-90]{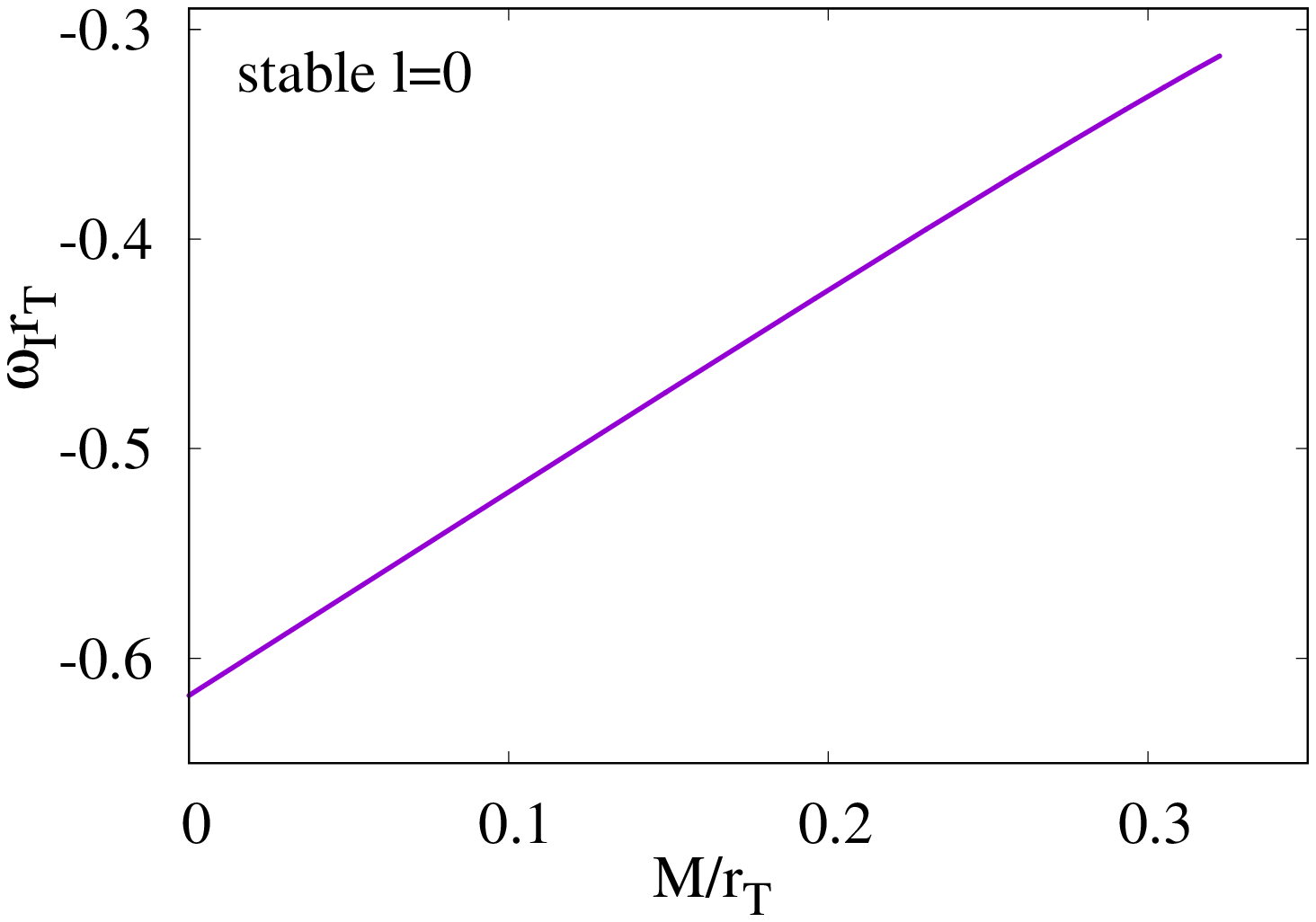}
    \includegraphics[angle=-90]{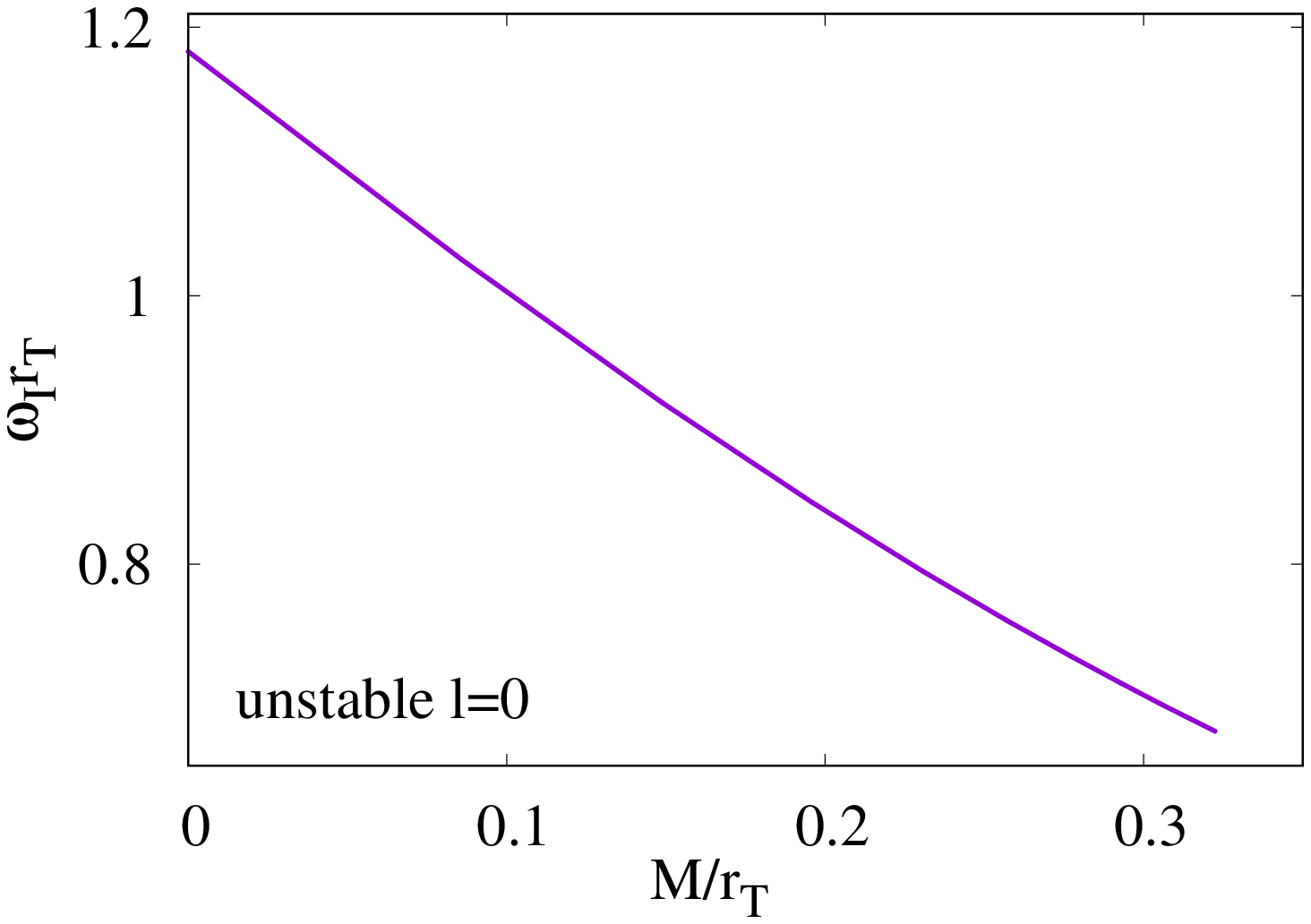}}}
\vspace*{-0.7cm}
\end{center}
\caption{
Radial modes for EB wormholes.
Stable (left) and unstable (right) branches $\text{Im}(\omega) r_T$ vs $M/r_T$.
}
\label{fig:radial_spectrum_massive_WH}
\end{figure*}

Let us discuss now the radial modes of the EB wormholes.
We show the imaginary part $\omega_I$ of the stable and unstable radial branches of the Ellis wormhole in Figure \ref{fig:radial_spectrum_massive_WH} (left panel and right panel, respectively). 
The stable branch increases with mass, analogous to the previous results.  
In contrast, the unstable branch monotonically decreases with mass \cite{Blazquez-Salcedo:2018ipc}. 
But it is far from reaching the stability limit within the range of masses studied.
Recall that, for larger values of the mass, the spectral method loses accuracy. 
We thus discontinued the calculations once the estimated numerical accuracy became worse than $10^{-3}$. 
We note that our calculations show only a single branch of stable modes for $l=0$ radial perturbations.

\subsection{Critical wormholes}

Now, we turn to the charged wormholes and start with the discussion of the QNMs of the critical wormholes, for which $\Lambda=0$.

\subsubsection{Axial modes}

\begin{figure*}[p!]
\begin{center}
\mbox{
\includegraphics[width=1.2\textwidth,angle=-90]{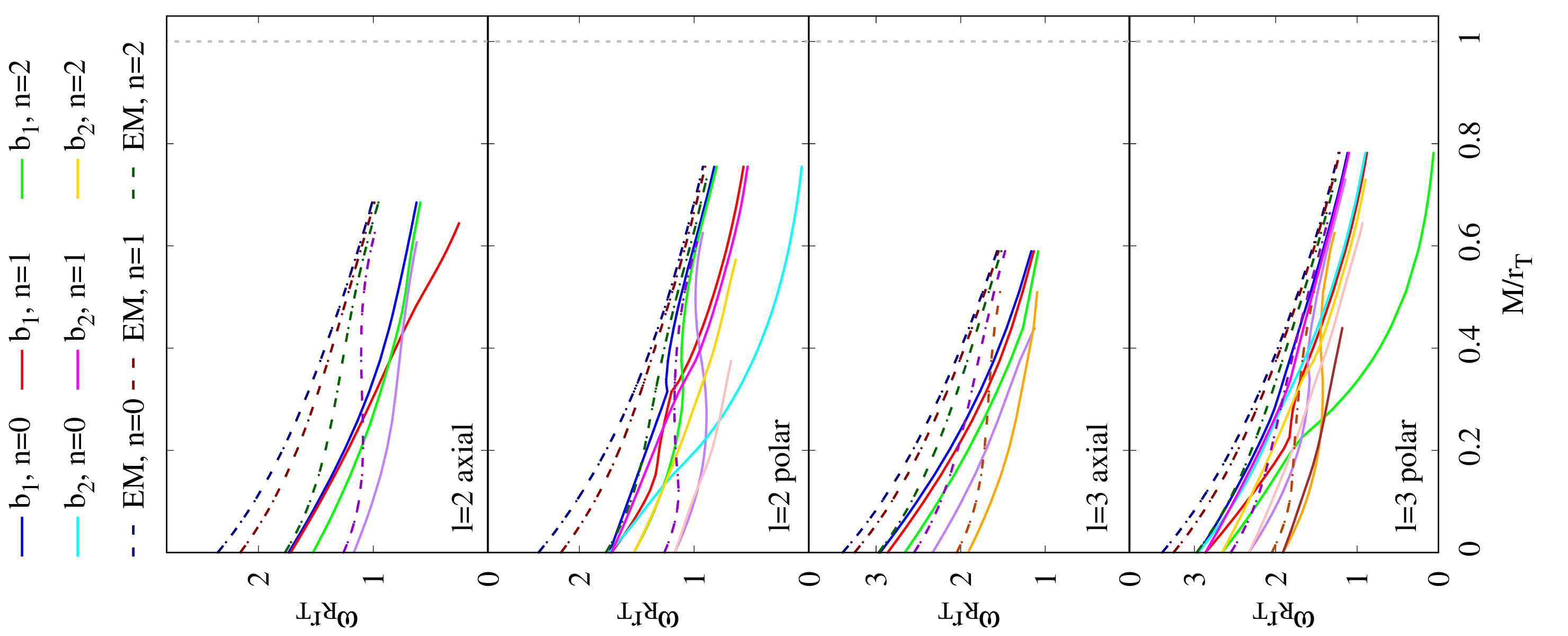}
\includegraphics[width=1.2\textwidth,angle=-90]{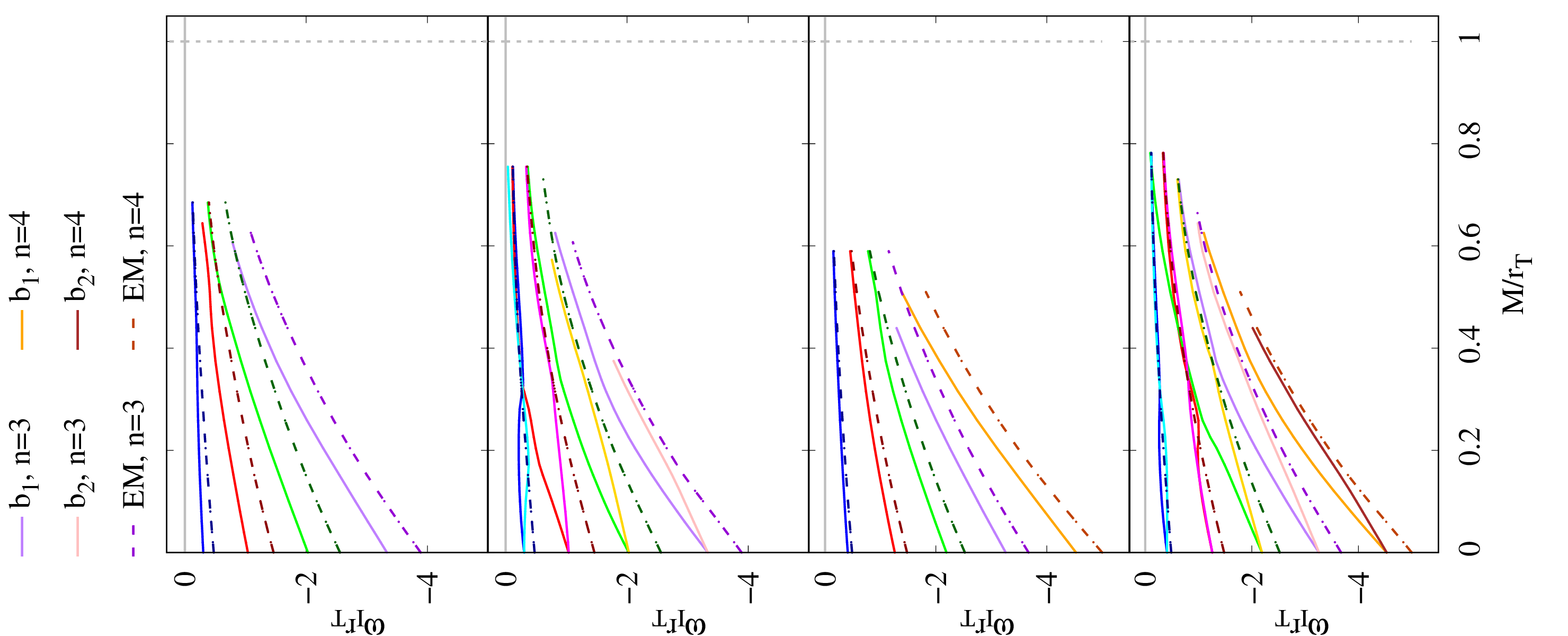}
}
\vspace*{-0.5cm}
\end{center}
\caption{
$l=2$ and $l=3$ axial and polar modes for critical wormholes. 
From top to bottom the rows correspond to: $l=2$ axial, $l=2$ polar, $l=3$ axial, $l=3$ polar modes. 
Left panels show the real part of the QNMs. 
Right panels exhibit the imaginary part. 
The gray vertical line denotes $\gamma_1^{\rm{max}}=2/\pi$.
}
\label{fig:axial_polar_charged_WH}
\end{figure*}

In Figure~\ref{fig:axial_polar_charged_WH}  we show $\omega r_T$ versus $M/r_T$ for the axial and polar modes of the critical wormholes. 
The left column shows the real part of $\omega$, while the right column shows its imaginary part.
The axial modes are displayed in rows 1 and 3 for the $l=2$ and $l=3$ multipoles, respectively. 
In each panel, solid lines represent the gravitational modes and dashed lines the electromagnetic modes.

Analogously to the EB wormhole case, the imaginary part $\omega_I$ increases with $M/r_T$ (meaning the damping time grows).
We note that the damping times of the gravitational modes are always greater than those of the corresponding EM ones.

Turning to the real part $\omega_R$ of the axial modes of the critical wormholes, we observe that the real part $\omega_R$ decreases for most of the modes with increasing mass. 
However, the 3rd excited EM $l=2$ mode shows a different behavior. 
After initially decreasing, this branch starts growing until $M/r_T\approx0.4$,  crossing all other gravitational branches. 
We also see that the EM modes tend to approach each other, meaning that the frequencies of the EM modes are almost degenerate, although the damping times are still very different, as can be seen in the right panels. 

\begin{figure*}[h!]
\begin{center}
\mbox{
\resizebox{\textwidth}{!}{
    \includegraphics[angle=-90]{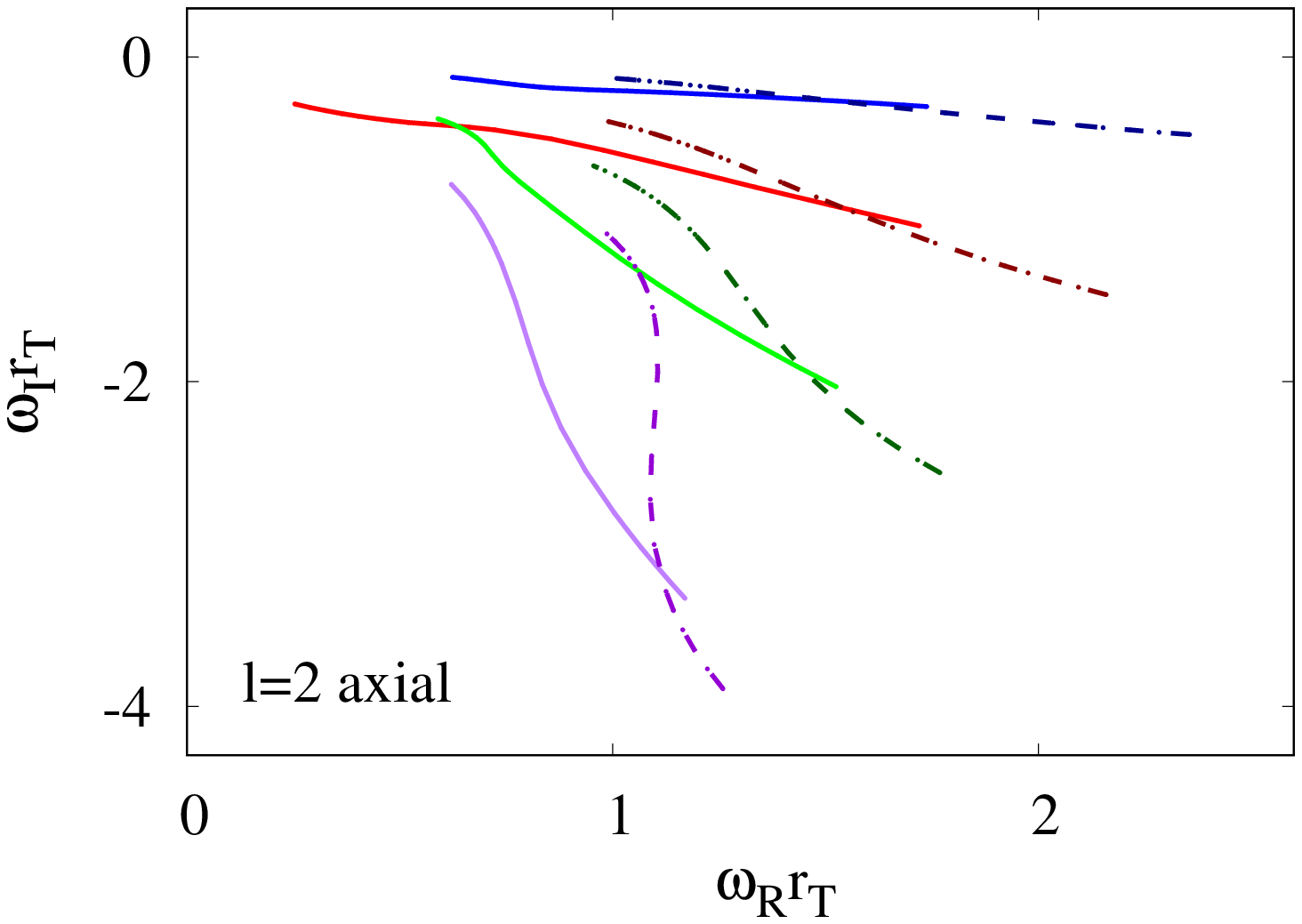}
    \includegraphics[angle=-90]{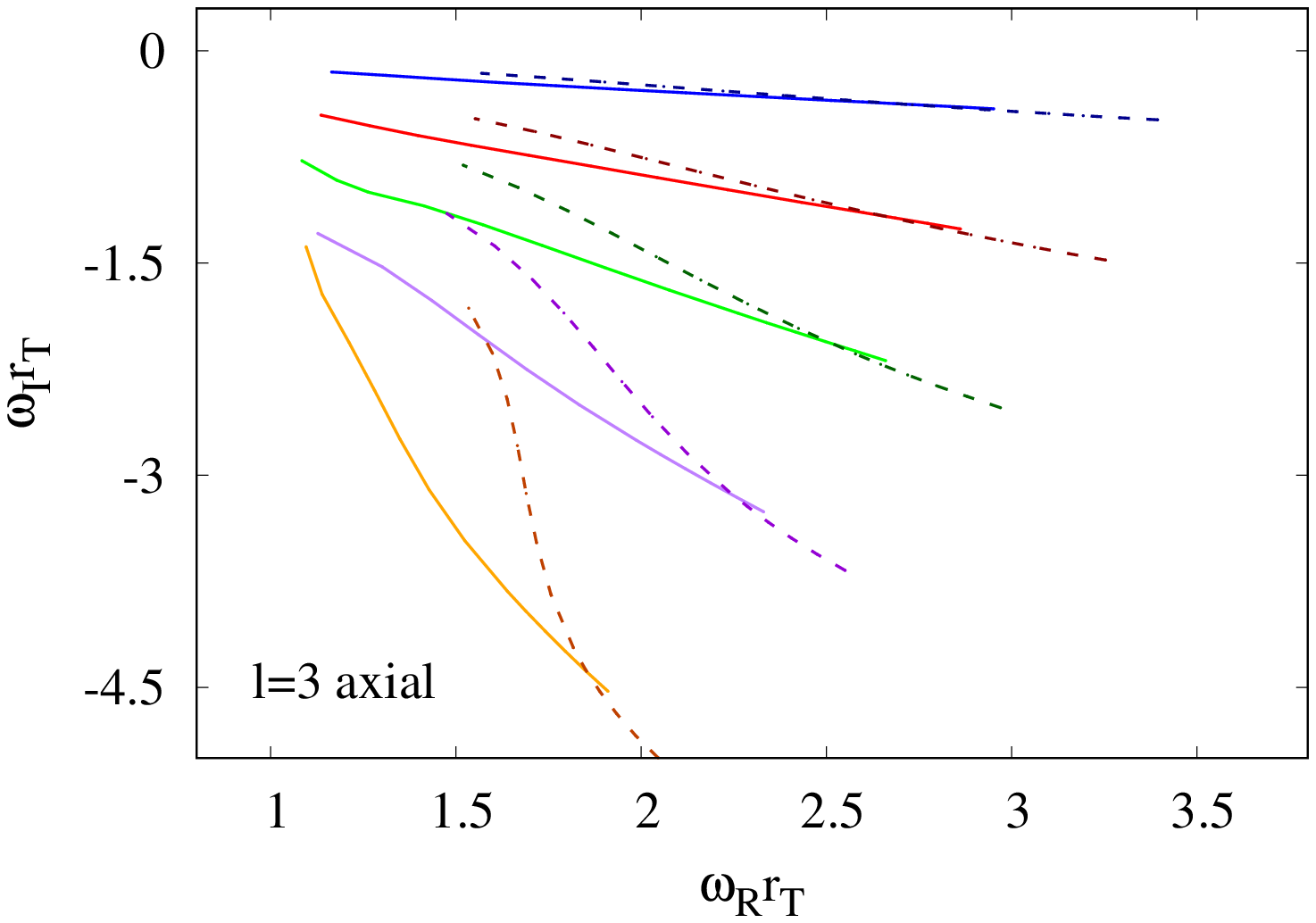}
}
}
\vspace*{-0.7cm}
\end{center}
\caption{
Axial modes for critical wormholes.
$\text{Im}(\omega)$ vs $\text{Re}(\omega)$ for $l=2$ (left) and $l=3$ (right).
}
\label{fig:axial_spectrum_charged_WH}
\end{figure*}

Finally, we show in Figure \ref{fig:axial_spectrum_charged_WH} the modes in the complex plane $(\omega_R,\omega_I)$. The calculated spectrum does not show any evidence of unstable modes.

\subsubsection{Polar modes}

We show the $l=2$ and the $l=3$ polar modes in the second and fourth rows of Figure \ref{fig:axial_polar_charged_WH} respectively.
Here we find again a pattern that is in many aspects similar to the one observed before for the polar modes of the EB wormholes.
We recognize the general trend of the imaginary part $\omega_I$ to increase with mass and the real part $\omega_R$ to decrease with mass.
Moreover, the imaginary part of the fundamental $l=2$ $b_2$ branch and the 2nd excited $l=3$ $b_1$ branch again exhibit a distinctive behavior.
Both branches cross the other branches with smaller $\omega_I$, to become the longest-lived modes beyond some critical value of the mass.
However, in contrast to the EB wormhole case, we do not observe an instability for the $l=2$ branch.
At the same time, the real parts of these two modes approach zero for large masses.

\begin{figure*}[h!]
\begin{center}
\mbox{
\resizebox{\textwidth}{!}{
    \includegraphics[angle=-90]{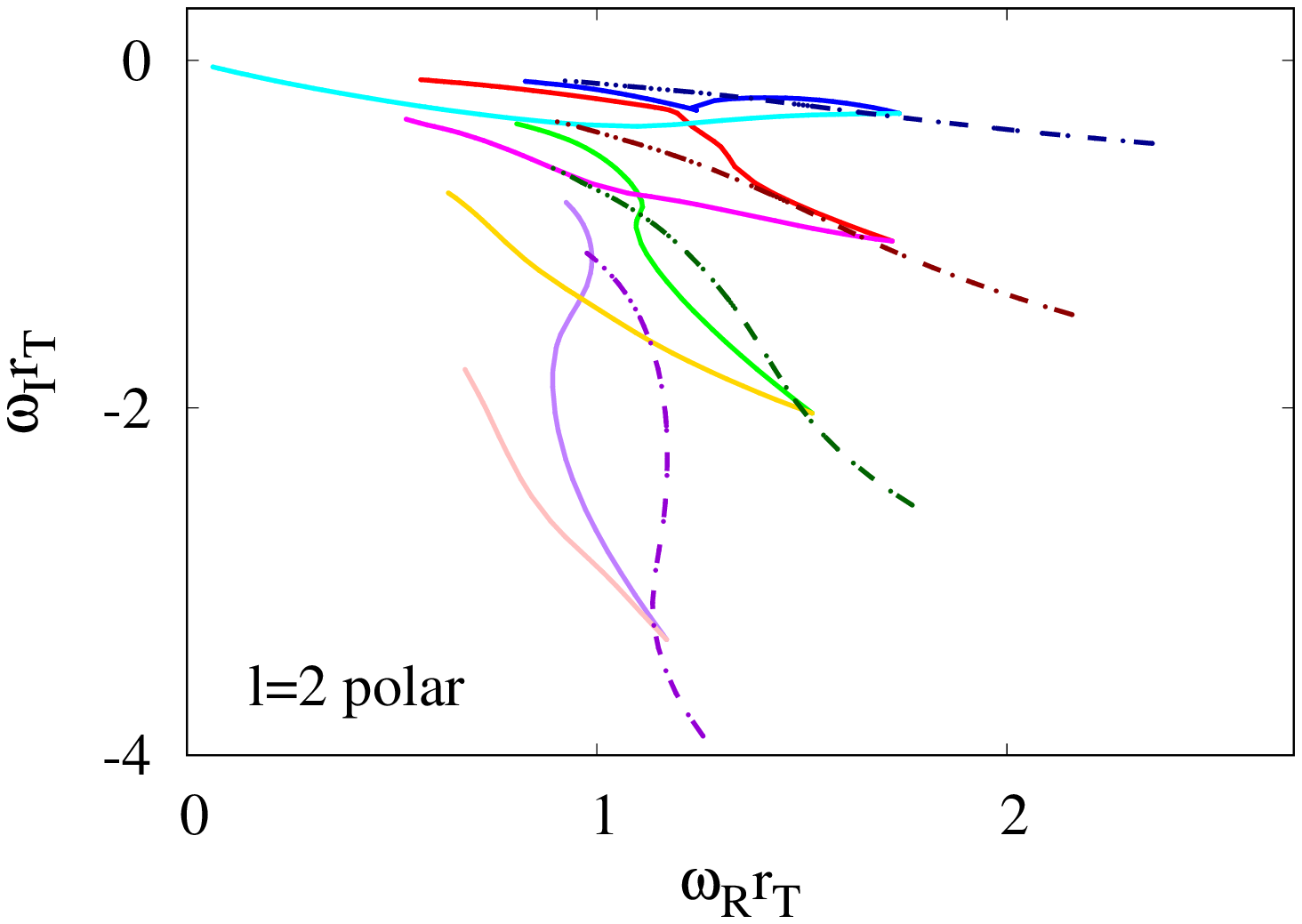}
    \includegraphics[angle=-90]{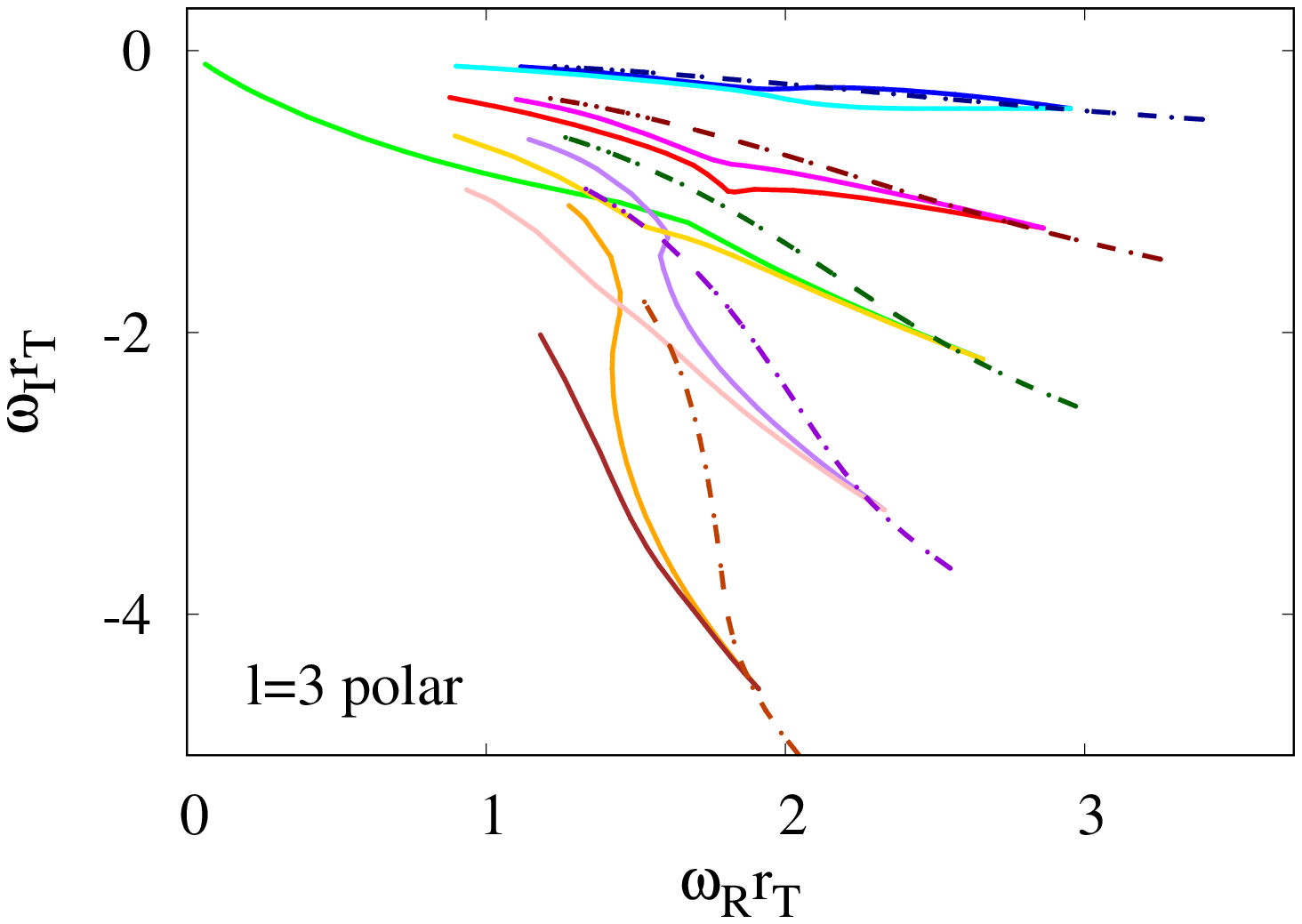}
}
}
\vspace*{-0.7cm}
\end{center}
\caption{
Polar modes for critical wormholes.
$\text{Im}(\omega)$ vs $\text{Re}(\omega)$ for $l=2$ (left) and $l=3$ (right).
}
\label{fig:polar_spectrum_charged_WH}
\end{figure*}

In Figure \ref{fig:polar_spectrum_charged_WH} we exhibit the polar spectrum in the complex plane.
Here the particular behavior of the fundamental $l=2$ $b_2$ branch and the 2nd excited $l=3$ $b_1$ branch can be better appreciated.
In addition, it is clearly seen that the degeneracy of the fundamental $b_1$ and $b_2$ modes of the wormholes at vanishing mass is broken by the presence of charge, yielding two distinct branches of modes.  
The EM branches are relatively isolated and behave similarly to the axial EM modes.

\begin{figure*}[h!]
\begin{center}
\mbox{
\resizebox{\textwidth}{!}{
    \includegraphics[angle=-90]{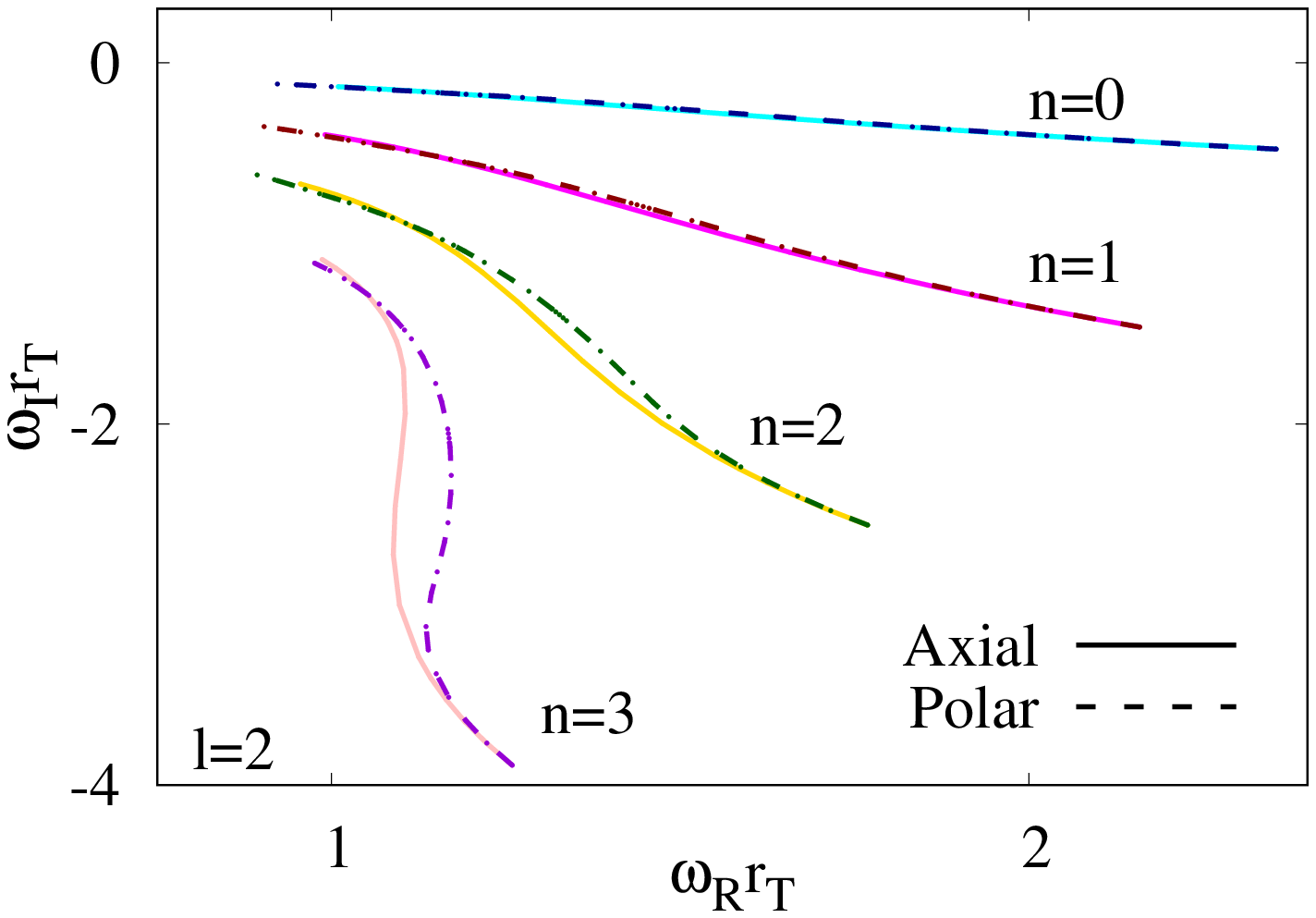}
    \includegraphics[angle=-90]{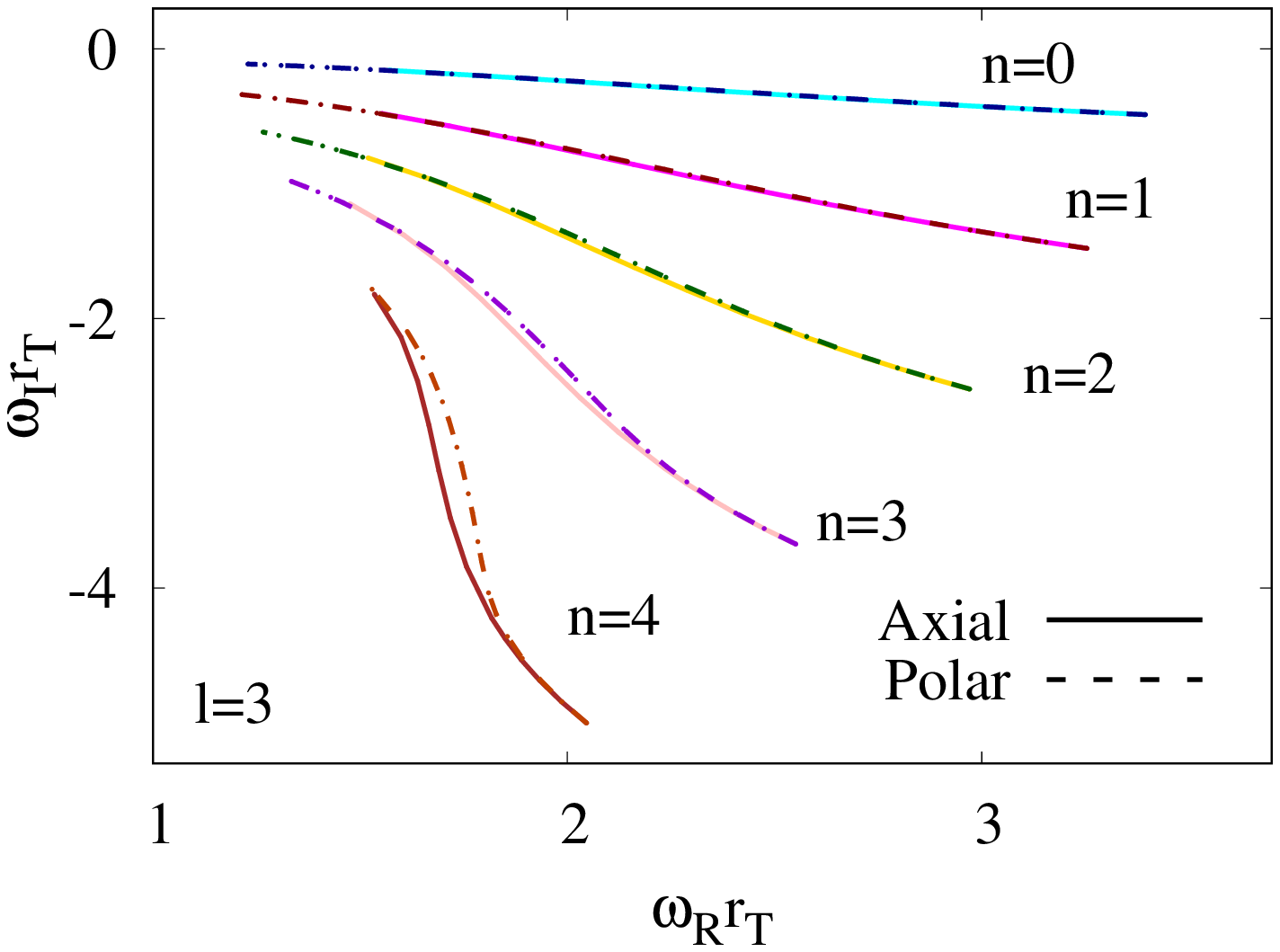}
}
}
\vspace*{-0.7cm}
\end{center}
\caption{
Comparison of EM modes for critical wormholes. Solid lines correspond to axial modes, and dashed lines to polar modes. We show the fundamental mode and several excitations in the left panel for $l=2$ and in the right panel for $l=3$.
}
\label{fig:EM_spectrum_critical_WH}
\end{figure*}

However, we note that there is no longer isospectrality of the axial and polar EM modes for these charged wormholes since the EM perturbations are no longer decoupled.
Comparing the EM modes of the critical wormholes in Figure \ref{fig:EM_spectrum_critical_WH} shows explicitly that the isospectrality of the uncharged case is broken once we increase the mass. However, the figure also suggests that, interestingly, the axial and polar modes seem to tend to each other again for large enough values of the mass.

\subsubsection{Radial modes}

\begin{figure*}[h!]
\begin{center}
\mbox{
\resizebox{\textwidth}{!}{
    \includegraphics[angle=-90]{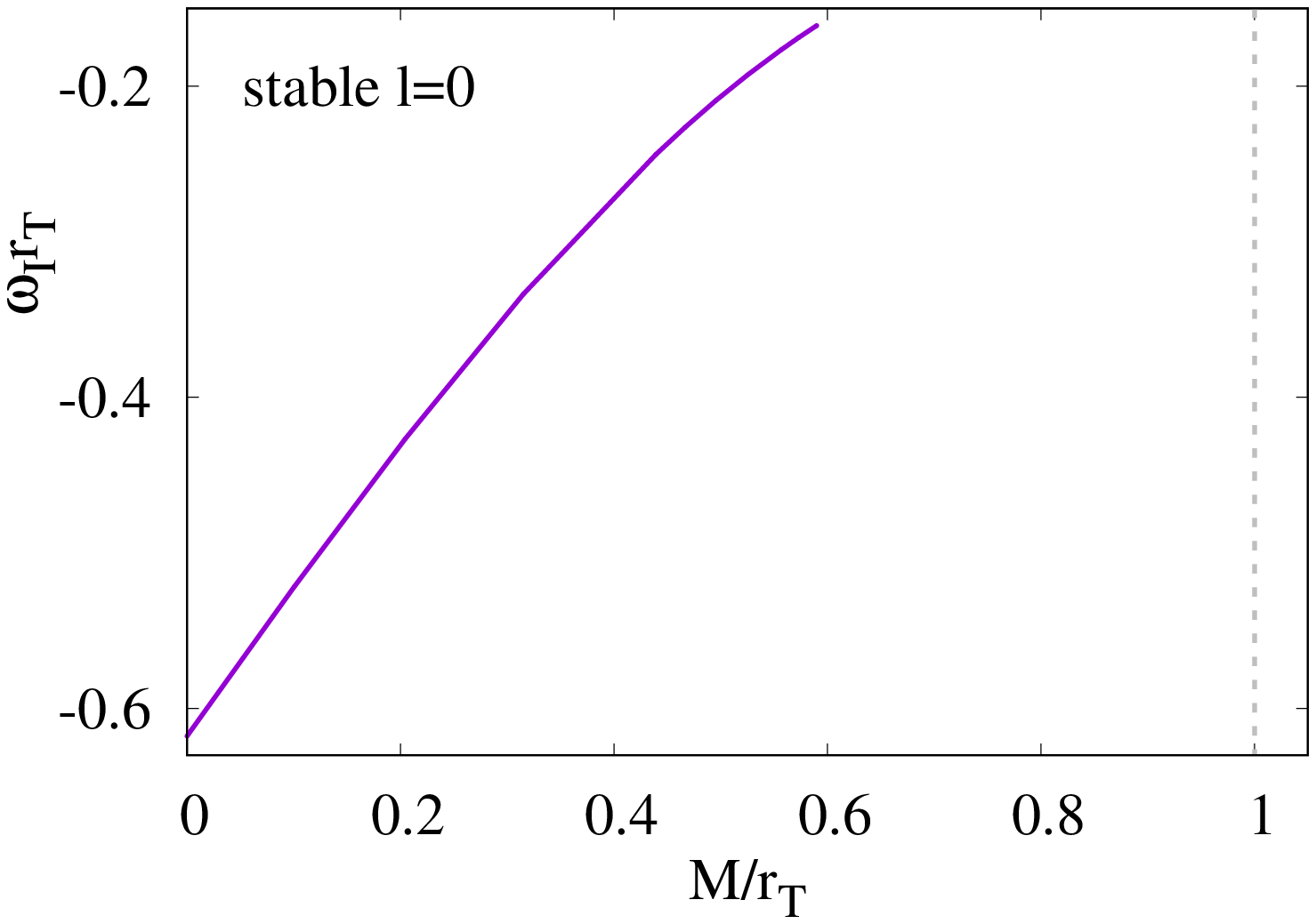}
    \includegraphics[angle=-90]{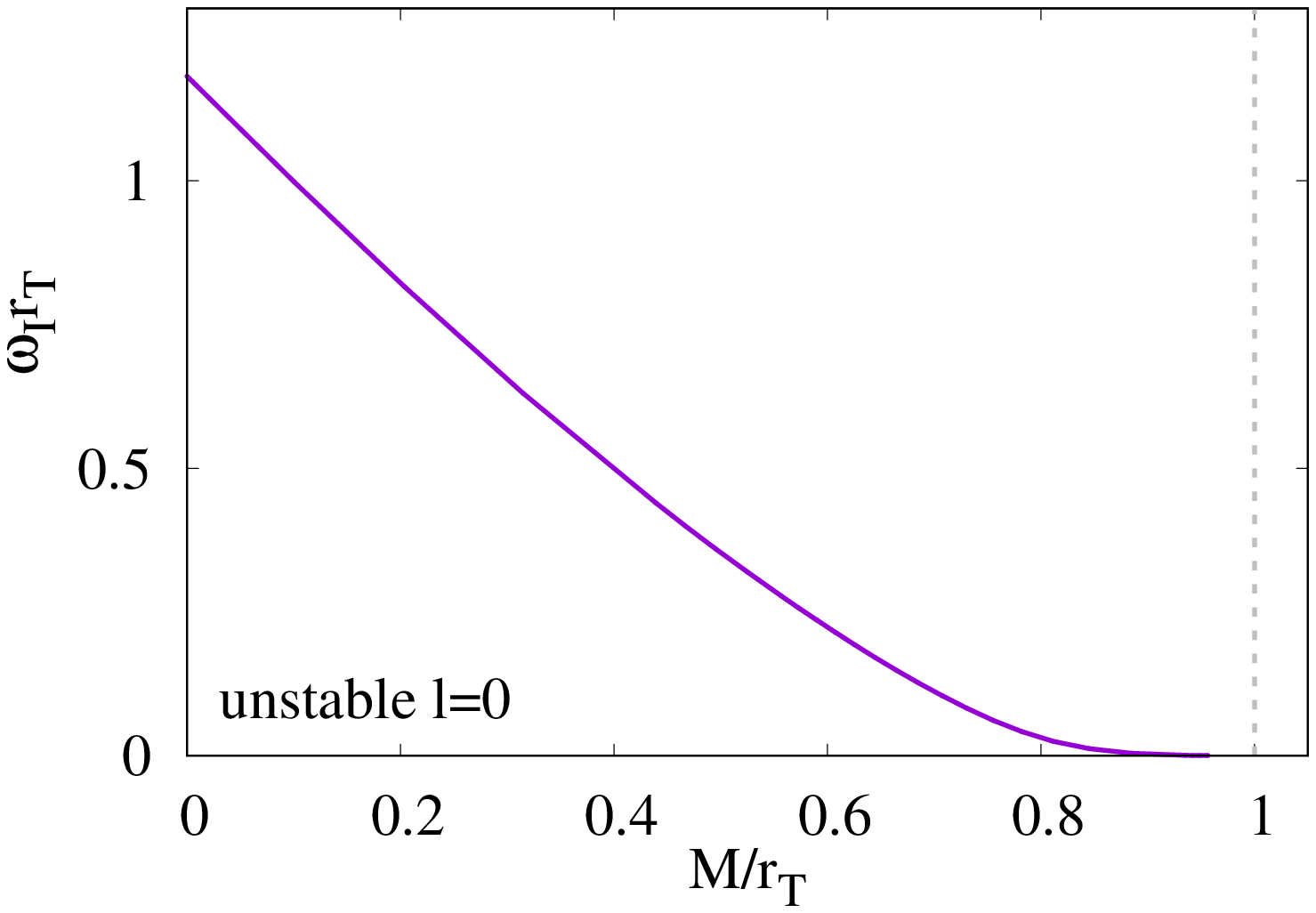}
}
}
\vspace*{-0.7cm}
\end{center}
\caption{
Radial modes for critical wormholes.
Stable (left) and unstable (right) branches $\text{Im}(\omega)r_T$ vs $M/r_T$. The gray vertical line indicates $\gamma_1^{\rm{max}}=2/\pi$.}
\label{fig:radial_spectrum_charged_WH}
\end{figure*}

We now consider the radial modes of the critical wormholes.
Again we find that the spectrum is composed of one unstable and one stable mode, like in the uncharged EB wormhole case.
As seen in Figure \ref{fig:radial_spectrum_charged_WH}, the imaginary part $\omega_I$ of the stable mode increases with increasing mass (i.e., increasing electric charge) in the same way as seen in the above results for the nonradial modes of the charged wormholes. 
The most interesting behavior is found for the unstable branch. 
Although the mode is still unstable, we see that $\omega_I$ decreases with increasing $M/r_T$. 
Thus, the instability relaxes as the limiting eRN solution is approached
(residing at $\gamma_1^{\rm{max}}=\frac{2}{\pi}$).
In fact, $\omega_I$ tends to zero very quickly in the limit, following $\omega_I \sim \left(1-M/r_T\right)^{3.1}$ \cite{Blazquez-Salcedo:2025dit}.

\subsection{Subcritical wormholes}

The next step is to address the QNMs of the subcritical wormholes (featuring $\Lambda\neq 0$, $\gamma_1\neq 0$).

\subsubsection{Axial modes}

\begin{figure*}[h!]
\begin{center}
\mbox{
\includegraphics[width=0.72\textwidth,angle=-90]{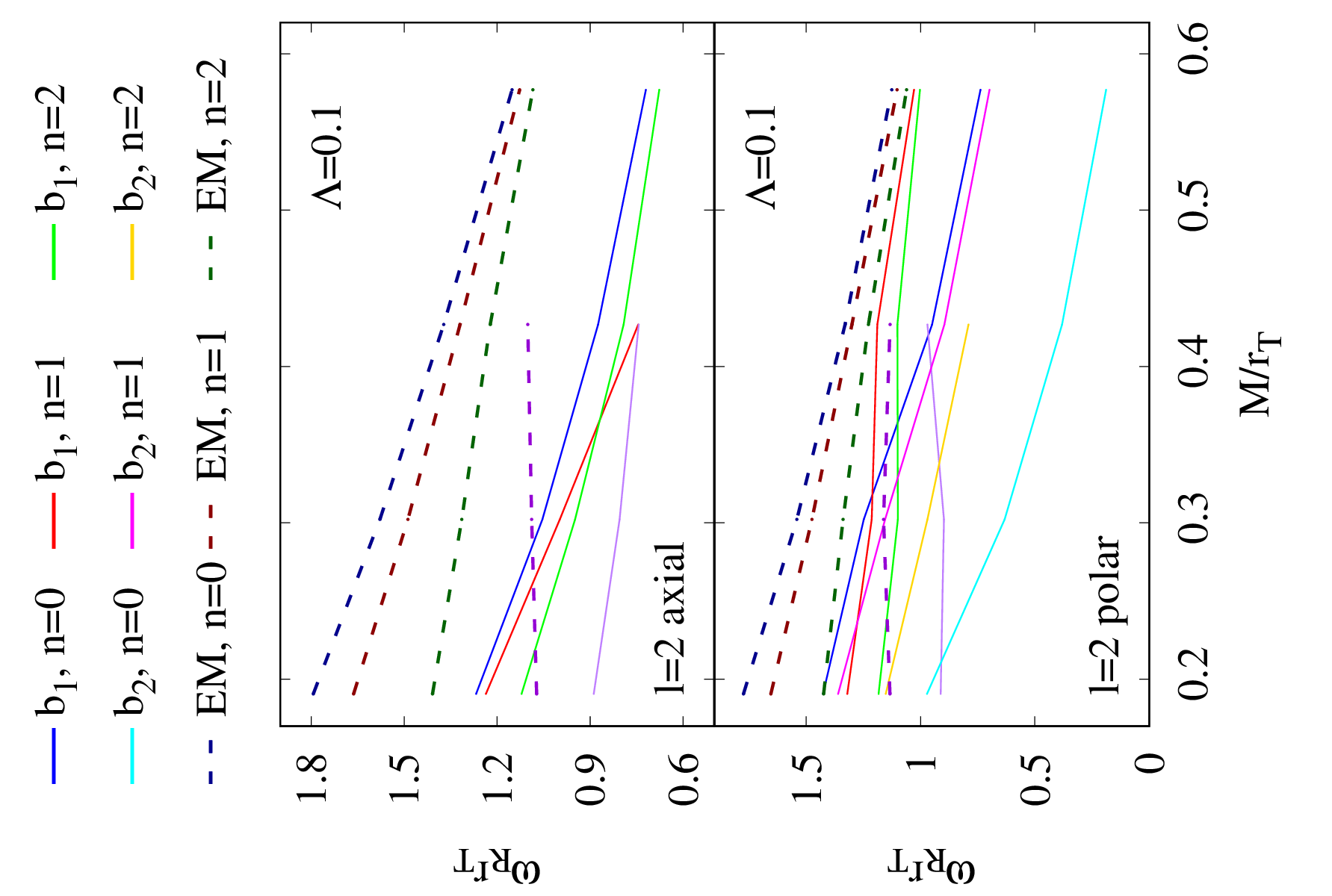}
\includegraphics[width=0.72\textwidth,angle=-90]{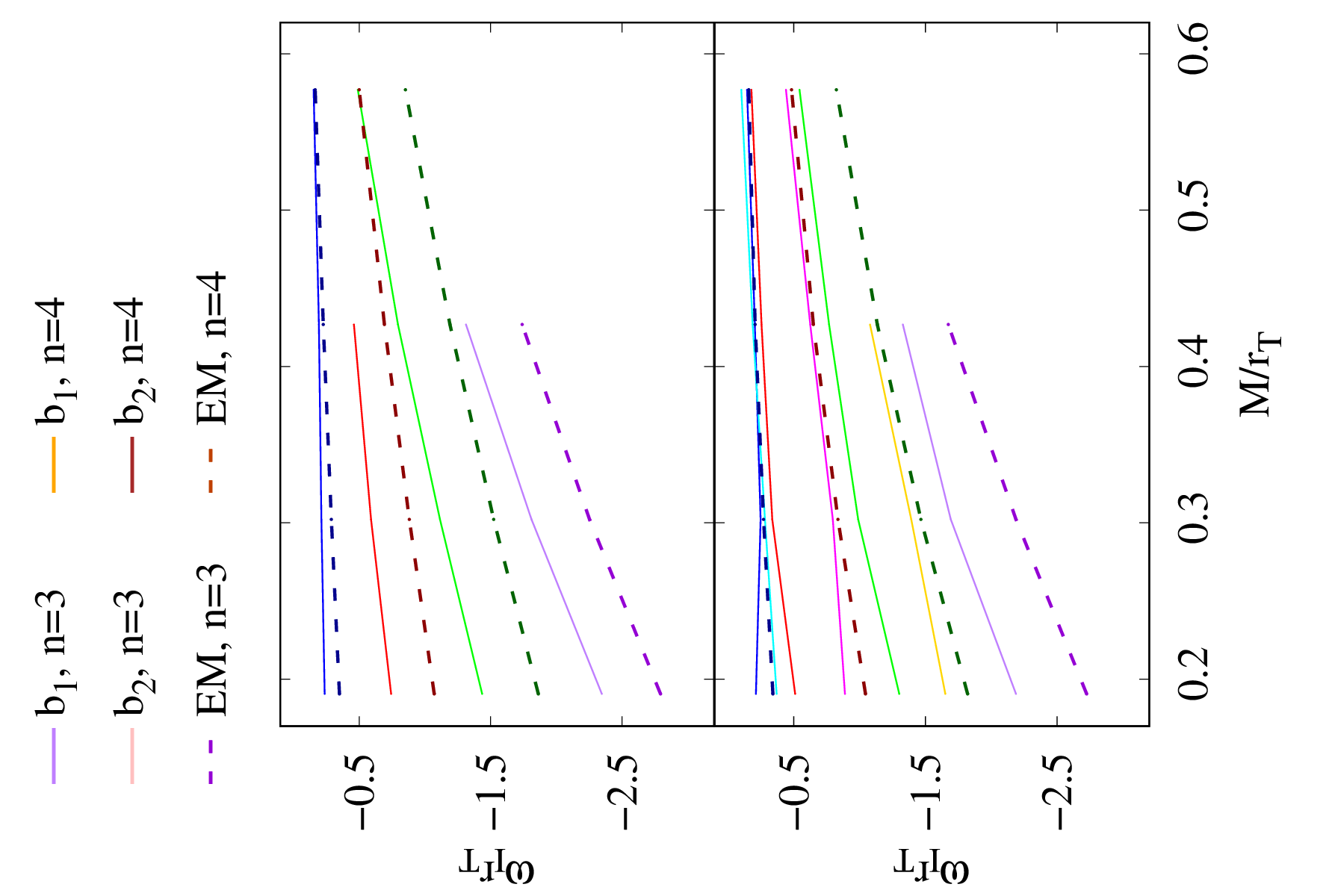}
}
\vspace*{-0.5cm}
\end{center}
\caption{
$l=2$ axial and polar modes for subcritical wormholes with $\Lambda=0.1$. 
From top to bottom the rows correspond to: $l=2$ axial, $l=2$ polar. 
Left panels show the real part of the QNMs. 
Right panels exhibit the imaginary part.
}
\label{fig:axial_polar_massive_charged_WH}
\end{figure*}

In Figure \ref{fig:axial_polar_massive_charged_WH} we show the QNMs of the subcritical wormholes with $\Lambda=0.1$, focusing on the $l=2$ modes. In the upper panels, we show the axial modes scaled to the throat radius as a function of $M/r_T$.

In the left panel we can see that $\omega_R$ decreases with mass for all modes except for the 3rd excited EM mode, which crosses some of the gravitational modes with higher $\omega_R$, tending to the other EM modes. 
The $\omega_R$ branches of the first and second excited gravitational modes also cross.
On the other hand, in the right panel we observe that, as expected, $\omega_I$ increases monotonically with $M/r_T$, but remains well below zero. 

\begin{figure*}[h!]
\begin{center}
\mbox{
\resizebox{\textwidth}{!}{
    \includegraphics[angle=-90]{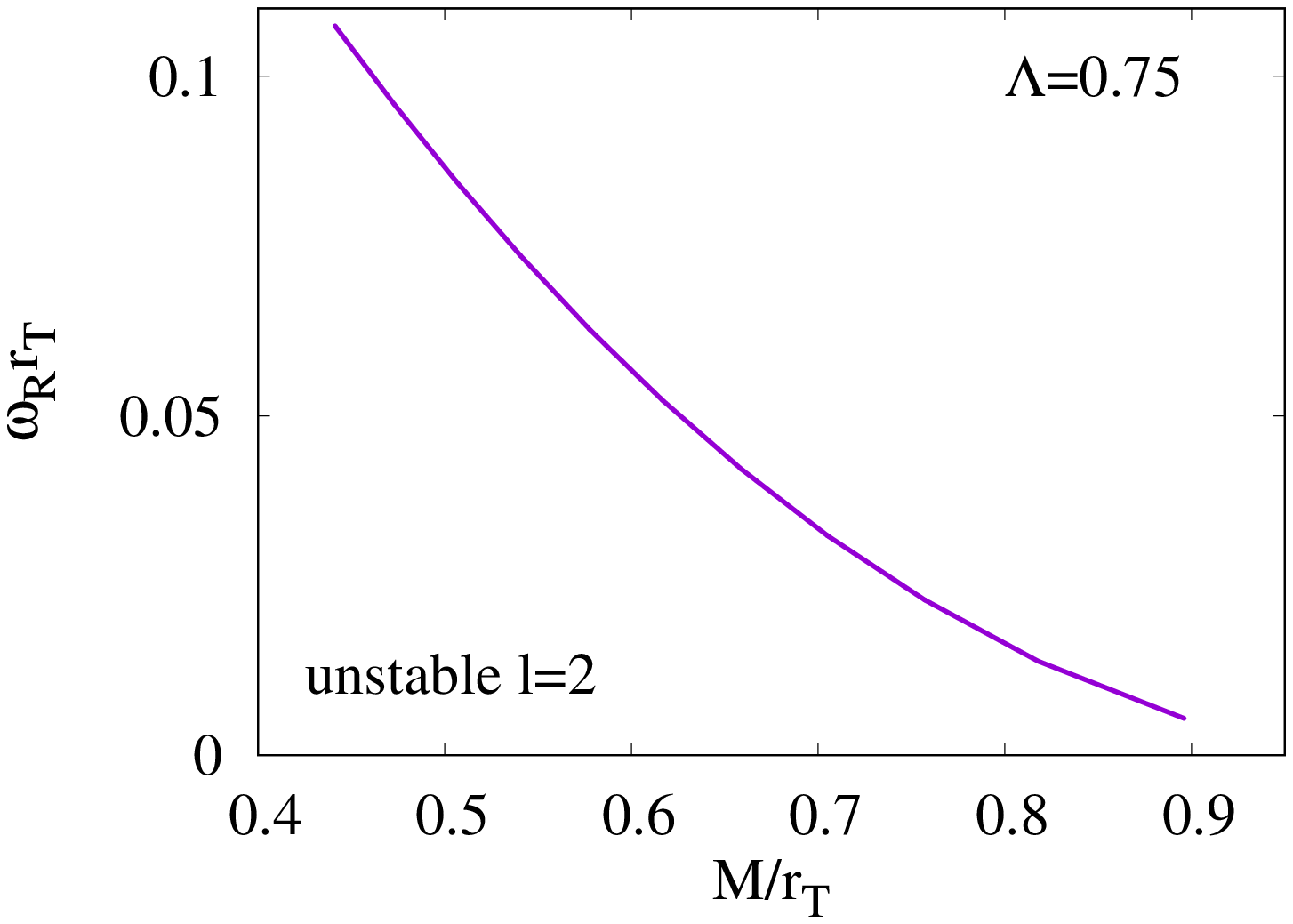}
    \includegraphics[angle=-90]{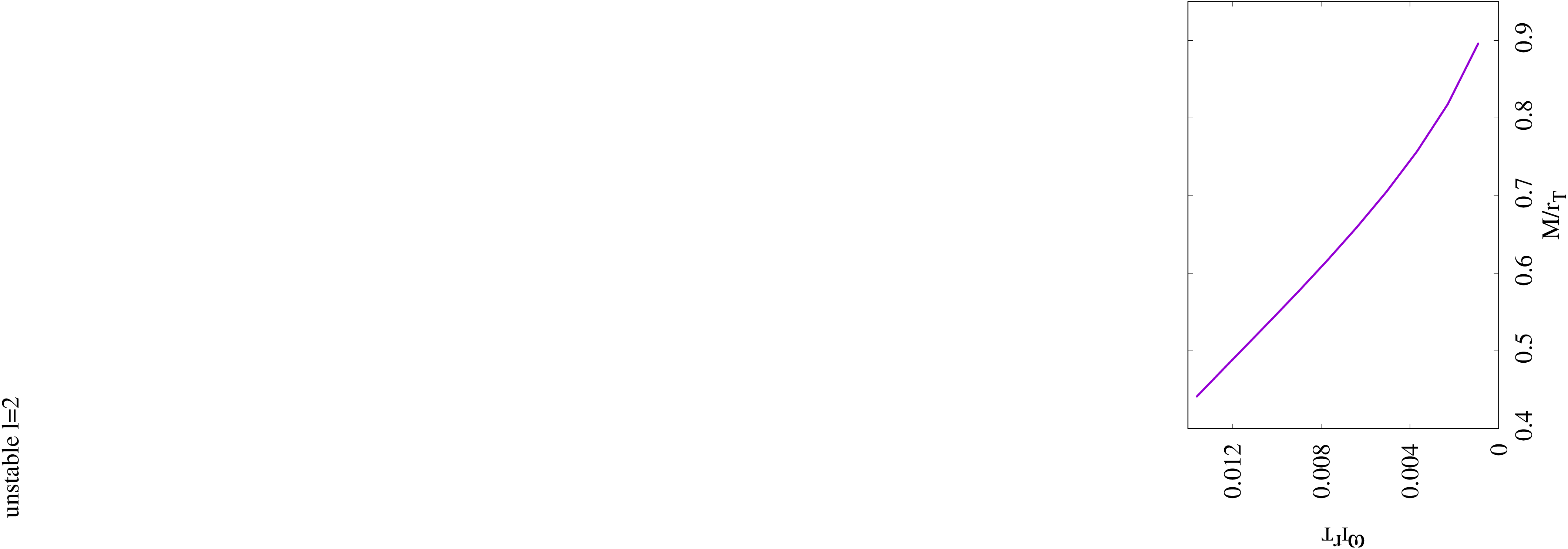}
}
}
\vspace*{-0.7cm}
\end{center}
\caption{
$l=2$ polar unstable $b_2$ branch for subcritical wormholes, for which $\Lambda$ is fixed to $0.75$. 
Left: Real part of the QNMs. 
Right: Imaginary part. 
}
\label{fig:unstable_polar_massive_charged_WH}
\end{figure*}

\subsubsection{Polar modes}

The second row of Figure \ref{fig:axial_polar_massive_charged_WH} exhibits the $l=2$ polar modes of subcritical wormholes, for the particular case $\Lambda=0.1$.
Here, a similar general pattern is found as above.
In particular, we recognize the behavior of the fundamental $l=2$ $b_2$ branch, with its strongly decreasing real part $\omega_R$ and its increasing imaginary part $\omega_I$.
Although this branch becomes the longest-lived mode beyond a critical value of the mass, the figure does not yet reveal an instability.

Recall that for the uncharged EB wormholes we observed that for $M/r_T\approx0.3$, an instability in the $l=2$ polar perturbations appeared (the fundamental $l=2$ $b_2$ branch of modes). 
Our results show that subcritical wormholes can also develop such an instability, but it appears only for sufficiently large values of $\Lambda$. 
As can be seen in Figure \ref{fig:axial_polar_massive_charged_WH}, for $\Lambda=0.1$ such an instability is not present. 
However, if we choose, for example, $\Lambda=0.75$, the fundamental $l=2$ $b_2$ branch of polar modes again becomes unstable. 
The above results show that the addition of electric charge suppresses the growth rate of this unstable branch. 
This is illustrated in Figure~\ref{fig:unstable_polar_massive_charged_WH}, where both the real and imaginary parts rapidly approach zero as the mass increases.

\subsubsection{Radial modes}

\begin{figure*}[h!]
\begin{center}
    \includegraphics[width=0.42\textwidth,angle=-90]{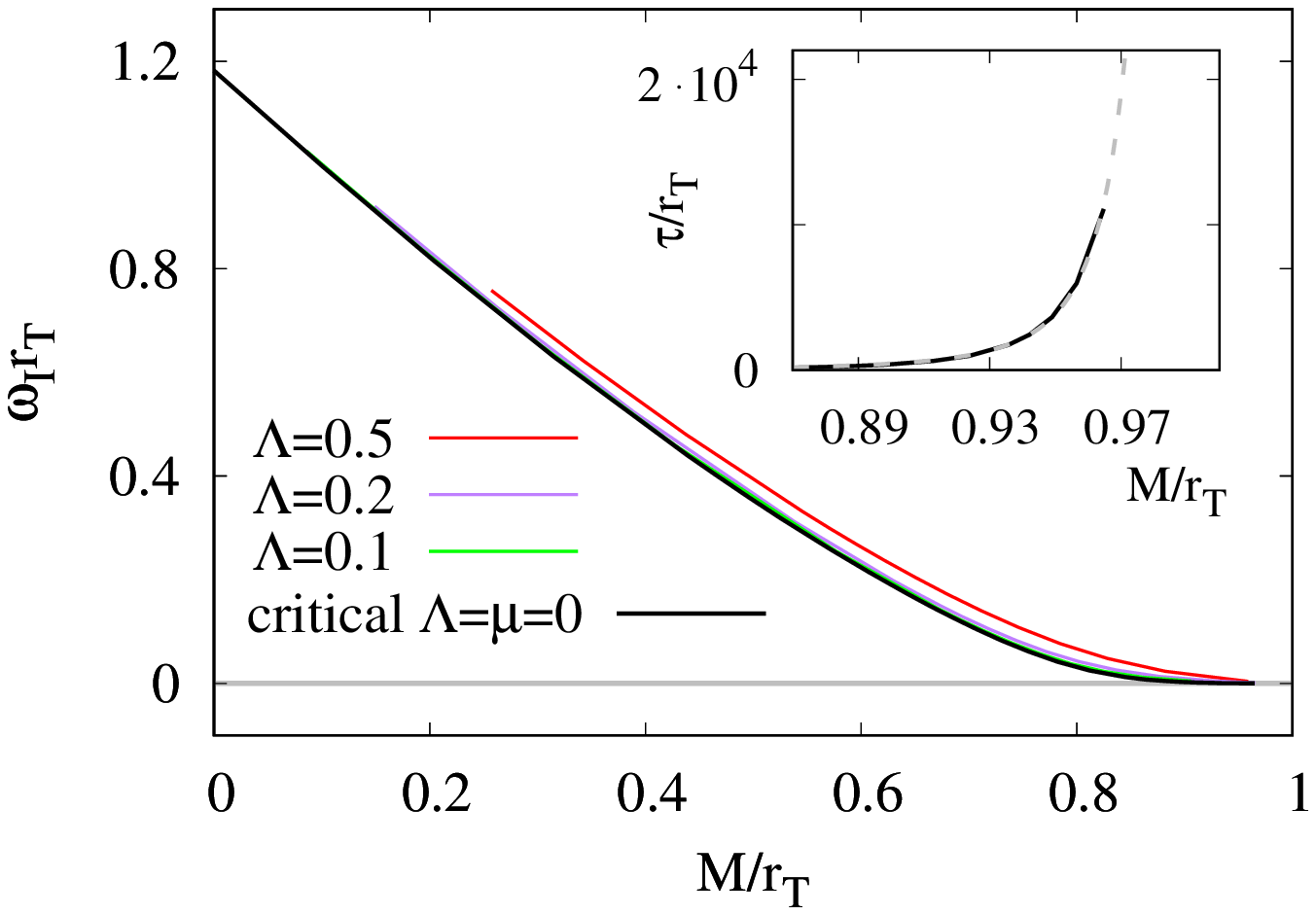}
\vspace*{-0.7cm}
\end{center}
\caption{
Radial unstable branch for subcritical wormholes with different values of $\Lambda$ (colored curves). For reference we also include the critical wormholes (black curve), and in the inset, the corresponding instability timescale $\tau$.  
}
\label{fig:unstable_radial_subcritical}
\end{figure*}

The fate of the radial instability of the subcritical wormholes is analogous to that of the critical wormholes.
This is seen in Figure \ref{fig:unstable_radial_subcritical}, where the eigenvalue $\omega_I$ of the unstable mode is shown for several values of $\Lambda$, including the critical one.
The figure also shows in the inset the increase of the instability timescale $\tau$ as the limiting eRN black hole is approached, diverging like $\tau \sim (1-M/r_T)^{-3.1}$. 
These results indicate that, although the instability does not seem to completely vanish as long as $M/r_T<1$, the instability timescale $\tau$ can be arbitrarily large for a sufficiently large mass.

\subsection{Supercritical wormholes}

Finally, we discuss the QNMs of the supercritical wormholes. 
Here we mainly explore the $\gamma_1=0$ branch which connects directly to the EB solution. 
Our results indicate that the results are qualitatively similar for other values of $\gamma_1$.
Note that for a fixed value of $\gamma_1$, the larger the value of $\mu$, the closer we approach the limiting eRN black hole.

\subsubsection{Axial modes}

\begin{figure*}[p!]
\begin{center}
\mbox{
\includegraphics[width=1.2\textwidth,angle=-90]{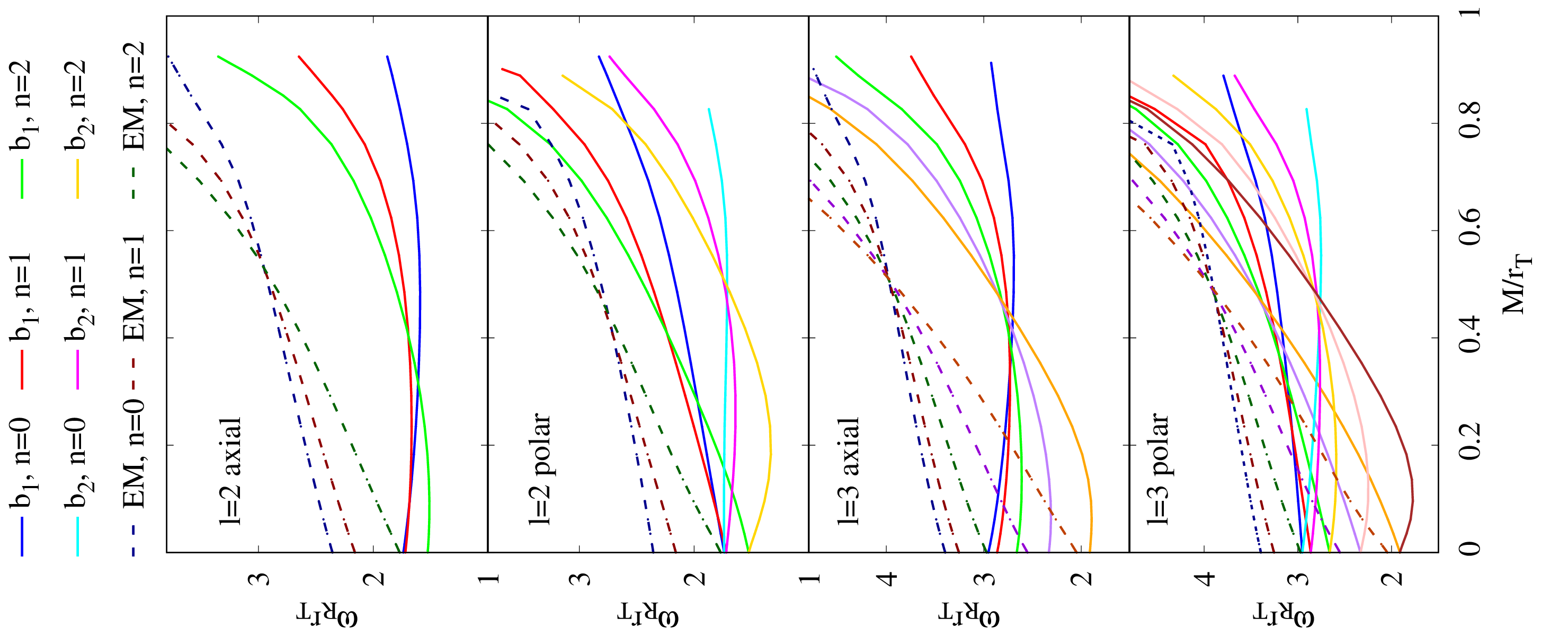}
\includegraphics[width=1.2\textwidth,angle=-90]{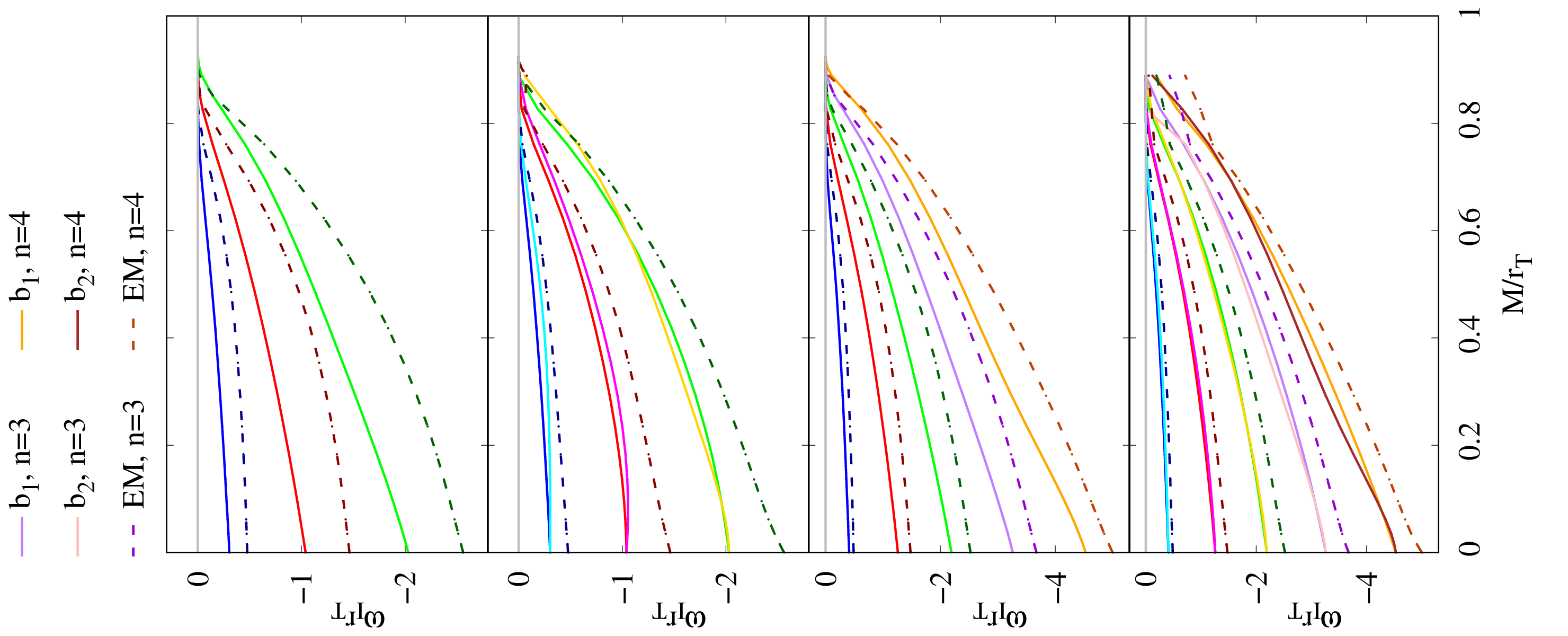}
}
\vspace*{-0.5cm}
\end{center}
\caption{
$l=2$ and $l=3$ axial and polar modes for supercritical wormholes with $\gamma_1=0$.
From top to bottom the rows correspond to: $l=2$ axial, $l=2$ polar, $l=3$ axial, $l=3$ polar modes. 
Left panels show the real part of the QNMs. 
Right panels exhibit the imaginary part. 
The limiting value of $\mu=1$ corresponds to $M/r_T=1$.
}
\label{fig:axial_polar_supercritical_WH}
\end{figure*}

We exhibit in Figure \ref{fig:axial_polar_supercritical_WH} the axial $l=2$ and $l=3$ modes (in rows 1 and 3, respectively) of the supercritical wormholes. 
The real parts $\omega_R$ of the $l=2$ and $l=3$ axial modes exhibit a similar behavior. 
In particular, we note that the EM modes tend to cross each other at a given point (roughly at $M/r_T\approx 0.5$). 
Similarly, the real parts of the gravitational modes also cross each other, but the crossings are not as close to each other as for the EM modes.

The general trend of both $\omega_R$ and $\omega_I$ is to grow with the mass, although for small masses the real part of the gravitational modes first decreases slightly.
Interestingly, the imaginary part $\omega_I$ of all the modes tends to zero for sufficiently high masses.
Thus, all branches accumulate at zero, making it difficult to resolve and decide which ones are the longer-lived.

\begin{figure*}[h!]
\begin{center}
\mbox{
\resizebox{\textwidth}{!}{
    \includegraphics[angle=-90]{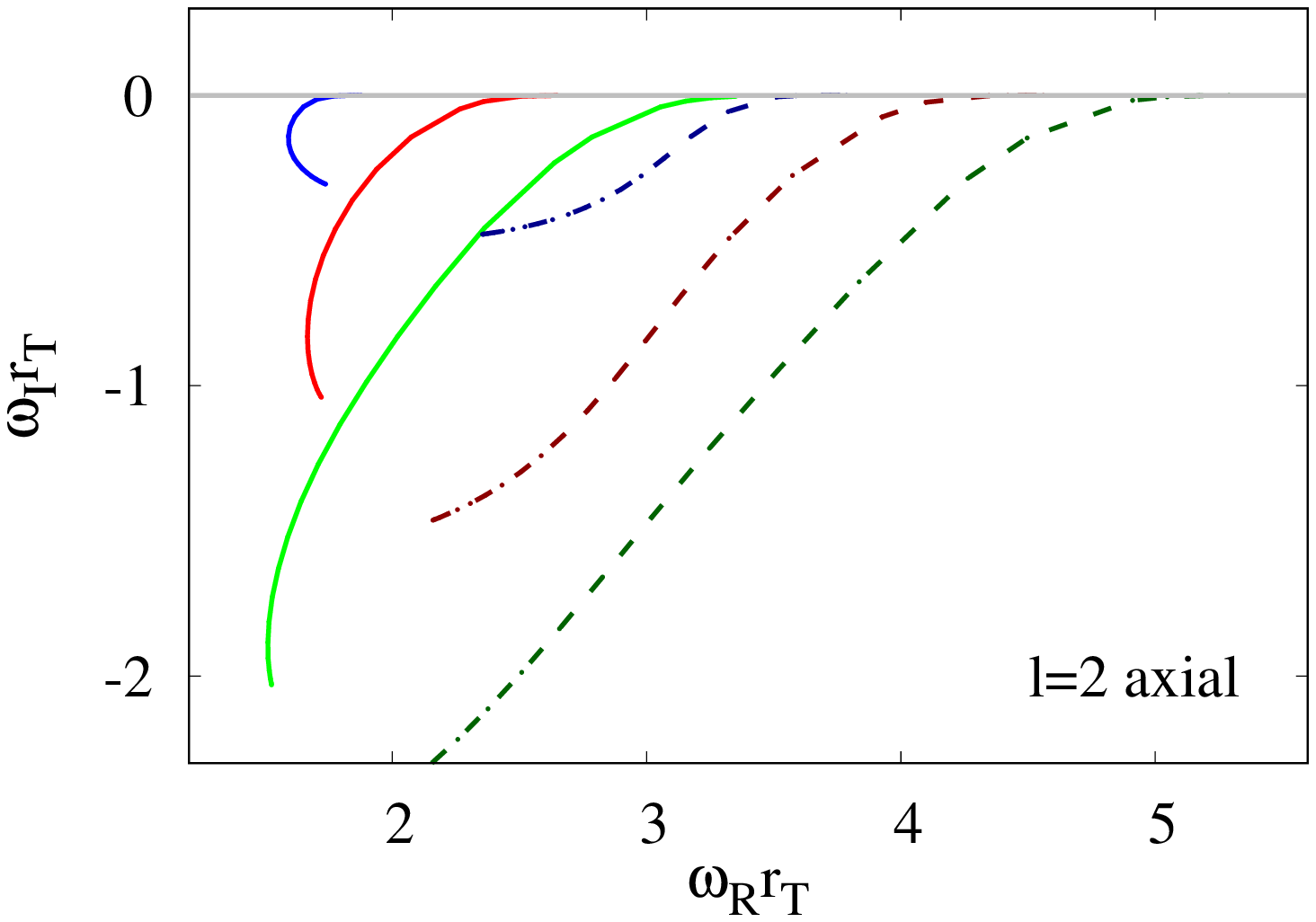}
    \includegraphics[angle=-90]{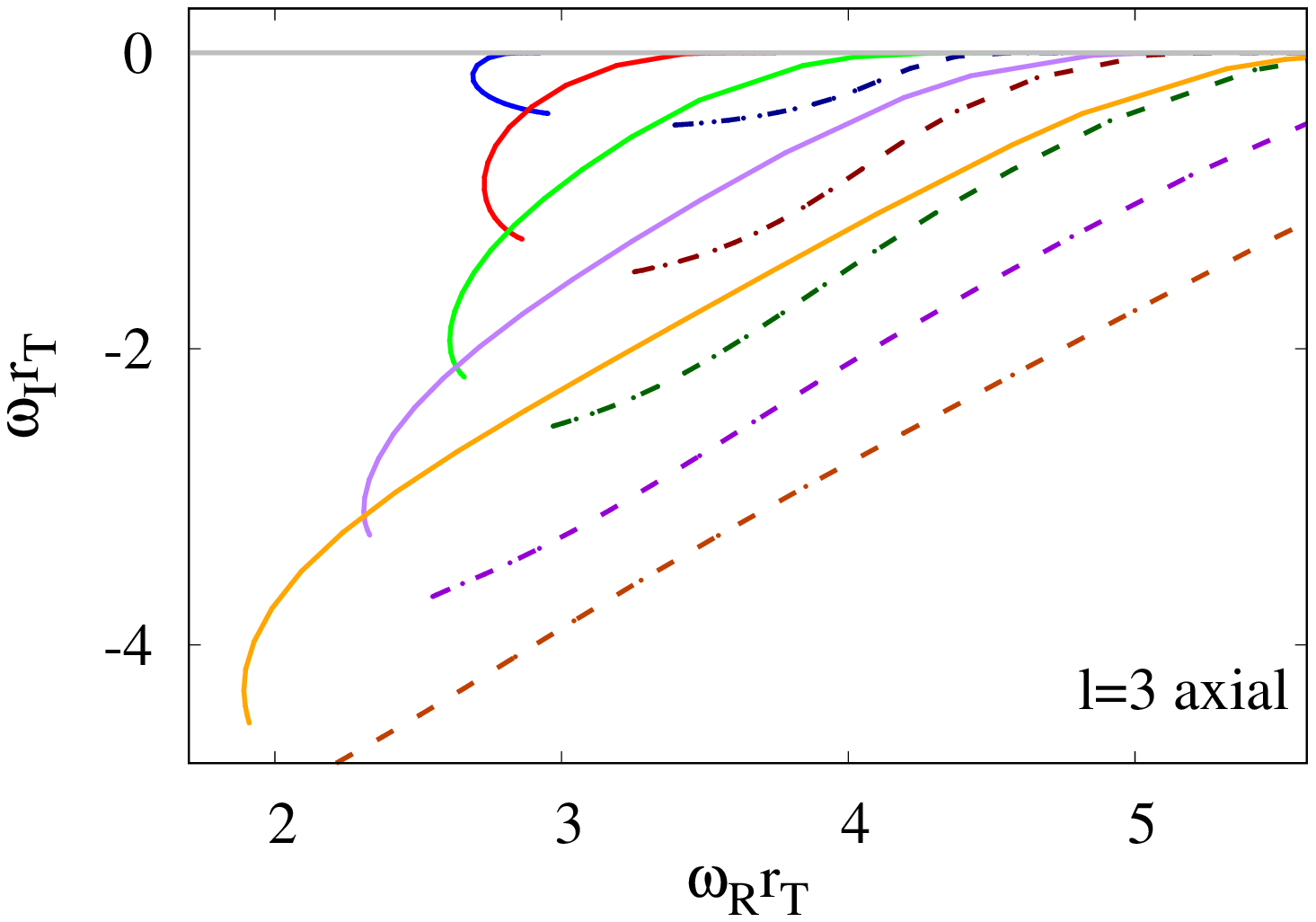}
}
}
\vspace*{-0.7cm}
\end{center}
\caption{
Axial modes for supercritical wormholes.
$\text{Im}(\omega)$ vs $\text{Re}(\omega)$ for $l=2$ (left) and $l=3$ (right).
}
\label{fig:axial_spectrum_supercritical_WH}
\end{figure*}

In Figure \ref{fig:axial_spectrum_supercritical_WH} we show the axial modes in the complex plane.
We note that the gravitational modes twist, while the EM modes have a monotonic behavior, allowing their clear identification.

\subsubsection{Polar modes}

The polar $l=2$ and $l=3$ modes of the supercritical wormholes are shown (in rows 2 and 4, respectively) in Figure \ref{fig:axial_polar_supercritical_WH}.
Here we observe behavior analogous to that seen in the axial spectrum. 
In addition to the $b_1$ branches, the polar sector contains the $b_2$ branches, owing to the presence of both gravitational and scalar perturbations. 

\begin{figure*}[h!]
\begin{center}
\mbox{
\resizebox{\textwidth}{!}{
    \includegraphics[angle=-90]{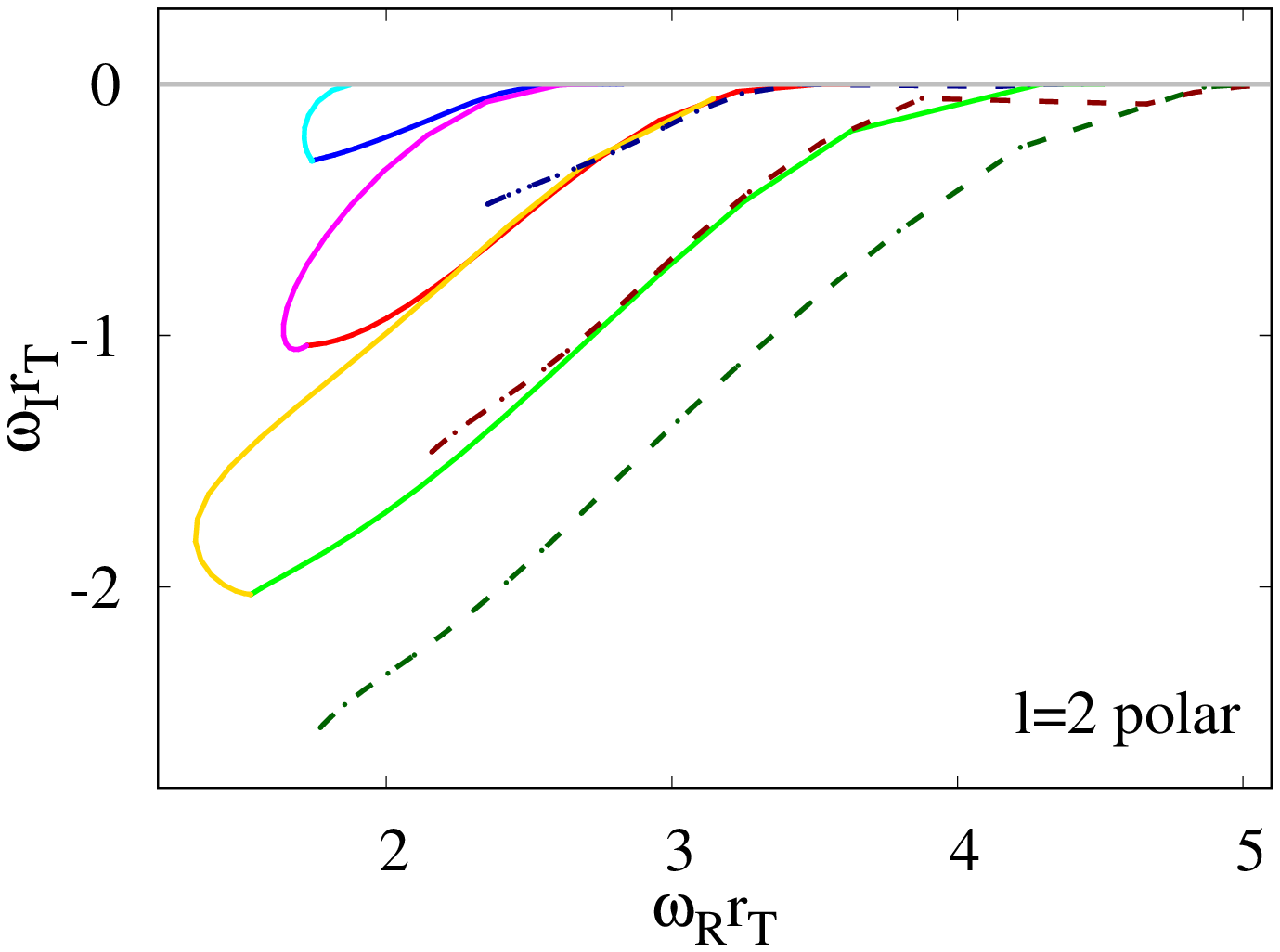}
    \includegraphics[angle=-90]{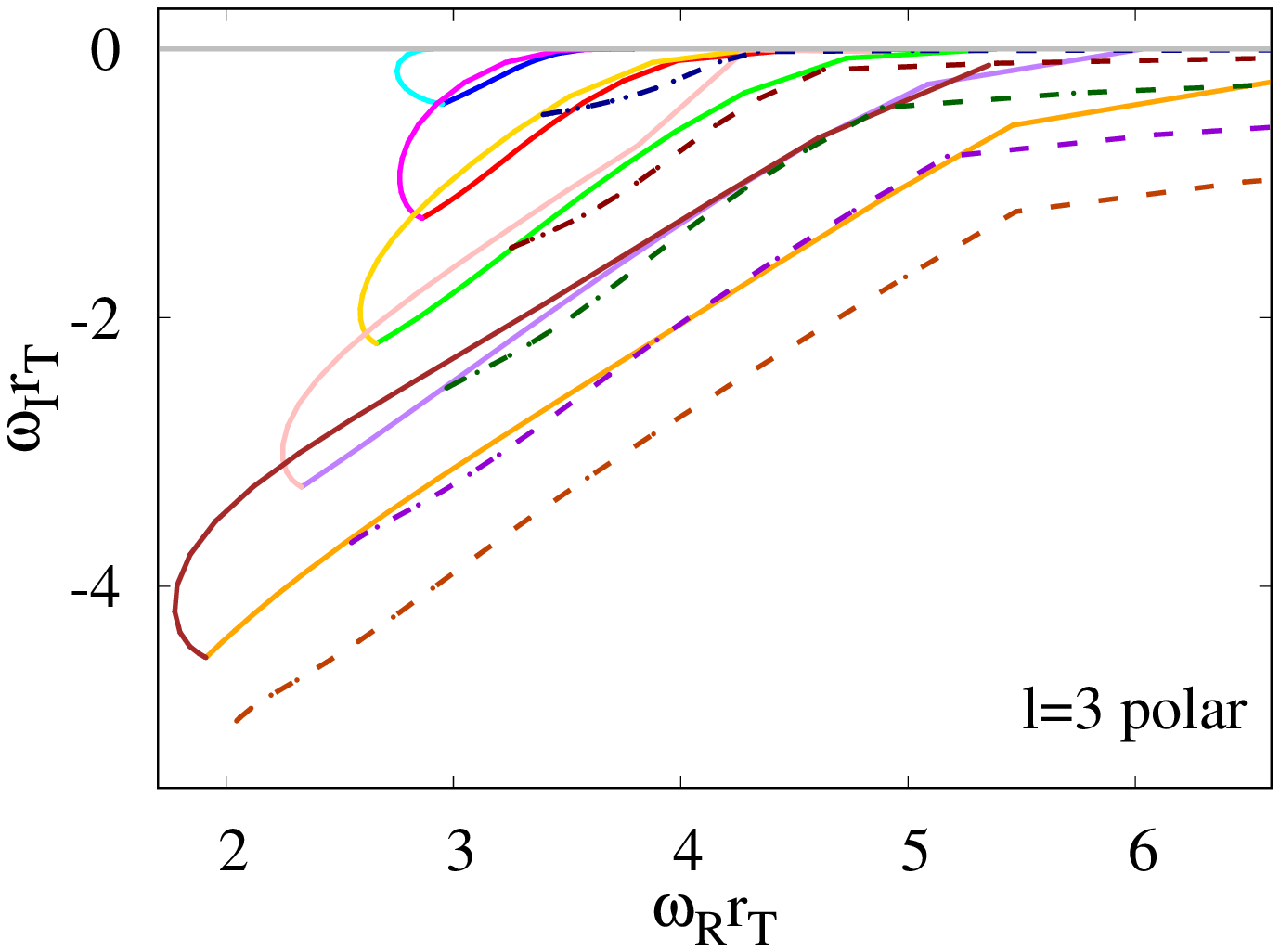}
}
}
\vspace*{-0.7cm}
\end{center}
\caption{
Polar modes for supercritical wormholes.
$\text{Im}(\omega)$ vs $\text{Re}(\omega)$ for $l=2$ (left) and $l=3$ (right).
}
\label{fig:polar_spectrum_supercritical_WH}
\end{figure*}

Again, the real part $\omega_R$ of the QNM branches exhibits crossings near $M/r_T\approx0.45$, that are distinct for the $b_1$ branches, the $b_2$ branches and the EM branches.
At the same time, the imaginary part $\omega_I$ of the QNM branches increases with increasing $M/r_T$, again so that the branches accumulate at zero. 
Also note that the imaginary part $\omega_I$ of the $l=2$ modes does not undergo any noticeable crossings between the different branches, in contrast to the subcritical wormholes. 
Figure \ref{fig:polar_spectrum_supercritical_WH} shows the polar spectrum in the complex plane, revealing behavior analogous to that of the axial spectrum. 
The figure also makes the breaking of isospectrality between the $b_1$ and $b_2$ branches particularly clear. 
Comparison with Figure \ref{fig:axial_spectrum_supercritical_WH} further shows that the axial and polar EM modes cease to be isospectral as the mass increases.

\subsubsection{Radial modes}

\begin{figure}[h!]
\centering
\mbox{ 
\resizebox{\textwidth}{!}{
    \includegraphics[angle=-90]{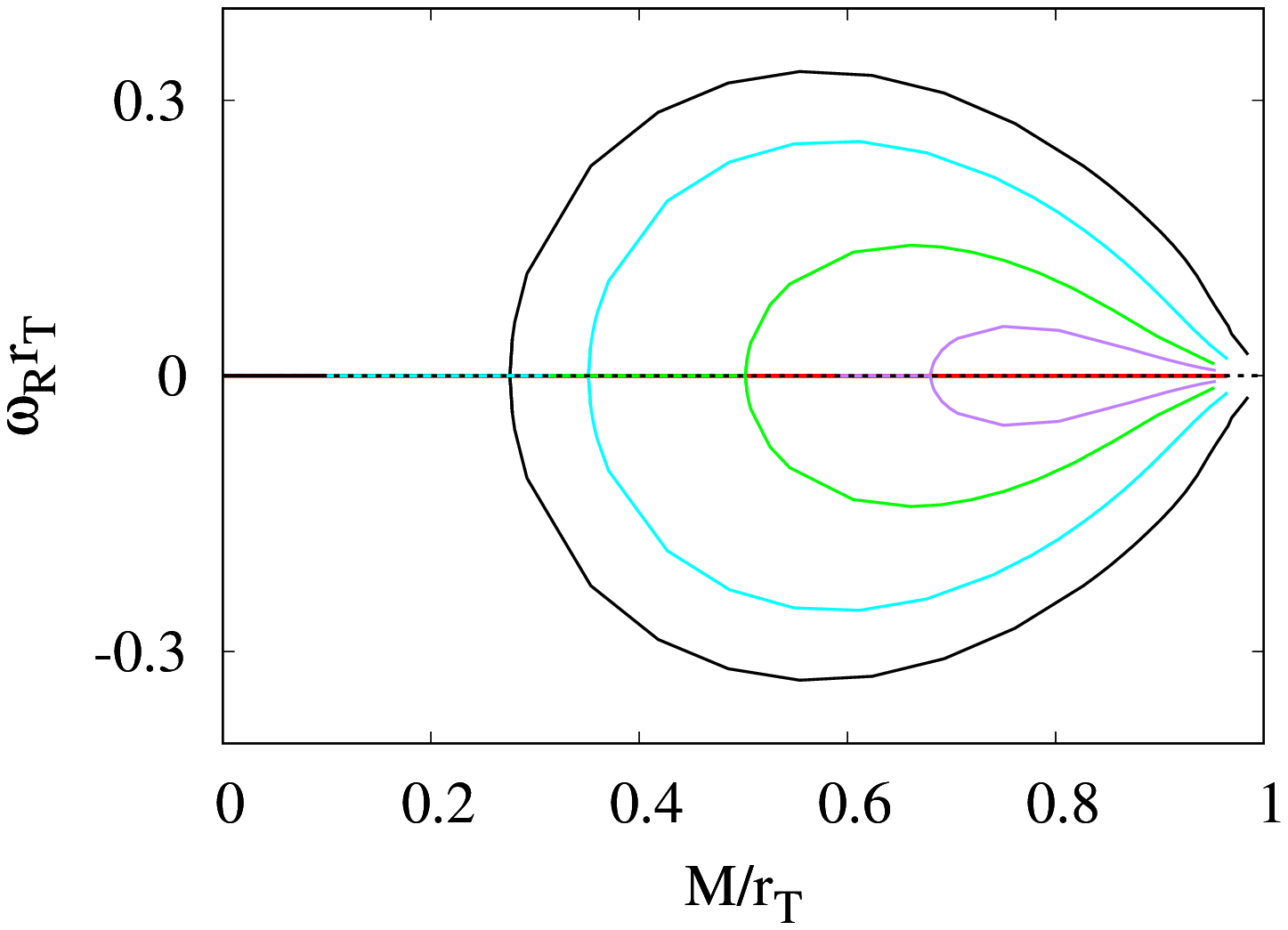}
    \includegraphics[angle=-90]{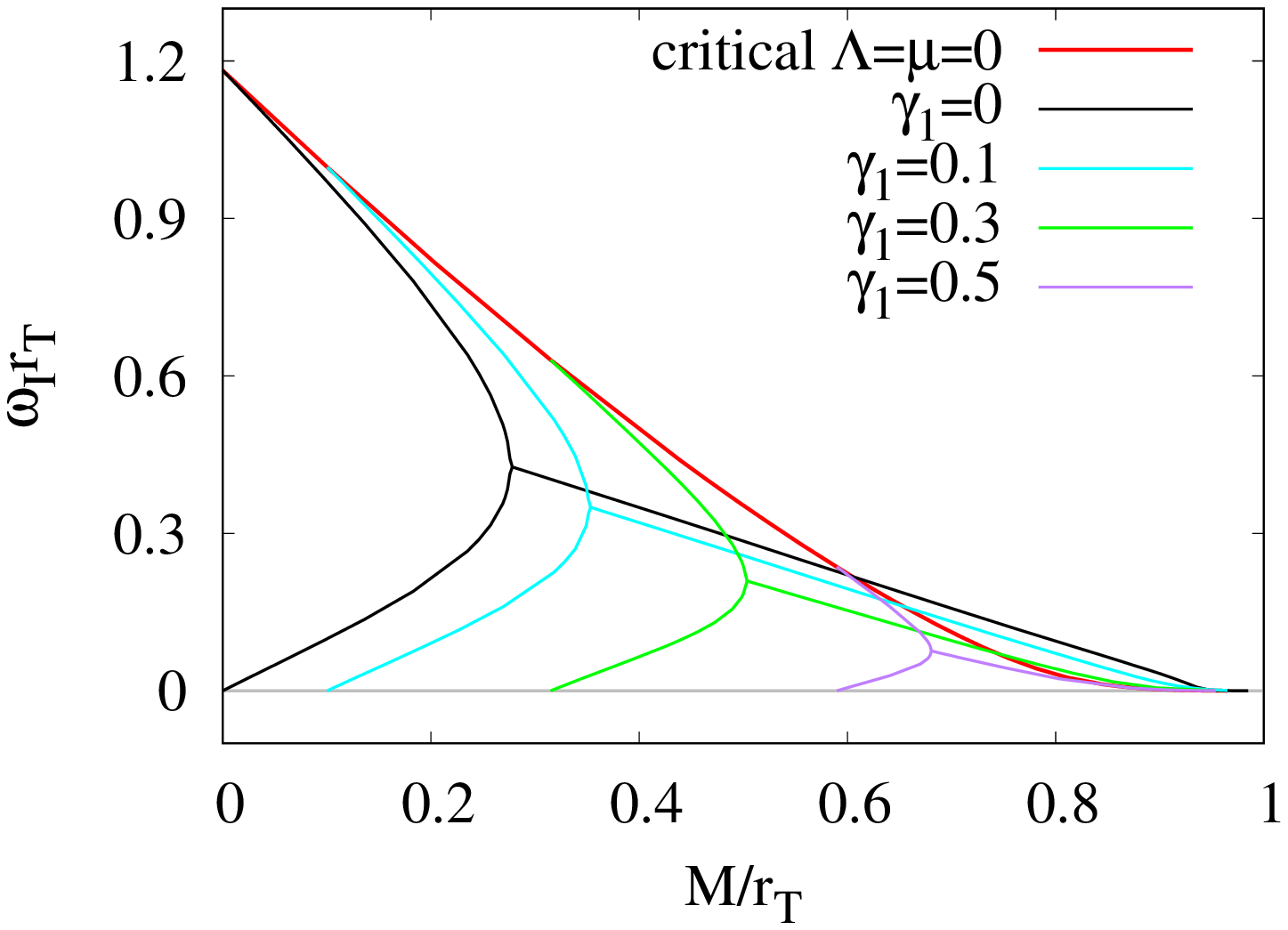}
}}
\caption{Unstable radial branch of supercritical wormholes.}
\label{fig:radial_spectrum_supercritical_WH}
\end{figure}

Finally, we turn to the radial modes of supercritical wormholes, arriving at one of the most interesting results. 
We exhibit in Figure \ref{fig:radial_spectrum_supercritical_WH} the modes of the unstable radial branch of the supercritical wormholes. 
We find that, given a value of $\gamma_1$, there are two unstable branches of purely imaginary modes at small masses that converge and merge with each other at some critical value of $M/r_T$.
After the merger, the branches remain degenerate in $\omega_I$, but acquire a real part $\omega_R$ with opposite signs \cite{Blazquez-Salcedo:2025dit}.
This is an interesting feature that is found neither for EB wormholes nor for critical and subcritical charged wormholes. 
In those cases, there was only a single unstable radial branch.
But as for critical and subcritical wormholes, the imaginary part $\omega_I$ of the modes tends rapidly to zero, when the limiting eRN black hole is approached.

\section{Conclusions}
\label{sec:Conclusions}

We have studied the QNM spectrum of static, spherically symmetric EB wormholes and their charged generalizations.
The latter form three classes, the critical, the subcritical, and the supercritical wormholes.
The EB wormholes have long been known to suffer from a radial instability.
In the presence of charge, this instability relaxes.
In fact, for all three classes, $\omega_I$ tends to zero when the branches approach the limiting eRN black hole solution.
While there is a single purely imaginary radial unstable mode for critical and subcritical wormholes, supercritical wormholes feature two radial unstable modes that merge and subsequently acquire real parts $\omega_R$ with opposite signs \cite{Blazquez-Salcedo:2025dit}.

The main focus of the study has been the $l=2$ and $l=3$ QNMs of these wormholes.
These comprise the axial and polar modes.
In the axial case, there is a single gravitational mode for a given pair $l,n$, where $n$ represents the excitation number.
In the polar case, in contrast, there are two modes since gravitational and scalar perturbations are present.
Since we cannot, in general, associate the modes with either being gravitational or scalar, we have denoted them by $b_1$ and $b_2$.
In addition, we have the EM modes that can be identified more easily.
Employing a numerical spectral scheme, we have obtained  up to four excited modes with good precision.

The general trend of the modes with the mass is to have an increasing imaginary part $\omega_I$ and a decreasing real part $\omega_R$.
However, some modes stand out by revealing a rather strong increase in $\omega_I$, crossing all other modes on their rise toward zero.
These are the fundamental $l=2$ $b_2$ polar mode and the 2nd excited $l=3$ $b_1$ polar mode.
At the same time, their real parts $\omega_R$ decrease strongly toward zero.

The most startling result of our analysis is the presence of a previously unknown instability in the polar sector.
In fact, it arises for the fundamental $l=2$ $b_2$  polar mode beyond a critical mass.
This instability is already present for the EB wormholes, and it is retained for the subcritical wormholes.
For these, the growth rate also appears to approach zero as the limiting eRN black hole is approached.
We did not find $l=2$ unstable modes for the critical solutions,
but it is possible that such an instability appears for large values of the parameters, for which our method loses accuracy.

An interesting behavior seen for the supercritical wormholes is the emergence of crossings of the real parts $\omega_R$ of a given type of mode.
The clustering of these crossings thus points to the identity of a mode as $b_1$, $b_2$ or EM.
At the same time, the imaginary part $\omega_I$ tends to zero for large mass for all the modes of the supercritical wormholes, leading to an accumulation of the modes close to zero.
Thus the supercritical wormholes do not seem to acquire another instability.

This is interesting for the case of rotating wormholes.
We recall that the calculation of the $l=2$ and $l=3$ modes of a set of rapidly rotating wormholes did not show any instabilities either \cite{Khoo:2024yeh}.
Moreover, in four dimensions, the radial instability of slowly rotating wormholes features two unstable modes that merge, analogous to the radial unstable modes of charged supercritical wormholes \cite{Azad:2023iju}
(as well as 5-dimensional rapidly rotating wormholes with equal angular momenta \cite{Dzhunushaliev:2013jja}).
By analogy, we conjecture that the radial instability of 4-dimensional rotating wormholes will therefore be relaxed for fast rotation as well.

There are various future directions to extend the present work.
Of course, the fate of the instability of rapidly rotating wormholes should be clarified.
For the rotating EB wormholes, the numerical techniques need to be developed further to better cope with the numerical background solutions.
But there are various rotating wormhole solutions known in closed form, whose QNM spectrum and (in)stability seem interesting to study \cite{Cisterna:2023uqf,Batic:2026vzt}. 
On the other hand, it will also be interesting to study the quasinormal mode spectrum of wormholes in GR that exist in the presence of Dirac particles without needing to resort to exotic matter \cite{Blazquez-Salcedo:2020czn,Konoplya:2021hsm,Blazquez-Salcedo:2021udn,Dzhunushaliev:2025ntr}.
Likewise, the QNM spectrum and (in)stability of wormholes in generalized theories of gravity represent relevant alternative directions of study, since these wormholes, too, may exist without the need for exotic matter (see e.g.~\cite{Kanti:2011jz,Harko:2013yb,Bakopoulos:2021liw,DeFalco:2023twb,Ilyas:2023rde,Bronnikov:2025zgw}).


\subsection*{Acknowledgments}

We gratefully acknowledge support by MICINN project PID2021-125617NB-I00 ``QuasiMode''.
JLBS gratefully acknowledges support from MICINN project CNS2023-144089 ``Quasinormal modes''.
FSK gratefully acknowledges support from ``Atracci\'on de Talento Investigador Cesar Nombela'' of the Comunidad de Madrid under the grant number 2024-T1/COM-31385.
PNM gratefully acknowledges support from Universidad Complutense de Madrid through ``Contratos predoctorales de personal investigador en formación CT25/24'' and IPARCOS under ``Ayudas de doctorado IPARCOS-UCM/2024''.


\section*{Appendix}

\addcontentsline{toc}{section}{Appendix}

Here, we present in Tables 
\ref{table1}$-$\ref{table12} some of the numerical values of the quasinormal modes obtained in this work. The data in these tables were obtained by setting
$r_0=1$. 
In a few cases, the calculated modes were not sufficiently accurate;
the corresponding cells are therefore marked with a dash.


\begin{table}[h!]
\centering
\renewcommand{\arraystretch}{1.3}
\begin{tabularx}{\textwidth}{|>{\centering\arraybackslash}X|>{\centering\arraybackslash}X|>{\centering\arraybackslash}X|>{\centering\arraybackslash}X|>{\centering\arraybackslash}X|}
\hline
\hline
$\Lambda$ &  $\omega_R^{b_1} r_0$ & $\omega_I^{b_1} r_0$ & $\omega_R^{EM} r_0$ & $\omega_I^{EM} r_0$ \\ \hline\hline
0   & 1.737 & -0.305 & 2.354 & -0.478 \\ \hline
0.1 & 1.295 & -0.232 & 1.755 & -0.356 \\ \hline
0.2 & 1.003 & -0.190 & 1.359 & -0.277 \\ \hline
0.3 & 0.806 & -0.164 & 1.090 & -0.222 \\ \hline
0.4 & 0.669 & -0.145 & 0.900 & -0.184 \\ \hline
0.5 & 0.570 & -0.130 & 0.761 & -0.156 \\ \hline
0.6 & 0.496 & -0.117 & 0.657 & -0.135 \\ \hline
0.7 & 0.439 & -0.105 & 0.577 & -0.119 \\ \hline
0.8 & 0.393 & -0.095 & 0.513 & -0.106 \\ \hline
0.9 & 0.355 & -0.087 & 0.461 & -0.095 \\ \hline
\end{tabularx}
\caption{$l=2$ axial fundamental modes for EB wormholes.}
\label{table1}
\end{table}

\begin{table}[h!]
\centering
\renewcommand{\arraystretch}{1.3}
\begin{tabularx}{\textwidth}{|>{\centering\arraybackslash}X|>{\centering\arraybackslash}X|>{\centering\arraybackslash}X|>{\centering\arraybackslash}X|>{\centering\arraybackslash}X|>{\centering\arraybackslash}X|>{\centering\arraybackslash}X|}
\hline
\hline
$\Lambda$ &  $\omega_R^{b_1} r_0$ & $\omega_I^{b_1} r_0$ &  $\omega_R^{b_2} r_0$ & $\omega_I^{b_2} r_0$ & $\omega_R^{EM} r_0$ & $\omega_I^{EM} r_0$ \\ \hline\hline
0   & 1.738 & -0.305 & 1.738 & -0.305 & 2.355 & -0.478 \\ \hline
0.1 & 1.367 & -0.205 & 1.190 & -0.292 & 1.755 & -0.356 \\ \hline
0.2 & 1.101 & -0.157 & 0.738 & -0.275 & 1.359 & -0.277 \\ \hline
0.3 & 0.907 & -0.131 & 0.454 & -0.181 & 1.090 & -0.222 \\ \hline
0.4 & 0.764 & -0.115 & 0.289 & -0.108 & 0.900 & -0.184 \\ \hline
0.5 & 0.655 & -0.104 & 0.189 & -0.057 & 0.761 & -0.156 \\ \hline
0.6 & 0.571 & -0.097 & 0.126 & -0.023 & 0.657 & -0.135 \\ \hline
0.7 & 0.504 & -0.091 & 0.083 & -0.001 & 0.577 & -0.119 \\ \hline
0.8 & 0.451 & -0.087 & 0.053 & 0.013  & 0.513 & -0.106 \\ \hline
0.9 & 0.407 & -0.084 & 0.029 & 0.022  & 0.461 & -0.095 \\ \hline
\end{tabularx}
\caption{$l=2$ polar fundamental modes for EB wormholes.}
\label{table2}
\end{table}

\begin{table}[h!]
\centering
\renewcommand{\arraystretch}{1.3}
\begin{tabularx}{\textwidth}{|>{\centering\arraybackslash}X|>{\centering\arraybackslash}X|>{\centering\arraybackslash}X|>{\centering\arraybackslash}X|>{\centering\arraybackslash}X|}
\hline
\hline
$\Lambda$ &  $\omega_R^{s} r_0$ & $\omega_I^{s} r_0$ &  $\omega_R^{u} r_0$ & $\omega_I^{u} r_0$ \\ \hline\hline
0   & 0.681 & -0.618 & 0 & 1.182 \\ \hline
0.1 & 0.507 & -0.459 & 0 & 0.882 \\ \hline
0.2 & 0.392 & -0.353 & 0 & 0.686 \\ \hline
0.3 & 0.313 & -0.280 & 0 & 0.552 \\ \hline
0.4 & 0.258 & -0.228 & 0 & 0.458 \\ \hline
0.5 & 0.217 & -0.191 & 0 & 0.390 \\ \hline
0.6 & 0.186 & -0.163 & 0 & 0.338 \\ \hline
0.7 & 0.163 & -0.142 & 0 & 0.298 \\ \hline
0.8 & 0.144 & -0.125 & 0 & 0.266 \\ \hline
0.9 & 0.129 & -0.112 & 0 & 0.239 \\ \hline
1   & 0.117 & -0.101 & 0 & 0.218 \\ \hline
\end{tabularx}
\caption{$l=0$ radial stable (s) and unstable (u) modes for EB wormholes.}
\label{table3}
\end{table}


\begin{table}[h!]
\centering
\renewcommand{\arraystretch}{1.3}
\begin{tabularx}{\textwidth}{|>{\centering\arraybackslash}X|>{\centering\arraybackslash}X|>{\centering\arraybackslash}X|>{\centering\arraybackslash}X|>{\centering\arraybackslash}X|}
\hline
\hline
$\gamma_1$ &  $\omega_R^{b_1} r_0$ & $\omega_I^{b_1} r_0$ & $\omega_R^{EM} r_0$ & $\omega_I^{EM} r_0$ \\ \hline\hline
0 & 1.738 & -0.305 & 2.355 & -0.478 \\ \hline
0.1    & 1.763 & -0.314 & 2.425 & -0.485 \\ \hline
0.15   & 1.797 & -0.326 & 2.516 & -0.496 \\ \hline
0.2    & 1.851 & -0.344 & 2.652 & -0.511 \\ \hline
0.25   & 1.932 & -0.372 & 2.843 & -0.532 \\ \hline
0.3    & 2.054 & -0.414 & 3.107 & -0.561 \\ \hline
0.35   & 2.241 & -0.477 & 3.477 & -0.600 \\ \hline
0.4    & 2.541 & -0.564 & 4.010 & -0.654 \\ \hline
0.45   & 3.036 & -0.672 & 4.831 & -0.737 \\ \hline
0.5    & 3.892 & -0.821 & 6.241 & -0.880 \\ \hline
0.55   & 5.708 & -1.140 & 9.241 & -1.204 \\ \hline
\end{tabularx}
\caption{$l=2$ axial fundamental modes for critical wormholes.}
\label{table4}
\end{table}

\begin{table}[h!]
\centering
\renewcommand{\arraystretch}{1.3}
\begin{tabularx}{\textwidth}{|>{\centering\arraybackslash}X|>{\centering\arraybackslash}X|>{\centering\arraybackslash}X|>{\centering\arraybackslash}X|>{\centering\arraybackslash}X|>{\centering\arraybackslash}X|>{\centering\arraybackslash}X|}
\hline
\hline
$\gamma_1$ &  $\omega_R^{b_1} r_0$ & $\omega_I^{b_1} r_0$ &  $\omega_R^{b_2} r_0$ & $\omega_I^{b_2} r_0$ & $\omega_R^{EM} r_0$ & $\omega_I^{EM} r_0$ \\ \hline\hline
0    & 1.738 & -0.305 & 1.738 & -0.305 & 2.355 & -0.478 \\ \hline
0.1  & 1.879 & -0.282 & 1.641 & -0.396 & 2.416 & -0.482 \\ \hline
0.15 & 1.976 & -0.291 & 1.573 & -0.490 & 2.495 & -0.487 \\ \hline
0.2  & 2.098 & -0.319 & 1.438 & -0.550 & 2.615 & -0.495 \\ \hline
0.25 & 2.251 & -0.377 & 1.327 & -0.560 & 2.787 & -0.506 \\ \hline
0.3  & 2.450 & -0.537 & 1.239 & -0.560 & 3.028 & -0.523 \\ \hline
0.35 & 2.923 & -0.659 & 1.165 & -0.558 & 3.373 & -0.549 \\ \hline
0.4  & 3.481 & -0.737 & 1.101 & -0.554 & 3.884 & -0.597 \\ \hline
0.45 & 4.276 & -0.823 & 1.045 & -0.551 & 4.689 & -0.687 \\ \hline
0.5  & 5.569 & -0.949 & 0.995 & -0.549 & 6.108 & -0.857 \\ \hline
0.55 & 8.242 & -1.259 & 0.950 & -0.547 & 9.151 & -1.206 \\ \hline
\end{tabularx}
\caption{$l=2$ polar fundamental modes for critical wormholes.}
\label{table5}
\end{table}

\begin{table}[h!]
\centering
\renewcommand{\arraystretch}{1.3}
\begin{tabularx}{\textwidth}{|>{\centering\arraybackslash}X|>{\centering\arraybackslash}X|>{\centering\arraybackslash}X|>{\centering\arraybackslash}X|>{\centering\arraybackslash}X|}
\hline
\hline
$\gamma_1$ &  $\omega_R^{s} r_0$ & $\omega_I^{s} r_0$ &  $\omega_R^{u} r_0$ & $\omega_I^{u} r_0$ \\ \hline\hline
0   & 0.681 & -0.618 & 0 & 1.182 \\ \hline
0.1 & 0.694 & -0.622 & 0 & 1.190 \\ \hline
0.2 & 0.737 & -0.635 & 0 & 1.213 \\ \hline
0.3 & 0.825 & -0.663 & 0 & 1.251 \\ \hline
0.4 & 1.003 & -0.720 & 0 & 1.298 \\ \hline
0.5 & 1.423 & -0.886 & 0 & 1.297 \\ \hline
0.6 & -     & -      & 0 & 0.587 \\ \hline
0.63 & -     & -      & 0 & 0.082 \\ \hline
\end{tabularx}
\caption{$l=0$ radial stable (s) and unstable (u) modes for critical wormholes.}
\label{table6}
\end{table}


\begin{table}[h!]
\centering
\renewcommand{\arraystretch}{1.3}
\begin{tabularx}{\textwidth}{|>{\centering\arraybackslash}X|>{\centering\arraybackslash}X|>{\centering\arraybackslash}X|>{\centering\arraybackslash}X|>{\centering\arraybackslash}X|}
\hline
\hline
$\gamma_1$ &  $\omega_R^{b_1} r_0$ & $\omega_I^{b_1} r_0$ & $\omega_R^{EM} r_0$ & $\omega_I^{EM} r_0$ \\ \hline\hline
0.2 & 1.357  & -0.254 & 1.918 & -0.375 \\ \hline
0.3 & 1.499 & -0.302 & 2.244 & -0.410 \\ \hline
0.4 & 1.836 & -0.407 & 2.883 & -0.477 \\ \hline
0.5 & 2.766 & -0.589 & 4.425 & -0.633 \\ \hline
\end{tabularx}
\caption{$l=2$ axial fundamental modes for subcritical wormholes ($\Lambda=0.1$).}
\label{table7}
\end{table}

\begin{table}[h!]
\centering
\renewcommand{\arraystretch}{1.3}
\begin{tabularx}{\textwidth}{|>{\centering\arraybackslash}X|>{\centering\arraybackslash}X|>{\centering\arraybackslash}X|>{\centering\arraybackslash}X|>{\centering\arraybackslash}X|>{\centering\arraybackslash}X|>{\centering\arraybackslash}X|}
\hline
\hline
$\gamma_1$ &  $\omega_R^{b_1} r_0$ & $\omega_I^{b_1} r_0$ &  $\omega_R^{b_2} r_0$ & $\omega_I^{b_2} r_0$ & $\omega_R^{EM} r_0$ & $\omega_I^{EM} r_0$ \\ \hline\hline
0.2 & 1.526 & -0.228 & 1.040 & -0.396 & 1.898 & -0.365 \\ \hline
0.3 & 1.774 & -0.354 & 0.899 & -0.401 & 2.191 & -0.385 \\ \hline
0.4 & 1.993 & -0.430 & 0.800 & -0.397 & 2.792 & -0.435 \\ \hline
0.5 & 2.834 & -0.558 & 0.725 & -0.392 & 4.324 & -0.612 \\ \hline
\end{tabularx}
\caption{$l=2$ polar fundamental modes for subcritical wormholes ($\Lambda=0.1$).}
\label{table8}
\end{table}

\begin{table}[h!]
\centering
\renewcommand{\arraystretch}{1.3}
\begin{tabularx}{\textwidth}{|>{\centering\arraybackslash}X|>{\centering\arraybackslash}X|>{\centering\arraybackslash}X|}
\hline
\hline
$\gamma_1$ &  $\omega_R r_0$ & $\omega_I r_0$ \\ \hline\hline
0.80 & 0.00848 & 0.06695 \\ \hline
0.81 & 0.00875 & 0.06691 \\ \hline
0.82 & 0.00902 & 0.06688 \\ \hline
0.83 & 0.00930 & 0.06686 \\ \hline
0.84 & 0.00958 & 0.06685 \\ \hline
0.85  & 0.00987 & 0.06685 \\ \hline
0.86 & 0.01016 & 0.06687 \\ \hline
0.87 & 0.01047 & 0.06689 \\ \hline
0.88 & 0.01077 & 0.06693 \\ \hline
0.89 & 0.01109 & 0.06699 \\ \hline
0.90 & 0.01141 & 0.06705 \\ \hline
\end{tabularx}
\caption{$l=2$ polar unstable branch for subcritical wormholes ($\Lambda=0.75$).}
\label{table9}
\end{table}



\begin{table}[h!]
\centering
\renewcommand{\arraystretch}{1.3}
\begin{tabularx}{\textwidth}{|>{\centering\arraybackslash}X|>{\centering\arraybackslash}X|>{\centering\arraybackslash}X|>{\centering\arraybackslash}X|>{\centering\arraybackslash}X|}
\hline
\hline
$\mu$ &  $\omega_R^{b_1} r_0$ & $\omega_I^{b_1} r_0$ & $\omega_R^{EM} r_0$ & $\omega_I^{EM} r_0$ \\ \hline\hline
0 & 1.738 & -0.305 & 2.355 & -0.478 \\ \hline
0.1 & 1.706 & -0.297 & 2.346 & -0.469 \\ \hline
0.2 & 1.617 & -0.274 & 2.315 & -0.444 \\ \hline
0.3 & 1.483 & -0.237 & 2.246 & -0.403 \\ \hline
0.4 & 1.318 & -0.191 & 2.128 & -0.346 \\ \hline
0.5 & 1.132 & -0.137 & 1.949 & -0.274 \\ \hline
0.6 & 0.935 & -0.082 & 1.704 & -0.190 \\ \hline
0.7 & 0.734 & -0.033 & 1.392 & -0.099 \\ \hline
0.8 & 0.526 & -0.004 & 1.026 & -0.017 \\ \hline
0.9 & 0.287 & -6.14$\cdot10^{-6}$ & 0.577 & -2.17$\cdot10^{-6}$ \\ \hline
\end{tabularx}
\caption{$l=2$ axial fundamental modes for supercritical wormholes ($\gamma_1=0$).}
\label{table10}
\end{table}

\begin{table}[h!]
\centering
\renewcommand{\arraystretch}{1.3}
\begin{tabularx}{\textwidth}{|>{\centering\arraybackslash}X|>{\centering\arraybackslash}X|>{\centering\arraybackslash}X|>{\centering\arraybackslash}X|>{\centering\arraybackslash}X|>{\centering\arraybackslash}X|>{\centering\arraybackslash}X|}
\hline
\hline
$\mu$ &  $\omega_R^{b_1} r_0$ & $\omega_I^{b_1} r_0$ &  $\omega_R^{b_2} r_0$ & $\omega_I^{b_2} r_0$ & $\omega_R^{EM} r_0$ & $\omega_I^{EM} r_0$ \\ \hline\hline
0 & 1.738 & -0.305 & 1.738 & -0.305 & 2.355 & -0.478 \\ \hline
0.1 & 1.730 & -0.298 & 1.717 & -0.302 & 2.338 & -0.466 \\ \hline
0.2 & 1.703 & -0.276 & 1.655 & -0.291 & 2.285 & -0.431 \\ \hline
0.3 & 1.650 & -0.240 & 1.550 & -0.271 & 2.194 & -0.379 \\ \hline
0.4 & 1.563 & -0.194 & 1.403 & -0.238 & 2.060 & -0.313 \\ \hline
0.5 & 1.435 & -0.140 & 1.218 & -0.190 & 1.878 & -0.239 \\ \hline
0.6 & 1.262 & -0.084 & 1.006 & -0.126 & 1.641 & -0.160 \\ \hline
0.7 & 1.044 & -0.033 & 0.784 & -0.056 & 1.348 & -0.080 \\ \hline
0.8 & 0.777 & -0.003 & 0.559 & -0.008 & 1.003 & -0.012 \\ \hline
0.9 & 0.432 & 3.47$\cdot10^{-6}$  & -   & -      & 0.649 & -0.002 \\ \hline
\end{tabularx}
\caption{$l=2$ polar fundamental modes for supercritical wormholes ($\gamma_1=0$).}
\label{table11}
\end{table}

\begin{table}[h!]
\centering
\renewcommand{\arraystretch}{1.3}
\begin{tabularx}{\textwidth}{|>{\centering\arraybackslash}X|>{\centering\arraybackslash}X|>{\centering\arraybackslash}X|>{\centering\arraybackslash}X|>{\centering\arraybackslash}X|}
\hline
\hline
$\mu$ &  $\omega_R^{1} r_0$ &  $\omega_R^{2} r_0$ & $\omega_I^{1} r_0$ & $\omega_I^{2} r_0$ \\ \hline\hline
0     & \multicolumn{1}{c|}{0} & \multicolumn{1}{c|}{0} & 1.182 & 0     \\ \hline
0.1 & \multicolumn{1}{c|}{0} & \multicolumn{1}{c|}{0} & 1.150 & 0.016 \\ \hline
0.2 & \multicolumn{1}{c|}{0} & \multicolumn{1}{c|}{0} & 1.053 & 0.061 \\ \hline
0.3 & \multicolumn{1}{c|}{0} & \multicolumn{1}{c|}{0} & 0.891 & 0.136 \\ \hline
0.4 & \multicolumn{1}{c|}{0} & \multicolumn{1}{c|}{0} & 0.640 & 0.266 \\ \hline
0.42 & \multicolumn{1}{c|}{0} & \multicolumn{1}{c|}{0} & 0.563 & 0.316 \\ \hline
0.44 & \multicolumn{1}{c|}{0.059}   & \multicolumn{1}{c|}{-0.059}     & \multicolumn{1}{c|}{0.425}   & \multicolumn{1}{c|}{0.425}     \\ \hline
0.45 & \multicolumn{1}{c|}{0.111}  & \multicolumn{1}{c|}{-0.111}     & \multicolumn{1}{c|}{0.418}    & \multicolumn{1}{c|}{0.418}     \\ \hline
0.5 & \multicolumn{1}{c|}{0.228}  & \multicolumn{1}{c|}{-0.228}     & \multicolumn{1}{c|}{0.379}     & \multicolumn{1}{c|}{0.379}     \\ \hline
0.6 & \multicolumn{1}{c|}{0.319}  & \multicolumn{1}{c|}{-0.319}     & \multicolumn{1}{c|}{0.294}      & \multicolumn{1}{c|}{0.294}     \\ \hline
0.7 & \multicolumn{1}{c|}{0.327}   & \multicolumn{1}{c|}{-0.327}     & \multicolumn{1}{c|}{0.206}     & \multicolumn{1}{c|}{0.206}     \\ \hline
0.8 & \multicolumn{1}{c|}{0.275}   & \multicolumn{1}{c|}{-0.275}     & \multicolumn{1}{c|}{0.119}   & \multicolumn{1}{c|}{0.119}     \\ \hline
0.9 & \multicolumn{1}{c|}{0.170}   & \multicolumn{1}{c|}{-0.170}     & \multicolumn{1}{c|}{0.040}  & \multicolumn{1}{c|}{0.040}     \\ \hline
0.95 & \multicolumn{1}{c|}{0.087}  & \multicolumn{1}{c|}{-0.087}     & \multicolumn{1}{c|}{0.002}    & \multicolumn{1}{c|}{0.002}     \\ \hline
\end{tabularx}
\caption{$l=0$ radial unstable branches for supercritical wormholes ($\gamma_1=0$).}
\label{table12}
\end{table}

\end{document}